\documentclass[acmsmall,screen]{acmart} 
\AtBeginDocument{%
  \providecommand\BibTeX{{%
    \normalfont B\kern-0.5em{\scshape i\kern-0.25em b}\kern-0.8em\TeX}}}

\setcopyright{acmcopyright}
\copyrightyear{2022}
\acmYear{2022}
\acmDOI{10.1145/1122445.1122456}

\acmJournal{JACM}
\acmVolume{37}
\acmNumber{4}
\acmArticle{1}
\acmMonth{1}

\usepackage[utf8]{inputenc}
\usepackage{caption,subcaption}
\usepackage{microtype}
\usepackage{graphics}
\usepackage{graphicx}
\usepackage[linesnumbered,ruled,lined,boxed]{algorithm2e}
\usepackage{booktabs}
\usepackage{geometry}
\usepackage{threeparttable}
\usepackage[marginal]{footmisc}
\usepackage{tcolorbox}
\usepackage{multirow}
\usepackage{color, colortbl}
\definecolor{darkgreen}{rgb}{0,0.5,0}
\definecolor{purple}{rgb}{1,0,1}
\definecolor{todocolor}{rgb}{0.9,0.1,0.1}
\definecolor{hycolor}{rgb}{0.7,0.7,0.3}
\definecolor{fixcolor}{rgb}{0.1,0.7,0.3}
\definecolor{Gray}{gray}{0.9}
\definecolor{LightGray}{rgb}{0.9,0.9,0.9}

\newcount\DraftStatus  
\usepackage{color}
\definecolor{darkgreen}{rgb}{0,0.5,0}
\definecolor{purple}{rgb}{1,0,1}
\definecolor{todocolor}{rgb}{0.9,0.1,0.1}
\definecolor{hycolor}{rgb}{0.7,0.7,0.3}
\definecolor{fixcolor}{rgb}{0.1,0.7,0.3}

\newcommand{\draftnote}[2]{\ifnum\DraftStatus=1
	\marginpar{
		\tiny\raggedright
		\hbadness=10000
		\def\baselinestretch{0.8}
		\textcolor{#1}{\textsf{\hspace{0pt}#2}}}
	\fi}

\usepackage{etoolbox}
\newcommand{\algorithmicdoinparallel}{\textbf{do in parallel}}
\makeatletter
\AtBeginEnvironment{algorithmic}{%
  \newcommand{\FORALLP}[2][default]{\ALC@it\algorithmicforall\ #2\ %
    \algorithmicdoinparallel\ALC@com{#1}\begin{ALC@for}}%
}
\makeatother

\begin{document}

\title[]{Reinforced MOOCs Concept Recommendation in Heterogeneous Information Networks}


\author{Jibing Gong}
\email{gongjibing@ysu.edu.cn}
\affiliation{%
  \department{School of Information Science and Engineering, the Key Lab for Computer Virtual Technology and System Integration, and the Key Laboratory for Software Engineering of Hebei Province}
  \institution{Yanshan University}
  \streetaddress{No. 438 West Hebei Avenue}
  \city{Qinhuangdao}
  \state{Hebei}
  \country{China}
  \postcode{066004}
}

\author{Yao Wan}
\email{wanyao@hust.edu.cn}
\authornote{Corresponding authors}
\affiliation{%
  \department{School of Computer Science and Technology}
  \institution{Huazhong University of Science and Technology}
  \streetaddress{Luoyu Road 1037}
  \city{Wuhan}
  \state{Hubei}
  \country{China}
  \postcode{430074}
}

\author{Ye Liu}
\email{yeliu@salesforce.com}
\affiliation{%
  \institution{Salesforce Research}
  \streetaddress{556 Xixi Road}
   \city{Chicago}
  \country{USA}}
  \postcode{430074}

\author{Xuewen Li}
\email{lixuewen@stumail.ysu.edu.cn}
\affiliation{%
  \department{School of Information Science and Engineering, the Key Lab for Computer Virtual Technology and System Integration, and the Key Laboratory for Software Engineering of Hebei Province}
  \institution{Yanshan University}
    \streetaddress{No. 438 West Hebei Avenue}
  \city{Qinhuangdao}
  \state{Hebei}
  \country{China}
   \postcode{066004}
}
\author{Yi Zhao}
\email{zhaoyi.hb@gmail.com}
\affiliation{%
  \department{School of Information Science and Engineering, the Key Lab for Computer Virtual Technology and System Integration, and the Key Laboratory for Software Engineering of Hebei Province}
  \institution{Yanshan University}
  \streetaddress{No. 438 West Hebei Avenue}
  \city{Qinhuangdao}
  \state{Hebei}
  \country{China}
   \postcode{066004}
}
\author{Cheng Wang}
\email{862741851@qq.com}
\affiliation{%
  \department{School of Information Science and Engineering, the Key Lab for Computer Virtual Technology and System Integration, and the Key Laboratory for Software Engineering of Hebei Province}
  \institution{Yanshan University}
  \streetaddress{No. 438 West Hebei Avenue}
  \city{Qinhuangdao}
  \state{Hubei}
  \country{China}
   \postcode{066004}
}
\author{Yuting Lin}
\email{yuting_lin2022@163.com}
\affiliation{%
  \department{School of Information Science and Engineering, the Key Lab for Computer Virtual Technology and System Integration, and the Key Laboratory for Software Engineering of Hebei Province}
  \institution{Yanshan University}
  \streetaddress{No. 438 West Hebei Avenue}
  \city{Qinhuangdao}
  \state{Hebei}
  \country{China}
   \postcode{066004}
}
\author{Xiaohan Fang}
\email{xiaohan_fang_ysu@163.com}
\affiliation{%
  \department{School of Information Science and Engineering, the Key Lab for Computer Virtual Technology and System Integration, and the Key Laboratory for Software Engineering of Hebei Province}
  \institution{Yanshan University}
  \streetaddress{No. 438 West Hebei Avenue}
  \city{Qinhuangdao}
  \state{Hebei}
  \country{China}
   \postcode{066004}
}

\author{Wenzheng Feng}
\email{fwz17@mails.tsinghua.edu.cn}
\affiliation{%
	\department{Department of Computer Science}
	\institution{Tsinghua University}
	\city{Beijing}
	\country{China}
    \postcode{100084}
}

\author{Jingyi Zhang}
\email{zhangjy@buaa.edu.cn}
\affiliation{%
	\department{School of Cyber Science and Technology}
	\institution{Beihang University}
	\streetaddress{37 Xueyuan Road}
	\city{Beijing}
	\country{China}
	\postcode{100191}
}

\author{Jie Tang}
\email{jietang@tsinghua.edu.cn}
\authornotemark[1]
\affiliation{%
	\department{Department of Computer Science}
	\institution{Tsinghua University}
	\city{Beijing}
	\country{China}
	\postcode{100084}
}

\renewcommand{\shortauthors}{Gong et al.}

\begin{abstract}
Massive open online courses (MOOCs), which offer open access and widespread interactive participation through the internet, are quickly becoming the preferred method for online and remote learning.
Several MOOC platforms offer the service of course recommendation to users, to improve the learning experience of users. 
Despite the usefulness of this service, we consider that recommending courses to users directly may neglect their varying degrees of expertise.
To mitigate this gap, we examine an interesting problem of concept recommendation in this paper, which can be viewed as recommending knowledge to users in a fine-grained way.
We put forward a novel approach, termed HinCRec-RL, for \underline{C}oncept \underline{Rec}ommendation in MOOCs, which is based on \underline{H}eterogeneous \underline{I}nformation \underline{N}etworks and \underline{R}einforcement \underline{L}earning.
In particular, we propose to shape the problem of concept recommendation within a reinforcement learning framework to characterize the dynamic interaction between users and knowledge concepts in MOOCs.
Furthermore, we propose to form the interactions among users, courses, videos, and concepts into a heterogeneous information network (HIN) to learn the semantic user representations better.
We then employ an attentional graph neural network to represent the users in the HIN, based on meta-paths.
Extensive experiments are conducted on a real-world dataset collected from a Chinese MOOC platform, \textit{XuetangX}, to validate the efficacy of our proposed HinCRec-RL.
Experimental results and analysis demonstrate that our proposed HinCRec-RL performs well when compared with several state-of-the-art models.
\end{abstract}



\keywords{HIN, reinforcement learning, MOOCs, concept recommendation}

\maketitle

\section{Introduction}\label{sec:introduction}

\begin{figure*}[t!]
	\centering
	\begin{subfigure}[b]{0.4\textwidth}
		\includegraphics[width=\textwidth]{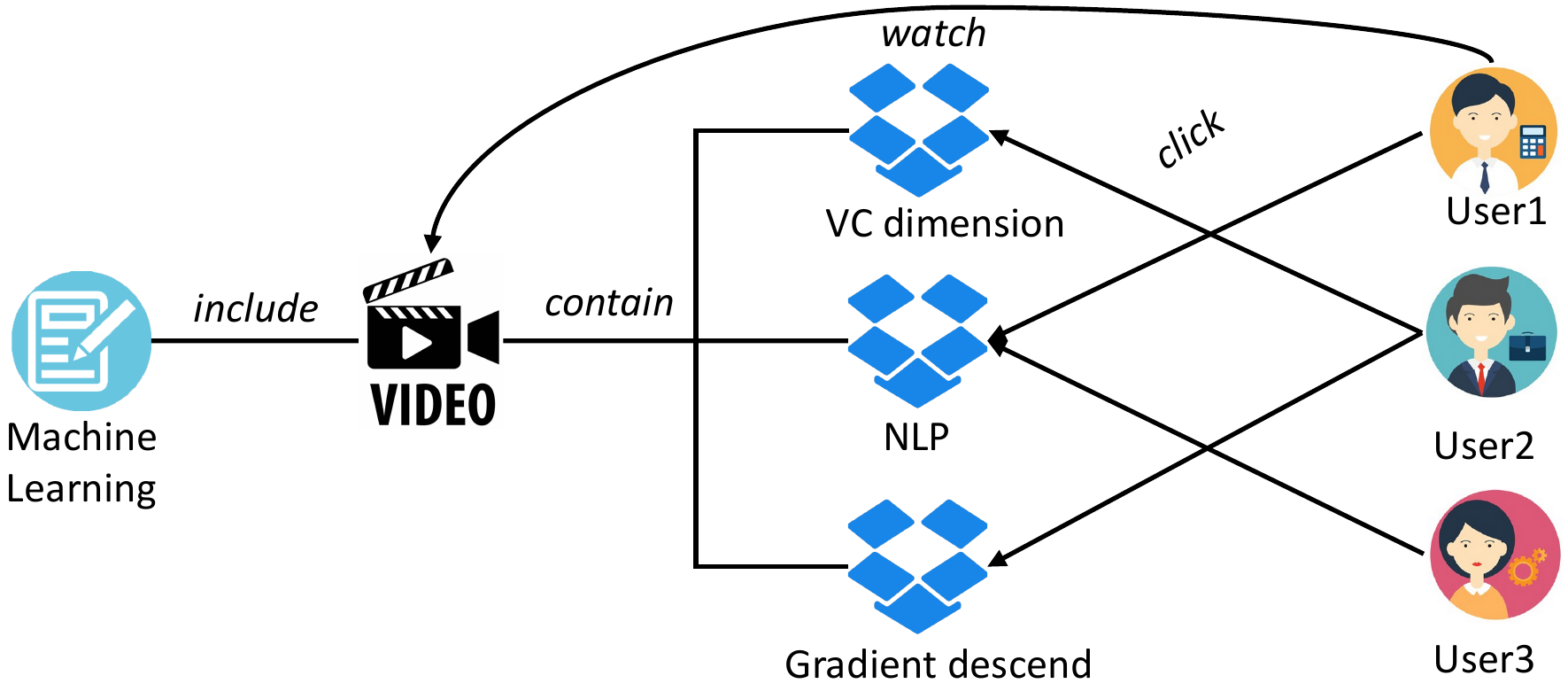}
		\caption{The heterogeneous network in MOOCs} 
		\label{fig:side:a}
	\end{subfigure}
	\hfill
	\begin{subfigure}[b]{0.28\textwidth}
		\includegraphics[width=\textwidth]{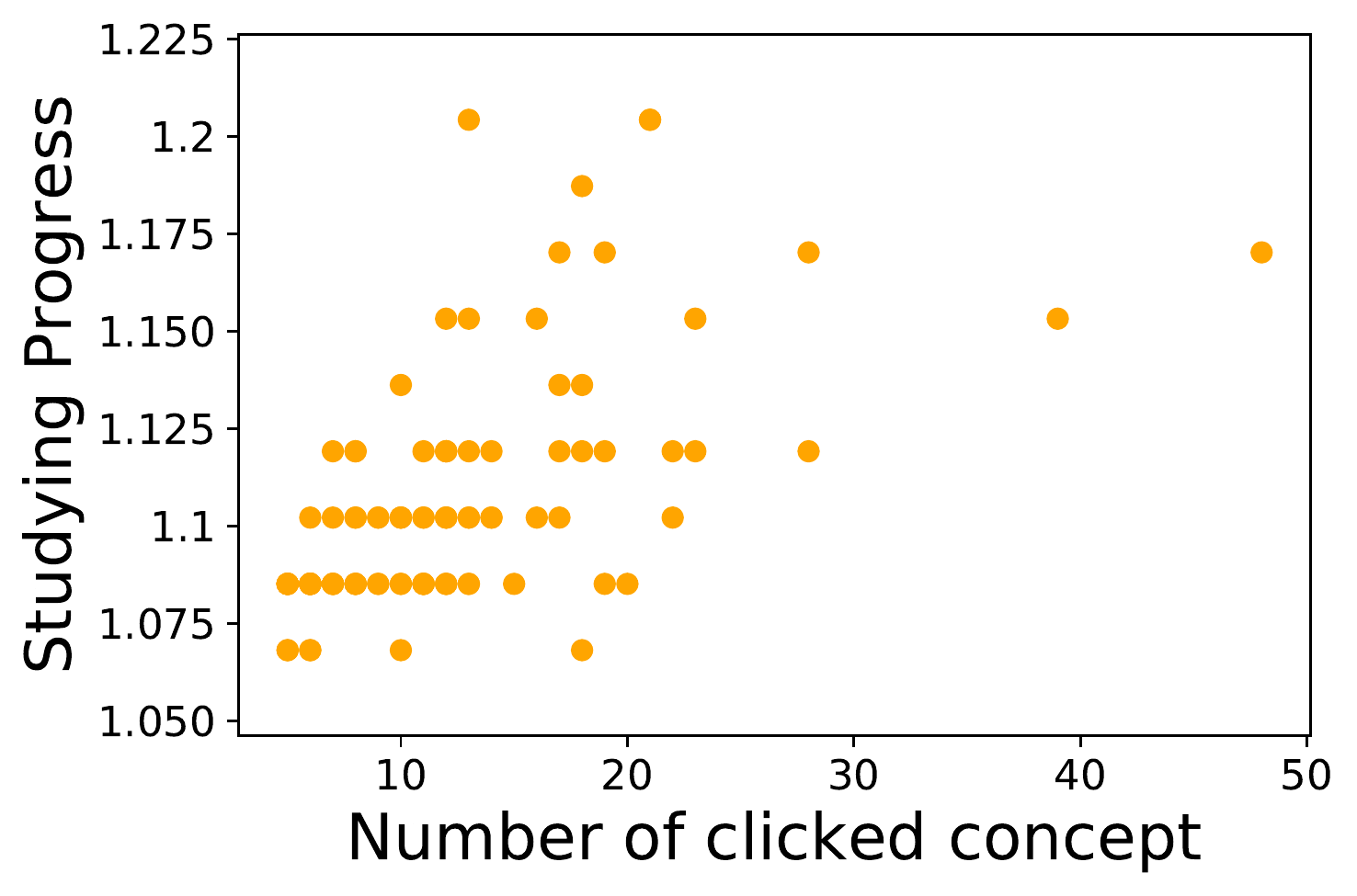}
		\caption{Studying progress}
		\label{fig:side:b}
	\end{subfigure}
	\hfill
	\begin{subfigure}[b]{0.28\textwidth}
		\includegraphics[width=\textwidth]{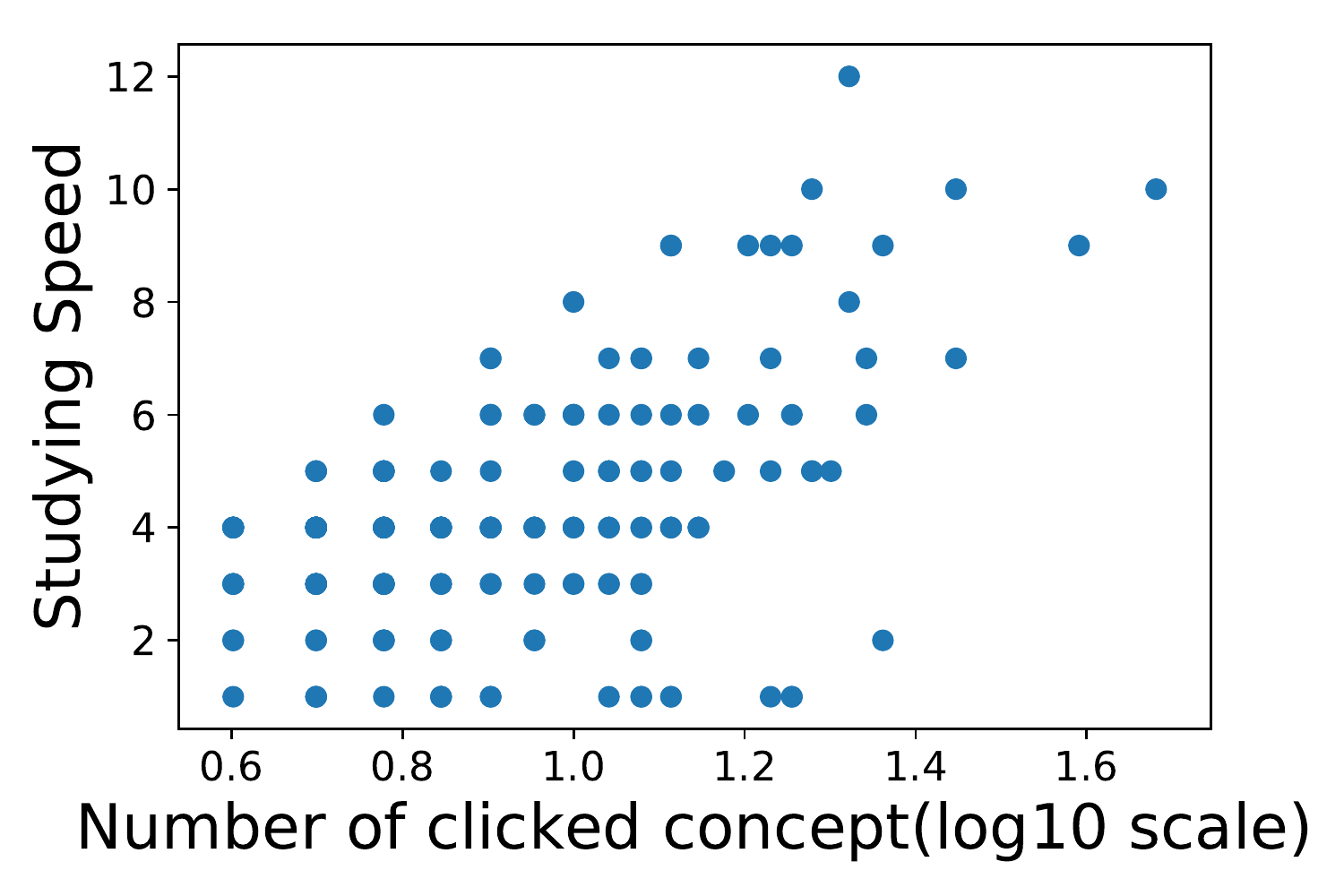}
		\caption{Studying speed}
		\label{fig:side:c}
	\end{subfigure}
	\caption{The motivation of concept recommendation. (a) The heterogeneous network in MOOCs. (b) and (c) denote the progress of studying, and the speed of studying, respectively.
	}\label{fig:introduction}
\end{figure*}
Massive Open Online Courses (MOOCs), which seek to offer widespread interactive participation and open access over the internet, 
are quickly becoming a reliable platform for online and remote education.
Coursera\footnote{https://www.coursera.org}, Udacity\footnote{https://www.udacity.com}, and edX\footnote{https://www.edx.org} are the three most well-known MOOC platforms, which have provided online access to millions of users to study from anywhere in the world. 
As one of China's leading MOOC platforms, \textit{XuetangX}\footnote{https://www.xuetangx.com} has more than 1,000 courses available and more than 6,000,000 users worldwide.
In MOOCs, \textit{course concepts} are used to describe the knowledge concepts and related subjects covered in the course videos. 
This can make it easier for users to find the information in the course videos more quickly.

Several approaches of data mining have been developed to discover user preferences and the relationships among users, course topics, and course videos, to enable users to have a better learning experience in MOOCs. 
These approaches have been used in various related tasks, e.g., behavior prediction~\cite{qiu2016modeling}, course recommendation~\cite{jing2017guess,zhang2019hierarchical}, and user intention understanding~\cite{feng2019understanding}.
The task most similar to ours is course recommendation, which seeks to recommend relevant courses to users based on their past behavior.
From our investigation, we observe that a course is always composed of a series of video lectures, each of which often conveys multiple specific knowledge concepts.
Consequently, we argue that recommending courses to users directly ignores the various knowledge levels of users.
Figure~\ref{fig:side:a} shows the heterogeneous information network existing in MOOCs.
For example, considering the \textit{Machine Learning} course taught by different instructors, the knowledge covered in each course will differ largely (e.g., some instructors will put more focus on the theory of machine learning, whereas others will put more focus on the applications of machine learning in the area of computer science).
If a student has a background in statistics theory, he/she may be familiar with several fundamental knowledge concepts, such as \textit{VC dimension} and \textit{gradient descent}, so that he/she will be interested in further applications of machine learning.
Additionally, a student who is a beginner in machine learning, he/she may be interested in several fundamental knowledge concepts of machine learning.

Based on this motivation, this paper examines an interesting problem of concept recommendation, with the aim of recommending related topics to users to help them with their online studies. 
To better illustrate the necessity of this research, we study the effects of clicked course concepts on the progress and speed of studying, as shown in Figure~\ref{fig:side:b} and Figure~\ref{fig:side:c}. 
The progress of studying is defined as the portion of videos the user has watched in a course, whereas the speed of studying is defined as the number of videos the user has watched in one week.
From these two figures, we can see that the number of clicked concepts has a positive correlation with the progress and speed of study. It confirms the fact that clicking relevant knowledge concepts can indeed help the users with their online studies.

\begin{figure}[!t]
	\centering
	\begin{subfigure}[b]{0.45\textwidth}
		\includegraphics[width=\textwidth]{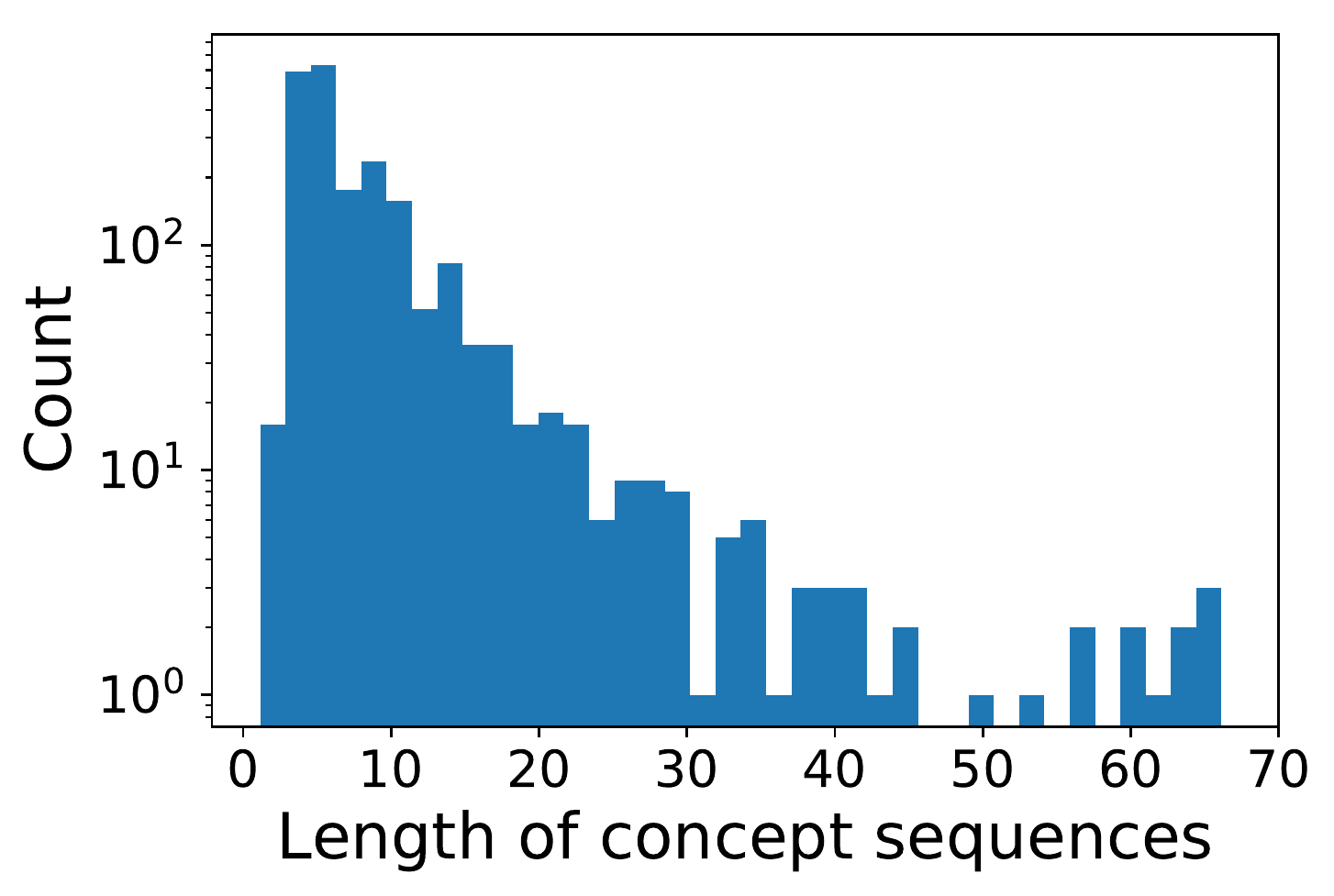}
		\caption{Concept length distribution}
		\label{data_analysis_4}
	\end{subfigure}
	\begin{subfigure}[b]{0.48\textwidth}
		\includegraphics[width=\textwidth]{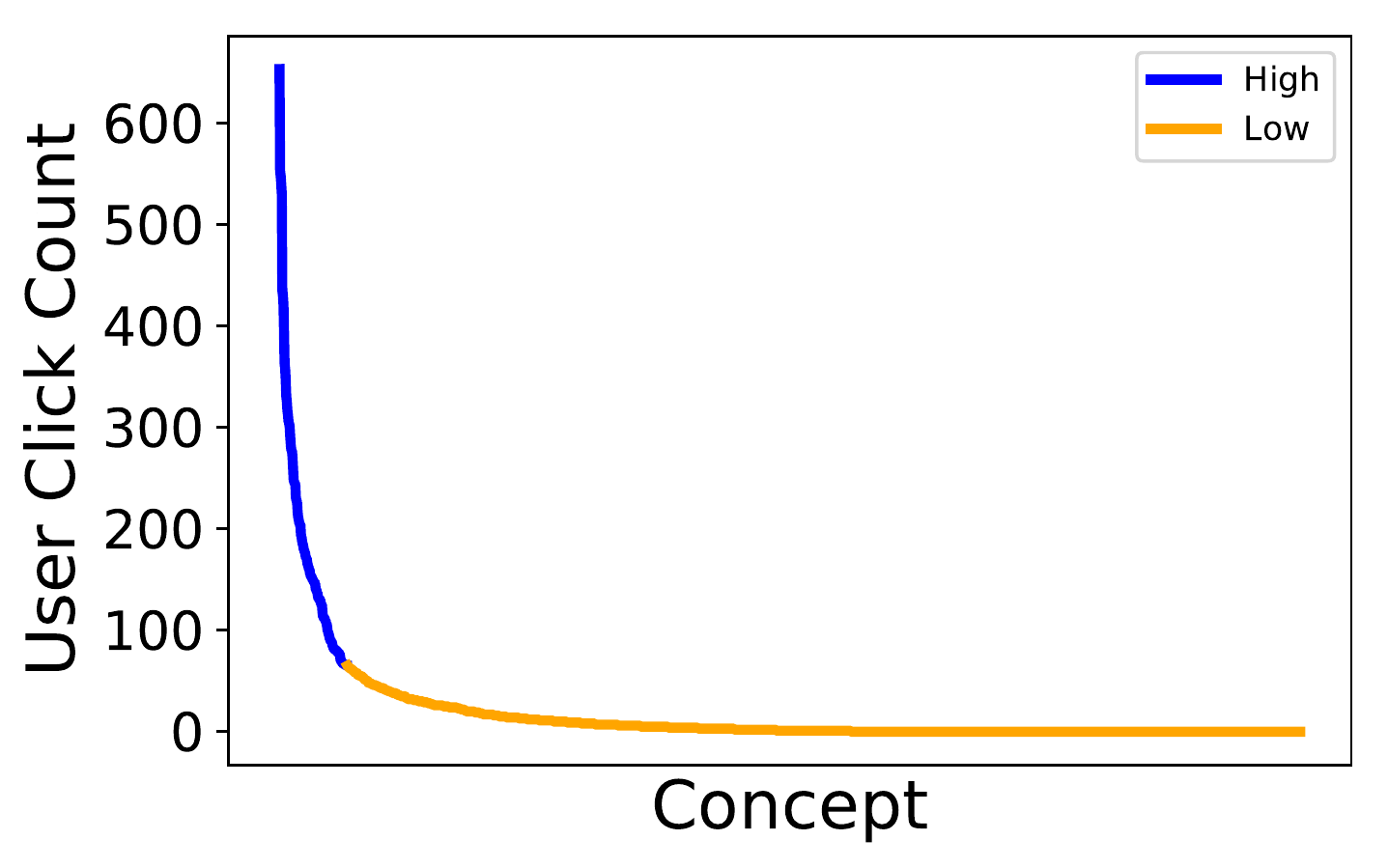}
		\caption{Concept click count}
		\label{data_analysis_1}
	\end{subfigure}
	\caption{A statistical analysis of \textit{XuetangX} dataset.}
	\label{fig_data_analysis}
\end{figure}

For the task of concept recommendation, one primary line of work is based on collaborative filtering.
The collaborative-filtering-based approaches provide recommendations according to the historical interactions among users and items. However, this kind of approaches has two major limitations. \textit{(1) Dynamic environment adaptation}.
The collaborative-filtering-based approaches consider the procedure of recommendation as a static process, and are intended to maximize the immediate reward of the recommendation, resulting in a fixed greedy recommendation strategy.
Therefore, this type of recommendation strategy is sub-optimal and cannot adapt to a dynamic environment where the preferences of users are always dynamic. 
On the other hand, when representing the historical dynamic preferences of users, it is a challenge to handle the long dependency issue existing in the past click history, as shown in Figure~\ref{data_analysis_4}. From Figure~\ref{data_analysis_4}, we can observe that the length of concepts clicked by users is mainly between 10 and 30, which is a relatively long sequential dependency when compared with traditional recommendation tasks, such as e-commerce recommendation.
\textit{(2) Data sparsity issue}. The interactions among users and concepts are always sparse, particularly as the number of users and concepts is increasingly large. 
Figure~\ref{data_analysis_1} shows the distribution of user click count for each concept. This distribution follows the long-tail distribution. 
That is, the majority of concepts are clicked by a few users, which confirms the necessity of concept recommendation.
To overcome this issue, several approaches are attempting to include additional supplementary information, such as the social network~\cite{wu2019neural,zhou2020interactive}, user/item attributes~\cite{wu2020joint,sun2020lara}, and contextual information~\cite{intayoad2020reinforcement,liu2022personalised}, as to enrich the interactions.

These aforementioned restrictions encourage us to develop a model to understand the intents of users, with the ability to recommend personalized concepts in a dynamic interactive environment.
We propose in this paper to shape the task of concept recommendation as a sequential recommendation problem. 
To preserve the dynamic interactions between users and concepts, a reinforcement-learning-based method is proposed by combining exploitation and exploration into a unified framework.
Our reinforcement-learning-based recommendation has two advantages: (1) The recommendation agent can adaptively update the trial-and-error strategy during interaction until convergence; 
(2) The reinforced strategy can produce more diverse recommended results, recommending those items with substantial long-term benefits rather than short-term benefits (i.e., small intermediate rewards).
To overcome the data sparsity issue, we observe that there exist heterogeneous relationships among users and concepts, which can be exploited to enrich the side information and boost the performance of recommendation in the following two perspectives: 
(1) Semantic relationships between knowledge concepts may be provided, which can aid in identifying latent interactions; (2) Users' interest can be understood by tracking the previous records of users alongside these relationships.

This paper puts forward a novel approach (termed HinCRec-RL) for \underline{C}oncept \underline{Rec}ommendation in MOOCs, based on \underline{H}eterogeneous \underline{I}nformation \underline{N}etworks (HINs) and \underline{R}einforcement \underline{L}earning.
We begin by shaping the concept recommendation as a Markov Decision Process (MDP).
We then form the interactions among users, courses, videos, and concepts, into an HIN.
We carry out extensive experiments on a real-world dataset obtained from a Chinese MOOC platform, \textit{XuetangX}, to validate the efficacy of the proposed model.
Note that this paper is related to our previous work ACKRec~\cite{gong2020attentional} for knowledge concept recommendation, which models the entities in MOOCs as an HIN, and proposes an attentional graph convolutional neural network to represent the entities in HIN.
To enable the situations of online dynamic interaction, we extend ACKRec by modeling the concept recommendation problem within a reinforcement learning framework.

The primary contributions of this paper include the following three aspects.
\begin{itemize}
	\item 
	Firstly, we investigate an interesting problem of concept recommendation, which can be considered as a fine-grained recommendation task of course recommendation in MOOCs. 
	\item Secondly, an end-to-end model based on reinforcement learning is proposed to cope with the dynamic interaction among users and concepts, providing a more diverse recommendation. To represent users with sparse data accurately, we propose a meta-path-based user embedding approach with hierarchical graph attention networks. 
	\item Thirdly, we validate the proposed model HinCRec-RL on a dataset collected from a popular Chinese MOOC platform, \textit{XuetangX}.
Experimental results and analysis demonstrate the effectiveness of HinCRec-RL when comparing it with several state-of-the-art models. 
\end{itemize}
We structure the rest paper as follows.
We first formulate the concept recommendation problem in Sec.~\ref{sec_preliminaries}.
We present the details of the proposed approach in Sec.~\ref{sec_methodology}. We analyze the \textit{XuetangX} dataset in Sec.~\ref{sec_experiments}, and show the experimental results with analyses in Sec.~\ref{sec_experimental_result}. 
We highlight some related work with a comparison in Sec.~\ref{sec_related_work}. 
We have a discussion of the mechanism of our proposed model in Sec.~\ref{sec_discussions}.
Finally, Sec.~\ref{sec_conclusion} concludes this paper with some potential directions suggested.

\section{Problem Formulation}\label{sec_preliminaries}
In MOOCs, suppose that we have $U$ users, $C$ courses, $V$ videos and $K$ concepts. Let $\mathcal{U} = \{u_1, u_2, \ldots, u_U\}$ be the set of users, $\mathcal{C} = \{c_1, c_2, \ldots, c_C\}$ be the set of courses, $\mathcal{V} = \{v_1, v_2, \ldots, v_V\}$ be the set of videos and $\mathcal{K} = \{k_1, k_2, \ldots, k_K\}$ be the set of concepts.
In reality, a course $c_i$ is always composed of multiple videos, each of which consists of multiple knowledge concepts. From the perspective of users, a user $u_j$ will take a course, watch the videos of the course, and click several related concepts to learn the knowledge presented in the course videos.
Consequently, the aim of concept recommendation is to enhance the learning experiences of users by automatically recommending relevant concepts to users based on their preferences acquired from historical experience.
\begin{figure}[!t]
	\centering
	\includegraphics[width=0.7\textwidth]{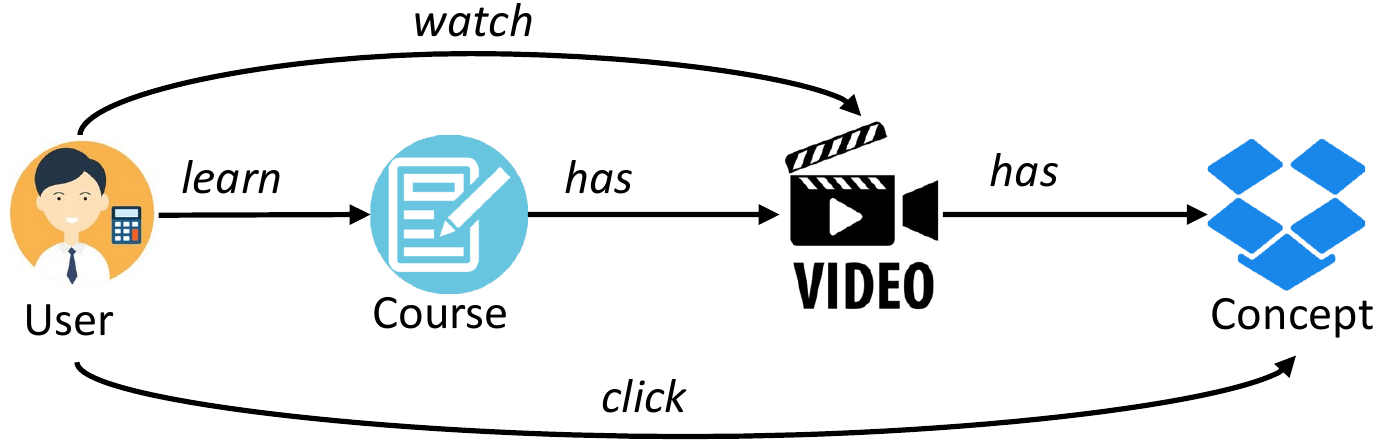}
	\caption{Network schema of HIN in \textit{XuetangX}.}\label{fig_schema}
\end{figure}
\subsection{Recommendation as an MDP}
As mentioned before, to adapt to the dynamic environment, we aim to learn a recommendation agent that interacts with users by sequentially recommending knowledge concepts at each time step. 
We formulate the problem of sequential recommendation as a Markov Decision Process (MDP) consisting of a tuple of five items $(\mathcal{S}, \mathcal{A}, P, R, \gamma)$:
\begin{itemize}
	\item \textbf{State space $\mathcal{S}$} denotes the space of user states. Specifically, the state $S_{t}$ denotes the past preference of user $u$, observed until time step $t$. 
	In our scenario, the past preference of user $u$ are the historical learned courses and clicked concepts. More specifically, the state in our scenario represents the user embedding in an HIN.
	\item \textbf{Action space $\mathcal{A}$} denotes the set of concepts to be selected, i.e., $\mathcal{A}=\mathcal{K}$. Without loss of generality, we sequentially recommend one concept to users at each time step.
	In each state $s_{t}$, an action $a_t$ can be drawn from the available action set $\mathcal{A}(s_{t})$, recursively defined as $\mathcal{A}(s_{t}) = \mathcal{A}(s_{t-1}) \backslash \{a_{t-1}\}$, for $t \neq 0$. 
    This indicates that the agent is not permitted to select those concepts that have been recommended in prior time steps.
	\item \textbf{Reward $R$}, formulated as $R^{a}_{ss'} = \mathbb{E}[r_{t+1}|s_{t}=s, a_{t}=a, s_{t+1}=s']$, 
represents the expected immediate reward produced by the environment when state $s$ changes to state $s'$ as a result of action $a$.
	In our scenario, the immediate reward of performing an action $a$ is determined by the operation of clicking or not, given by user $u$. Therefore, we define $R^{a}_{ss'} = R_{ua}$.
	\item \textbf{Transition function $P$}, formulated as $P^{a}_{ss'} = P[s_{t+1} = s' | s_{t} = s, a_{t} = a]$, denotes the likelihood that the environment would transit from state $s$ to state $s'$ after performing an action $a$. 
	\item \textbf{Discount factor $\gamma$} 
	is used to measure the proportion of reward the reinforced agent considers in the distant future versus those in the immediate future, with a range of $[0, 1]$.
\end{itemize}
The concept recommendation task can be formally characterized as follows, using the aforementioned notations and terminologies.
Given an MDP $(\mathcal{S}, \mathcal{A}, P, R, \gamma)$, the aim of concept recommendation is to find a policy $\pi: \mathcal{\pi} \rightarrow \mathcal{A}$ that
maximizes the cumulated reward from users, i.e., increasing the clickthrough rate of concepts.

\subsection{Construction of Heterogeneous Information Network}

\begin{figure}[!t]
	\centering
	\includegraphics[width=0.7\textwidth]{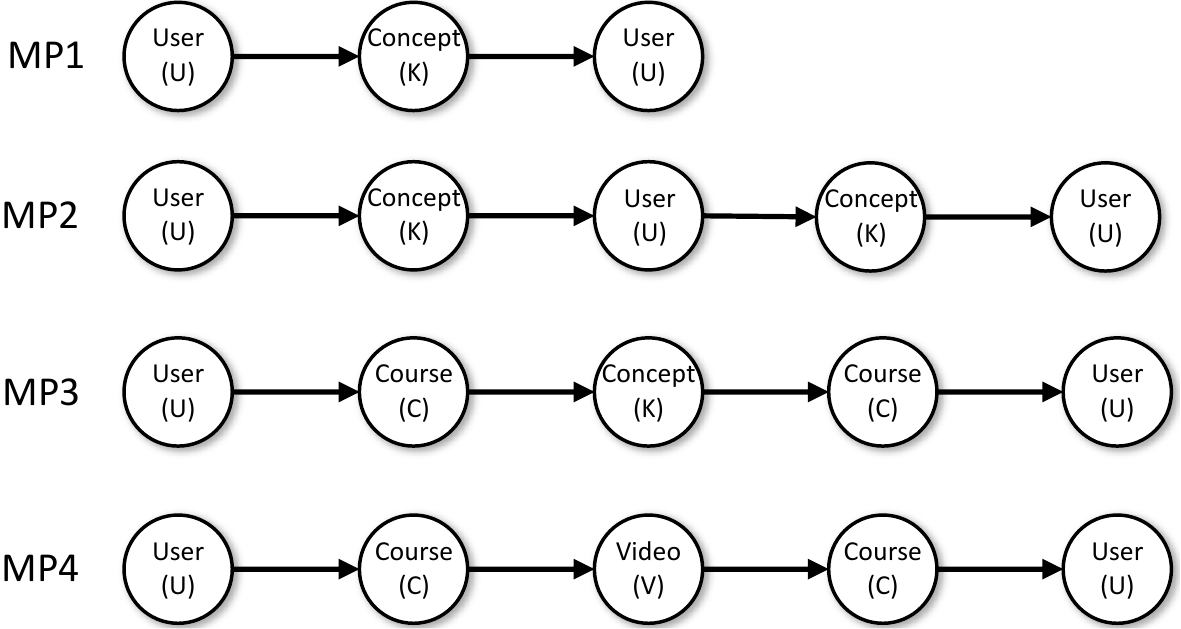}
	\caption{Defined meta-paths in \textit{XuetangX}.}\label{SelectedMetaPaths}
\end{figure}
In order to model different types of entities (e.g., course, video, concept, and user) and their complex relationships in MOOCs, we resort to considering the MOOC networks as typical heterogeneous information networks (HINs). First, we introduce some preliminary concepts about HINs.

\paragraph{Heterogeneous information network (HIN)} 
Let $\mathcal{G} = \{\mathcal{V}, \mathcal{E}\}$ denote a graph, which is composed of a collection of nodes $\mathcal{V}$ and edges $\mathcal{E}$. 
The edges in a graph can describe the relationships/connections among nodes. 
We refer to the graph as an HIN when the types of nodes and edges are not unique~\cite{sun2012mining}. 
In our scenario, there are four entities of the HIN in \textit{XuetangX} (i.e., \textit{course} ($\mathcal{C}$), \textit{video} ($\mathcal{V}$), \textit{user} ($\mathcal{U}$), and \textit{knowledge concept} ($\mathcal{K}$)). The relationships among these entities  include \textit{learn}, \textit{click}, \textit{watch}, and \textit{has}. 
Figure~\ref{fig_schema} describes the schema of network existing in \textit{XuetangX} MOOC.
We can extract the meta-path from the network schema, which has been frequently utilized to represent the semantic paths between a pair of entities.
\paragraph{Meta-path}
A meta-path is a specific path defined based on the network schema.  
That is a  sequence of relations between two node types,
in the form of $A_1\overset{R_1}{\rightarrow} A_2\overset{R_2}{\rightarrow} \cdots \overset{R_l}{\rightarrow}  A_{l+1}$, where $A_i$ denotes the node type~\cite{sun2011pathsim}. 
In our scenario, we manually identify four meta-paths as follows to connect users in \textit{XuetangX} MOOC, which will also enrich the semantic representation of each user.
\begin{itemize}
	\item \textbf{MP1 ($\mathbf{U}_1$-$\mathbf{K}$-$\mathbf{U}_2$)} 
indicates that if two users click the same concept, they are considered to be connected. 
	\item \textbf{MP2 ($\mathbf{U}_1$-$\mathbf{K}_1$-$\mathbf{U}_2$-$\mathbf{K}_2$-$\mathbf{U}_3$)}
	indicates that the two concepts clicked by users 1 and 2, respectively, are both clicked by user 3.
	\item \textbf{MP3 ($\mathbf{U}_1$-$\mathbf{C}_1$-$\mathbf{K}$-$\mathbf{C}_2$-$\mathbf{U}_2$)} denotes that users who take the courses that contain the same concepts will be connected.
	\item \textbf{MP4 ($\mathbf{U}_1$-$\mathbf{C}_1$-$\mathbf{V}$-$\mathbf{C}_2$-$\mathbf{U}_2$)} shows that users are related through paths of courses sharing some common videos. 
\end{itemize}
We also show these four meta-paths in Figure~\ref{SelectedMetaPaths}. In Sec.~\ref{sec_meta_analysis}, we will investigate the effect of each meta-path and their combinations.

The notations used throughout this paper are listed in Table~\ref{table_notations}.

\begin{table}[!t]
	\centering
	\caption{A list of notations used in this paper.}
	\label{table_notations}
	\begin{tabular}{l|l}
		\toprule
		Notation&Description\\
		\midrule
		$\mathcal{G}$ & the heterogeneous information network \\
		$\mathcal{V}$ & the entity set \\
		$\mathcal{E}$ & the relationship set \\
		$\mathcal{SM}$ & the network schema \\
		$\mathcal{MP}$ & the meta-path set \\
		$\mathcal{P}$ & the set of paths sampled based on meta-paths\\
		$\mathcal{S}$ & the state of MDP \\
		$R_{ss'}^{a}$ & the reward obtained when transferring state $s$ to $s'$ by taking action $a$ \\
		$\gamma$ & the discount factor \\
		$\pi$ & the recommendation policy \\
		$\lambda$ & the weight to regularize the entropy  \\
		$\Phi$ & meta-path \\
		${\mathcal{N}}_{i}^{\phi}$ & meta-path-based neighbors of node ${i}$ \\
		$\mathbf{h}$ & the initial node feature \\
		$\mathbf{u}$ & the user embedding \\
		$\mathbf{c}$ & the concept embedding \\
		\bottomrule
	\end{tabular}
\end{table}

\section{Proposed Approach}\label{sec_methodology}
We describe our proposed methodology in depth in this section, beginning with an overview of the entire framework. To learn the user embedding, we present our graph neural network to represent the introduced HIN. Subsequently, we present our reinforcement learning strategy for recommending concepts to users. 
\subsection{Overview}
\begin{figure*}[t!]
	\centering
	\includegraphics[width=0.99\textwidth]{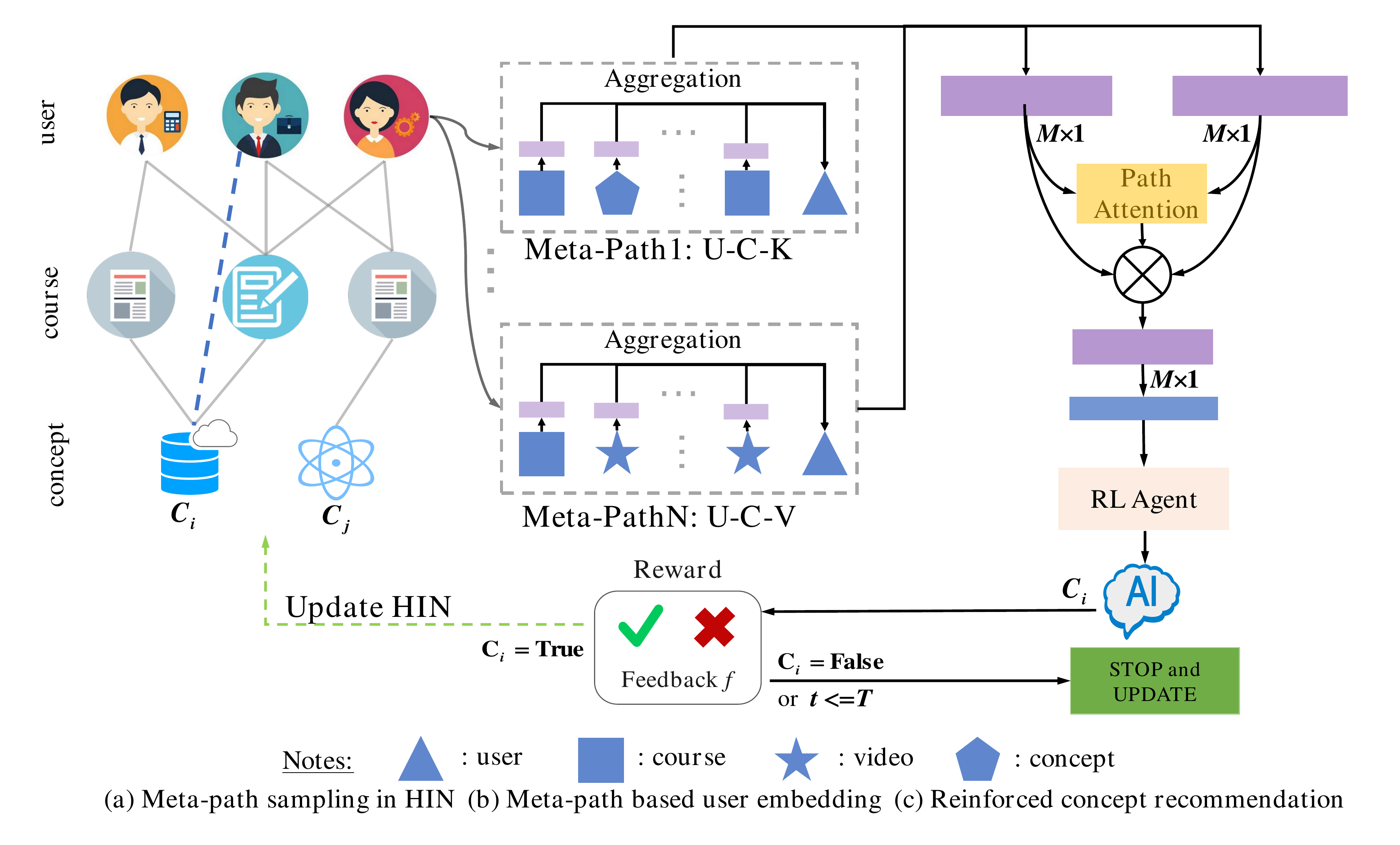}
	\caption{Overview of the proposed method. (a) Meta-path sampling in the HIN of MOOCs. (b) Meta-path based user embedding with hierarchical attention networks. (c) Reinforced concept recommendation.}
	\label{fig_framework}
\end{figure*}

Figure~\ref{fig_framework} depicts the structure of our proposed neural network, which is composed of
three sub-modules. (a) Meta-path sampling (Sec.~\ref{sub_path_sampling}). 
In this module, we first build an HIN among users, courses, and concepts. 
Subsequently, we sample meta-paths for each given user using a random walk in the network. 
(b) Meta-path-based user embedding with a hierarchical attention network (Sec.~\ref{han_node_level} and Sec.~\ref{han_semantic_level}). In this module, a hierarchical attention network is introduced to embed each math-path into a hidden space. On the node-level, we employ a self-attention layer to represent the user with its neighbors. On the path-level, we apply another attention layer to aggregate semantic representations along each path to the user. 
(c) Reinforced concept recommendation (Sec.~\ref{sec_reinforced_recommender}). 
In this module, we propose to recommend concepts to users within a reinforcement learning framework. 
Comparing with other HIN-based representation learning approaches for node classification and recommendation, this is the first time that the HIN and reinforcement learning have been combined to recommend concepts to users in MOOCs.

\subsection{Meta-Path Based User Embedding}\label{sec_encoder}
We represent the user based on meta-paths. Our intuition is that the representation of users can be enriched by the information aggregated from multi-hop neighbors along the meta-paths. For a given user $u_i$, we first collect multi-hop neighbors by meta-path sampling, starting from this node. We then design a heterogeneous attention network to aggregate the information along each sampled path.

\subsubsection{Meta-Path Sampling}\label{sub_path_sampling}
To index the network structure of $\mathcal{G}$, we propose to sample paths on $\mathcal{G}$ based on random walk. 
Inspired by DeepWalk~\cite{perozzi2014deepwalk}, we start random walk from each user in our training dataset $\mathcal{D}$ for whom a concept will be recommended. 
Specifically, we first employ a depth-first search, starting from the user node, to sample instances under the restrictive scheme of each meta-path, and set a limited walk length $l$ determining at most how many neighbors would be searched along the sampled path and controlling the size of its neighborhood. 
Then we utilize a breadth search to repeat the aforementioned procedure $N$ times in parallel for each meta-path, as each path potentially focuses on different input neighbors, which makes it possible to capture rich semantics. 
Thus, for each user $u_i\in \mathcal{U}$, we sample $|\mathcal{MP}| \times N$ paths. In total, we have $|\mathcal{U}| \times |\mathcal{MP}| \times N$ paths over the HIN. We denote the meta-path as $\mathcal{MP}=\{\Phi_1,\Phi_2,\ldots, \Phi_P\}$ and denote the path corpus as $\mathcal{P}=\{\rho_{\Phi_1},\rho_{\Phi_2},\ldots,\rho_{\Phi_P}\}$, where $\rho_{\Phi_k}$ denotes a path set sampled based on the meta-path $\Phi_k$.

\subsubsection{Node-level Attention}\label{han_node_level}
In order to obtain the semantic-specific user embedding, we design a module of node-level attention which first determines the weights of neighbors along a meta-path, and then combines them.
As the node types are heterogeneous and various node types have different feature spaces, a simple semantic feature space alignment should be designed. We first project the node features into a common space. For node $i$, let $\mathbf{h}_i$ denote its corresponding feature. The projected feature $\mathbf{h'}_i$ can be calculated as follows:
\begin{equation}
	\mathbf{h'}_i = \mathbf{M}_{\phi_i} \cdot \mathbf{h}_i\,,
\end{equation}
where $\mathbf{M}_{\phi_i}$ is the projection matrix of type $\phi_i$.
We argue that different nodes for a given meta-path should contribute different weights to the final user embedding. 
Thus, an attention mechanism is designed to calculate the weight of node $i$ to its meta-path-based neighbors $\mathcal{N}_i^{\Phi}$.
Formally, for a given meta-path $\Phi$, the node-level attention weight $\mathbf{a}_\Phi$ between nodes $i$ and $j$ can be calculated as follows:
\begin{equation}\label{eq_attn}
	\alpha_{i j}^{\Phi}=\frac{\exp \left(\sigma\left(\mathbf{a}_{\Phi}^{\mathrm{T}} \cdot\left[\mathbf{h}_{i}^{\prime} \| \mathbf{h}_{j}^{\prime}\right]\right)\right)}{\sum_{k \in \mathcal{N}_{i}^{\Phi}} \exp \left(\sigma\left(\mathbf{a}_{\Phi}^{\mathrm{T}} \cdot\left[\mathbf{h}_{i}^{\prime} \| \mathbf{h}_{k}^{\prime}\right]\right)\right)}\,, 
\end{equation}
where $||$ denotes the concatenation operator, and $\sigma$ denotes the activation function.
Note that, for the sake of simplicity, we randomly select one sampled path of meta-path-based neighbors to calculate the node-level attention.
Following~\cite{DBLP:conf/www/WangJSWYCY19}, we then aggregate the node embedding $\mathbf{u}_i$ for node $i$ along a meta-path $\Phi$.
Furthermore, the node-level attention is repeated $L$ times and the attention embeddings are concatenated through a layer of linear projection.
Consequently, the user embedding $\mathbf{u}_{i}^{\Phi}$ along a meta-path $\Phi$ is formulated as follows:
\begin{equation}
	\mathbf{u}_{i}^{\Phi}= \|_{l=1}^{L} \sigma\left(\sum_{j \in \mathcal{N}_{i}^{\Phi}} \alpha_{i j}^{\Phi} \cdot \mathbf{h}_{j}^{\prime}\right)\,,
\end{equation}
where $||_{l=1}^{L}$ represents the concatenation operation.
For a set of meta-paths $\{\Phi_1, \Phi_2,\ldots, \Phi_P \}$, we denote the user embedding along the meta-path set as $\mathbf{u}_{i}^{\Phi_1}, \mathbf{u}_{i}^{\Phi_2}, \ldots, \mathbf{u}_{i}^{\Phi_P}$. 

\subsubsection{Path-level Attention}\label{han_semantic_level}
Similar to node-level attention, we use attention at the path level. 
We assign different weights to meta-paths via path-level attention, to obtain the final user representation. 
The importance of meta-path $w_{\Phi_{k}}$ can be calculated as follows:
\begin{equation}
	w_{\Phi_{k}}=\frac{1}{|\mathcal{P}|} \sum_{p=1}^{|\mathcal{P}|} \mathbf{q}^{\mathrm{T}} \cdot \tanh \left(\mathbf{W} \cdot \mathbf{u}_{i}^{\Phi_k}+\mathbf{b}\right)\,,
\end{equation}
where $\mathbf{q}$ represents the path-level attention vector and $|\mathcal{P}|$ denotes the number of meta-path-based sampled paths. That is, we average all the sampled paths for one meta-path to calculate the path-level attention, which expresses sophisticated functions beyond the simple meta-path-based neighbors average.
Consequently, the importance of meta-path $w_{\Phi_{k}}$ can be normalized using a softmax layer, i.e., $\beta_{\Phi_{k}}=\frac{\exp \left(w_{\Phi_{k}}\right)}{\sum_{k=1}^{|\mathcal{MP}|} \exp \left(w_{\Phi_{k}}\right)}$.
The final embedding of user $i$ can be reformulated as the following formula, according to the normalized importance of the meta-path.
\begin{equation}
	\mathbf{u}_i=\sum_{k=1}^{|\mathcal{MP}|} \beta_{\Phi_{k}} \cdot \mathbf{u}_{i}^{\Phi_{k}}\,.
	\label{uembd}
\end{equation}

The two hierarchical attention modules (i.e., node-level attention and path-level attention) to assign different weights to the node neighbors and different paths have been verified in the node classification task in~\cite{DBLP:conf/www/WangJSWYCY19,wang2021self}. 
We have adopted them in this study with a slight modification. An important difference is that the HIN in our scenario is dynamically updated according to the feedback from the users. That is, a link will be assigned if a user clicks a recommended concept.

\subsection{Reinforced Concept Recommendation}\label{sec_reinforced_recommender}
As previously mentioned, to simulate the dynamic interactions between users and concepts, we frame the concept recommendation within a reinforcement learning framework.
Based on a predetermined loss function, we aim to develop a model to reduce the discrepancy between the model predictions and immediate user responses.

We begin by introducing the simplified setting of concept recommendation within a typical reinforcement learning framework, the goal of which is to find a recommendation policy $\pi_{\theta}$ that maximizes the expected reward:
\begin{equation}
	\mathcal{L}_{rl}(\theta) = \mathbb{E}_{c_{t}\sim \pi_{\theta}(\cdot| \mathbf{u}_{t})}\sum_{t=1}^{T}r(c_{t}|\mathbf{u}_{t})\,,
\end{equation}
where $r(c_{t}|\mathbf{u}_{t})$ is the immediate reward. 
Note that, in contrast to the previous section, the dynamic embedding of user $u$ at time step $t$ is indicated by the subscript $t$ in $\mathbf{u}_t$ in this section.
Given a particular user $u$, if the predicted concept $c_{t}$ is true, then the reward will be set to 1, otherwise, it will be -1. In particular, $R(c_{t}|\mathbf{u}_{t})$ is the cumulated reward, which can be defined as follows:
\begin{equation}
	R(c_{t}|\mathbf{u}_{t}) = \gamma^{0}R(c_{t}|\mathbf{u}_{t}) + \cdots + \gamma^{T-t}R(c_{T}|\mathbf{u}_{T})\,,
\end{equation} 
where $\gamma$ is the discount factor, which is set in the range from 0 to 1.
If the recommended concept $c_{t}$ is correct, an edge will be established in the current HIN to connect the user and the correct recommended concept. 
Therefore, we will obtain the new user embedding through the previous HIN embedding step. If the recommended concept is correct, HinCRec-RL will continually recommend a new concept, until it incorrectly recommends a concept or arrives at the max time step $T$.

However, if the recommended concept is incorrect, no new relation is added, and the structure of HIN $\mathcal{G}$ remains unchanged.
Consequently, the user embedding will remain unchanged.
As the Q-learning network~\cite{van2016deep} requires the next step transition $s_{t+1}$ of current step $s_{t}$ to update, if the recommendation is incorrect, then the predicted Q-value $Q_{t+1}$ will be the same as the target $Q_{t}$, and consequently, Q-learning cannot be used in our setting. Therefore, we seek to use the policy gradient to update the model. 

We can then learn the optimal policy by using the policy gradient method called REINFORCE~\cite{williams1992simple,sutton2000policy}, a popular learning approach in reinforcement learning.
The gradient of the expected cumulative reward can be determined using the following formula, according to the policy gradient method:
\begin{equation}
	\nabla_{\theta} \mathcal{L}_{rl}(\theta) = \sum_{t=1}^{T}\left[\nabla_{\theta}\log \pi_{\theta}(c_{t}|\mathbf{u}_{t}) R(c_{t}|\mathbf{u}_{t})\right]\,.
	\label{policy_gradient}
\end{equation}
In practice, we often observed a collapse onto a suboptimal deterministic policy. To prevent the model from not being able to explore new concepts that could lead to a better recommendation, we add an entropy regularization term to the objective function, as follows:
\begin{equation}
	\mathbf{H}[\pi_{\theta}(c|\mathbf{u})] = \sum_{t=1}^{T} \sum_{c_{t} \in \mathcal{C}} \log\pi_{\theta}(c_{t}|\mathbf{u}_{t})\pi_{\theta}(c_{t}|\mathbf{u}_{t})\,.
\end{equation}
Therefore, the ultimate goal of our model is reformulated as follows:
\begin{equation}
	\mathcal{J} = \mathbb{E}_{c \sim \pi_{\theta}(\cdot|\mathbf{u})}\mathcal{L}_{rl}(\theta) + \lambda \mathbf{H}[\pi_{\theta}(c|\mathbf{u})]\,.
	\label{pg}
\end{equation}
Here, $\lambda$ is the regularization weight. The policy gradient directly updates policy $\pi_{\theta}$ to increase the probability of $c_{t}$ given state $s_{t}$ when the reward is positive, and vice versa. 

Algorithm~\ref{p_gradient} depicts the HinCRec-RL algorithm with reinforcement learning.
Given the training set $\mathcal{U}_{train}$, clicking data $R$, the number of time steps $T$, the number of episodes $E$, discount factor $\gamma$, and $\epsilon$-greedy parameter $\epsilon$, our goal is to learn the recommendation policy $\pi_{\theta}$. We first conduct meta-path sampling and initialize the recommendation policy $\pi_{\theta}$, which will be used later.
At the reinforcement learning phase, for each episode, we first pick a user and learn the corresponding embedding based on Eq.~\ref{uembd} (see line 4-6). We then dynamically update the policy through reinforcement learning (line 8-15).
\subsection{Time Complexity of HinCRec-RL}\label{time_analysis}
Here, we analyze the time complexity of HinCRec-RL, as shown in Algorithm~\ref{p_gradient}.
The total complexity for a training episode is $O({L}\sum_{{v}\in{V}}{F}_{v}{F}_{1}+{L}{T}\sum_{i=1}^{|\mathcal{MP}|}(|{{E}_{{\Phi}_{i}}}|{F}_{1}+|{V}_{{\Phi}_{i}}|{F}_{1}{F}_{2}))$, where $L$ is the number of attention heads, $T$ is the number of time steps of an episode, ${|\mathcal{MP}|}$ is the number of hand-crafted meta-paths, and ${F}_{v}$ is the input dimension of the specific type node. ${F}_{1}$ and ${F}_{2}$ are the output dimensions of node type projection and path-specific transformation, respectively. 
$|{{E}_{{\Phi}}}|$ and $|{{V}_{{\Phi}}}|$ are the numbers of sampled-path-based node pairs and user nodes, respectively. 
More specifically, the time complexity of a single node-level attention head can be expressed as $O(|{V}|+|{E}|) = O(\sum_{{v}\in{V}}{F}_{v}{F}_{1}+|{{E}_{\Phi}}|{F}_{1})$, including input feature transformation, the computation of sampled path attention coefficients, and node-level aggregation.
The time complexity of the path-level attention can be expressed as $O(|{V}|) = O({|{V}_{\Phi}|}{F}_{1}{F}_{2})$, including path-specific transformation and meta-path based sampled path aggregation. 
Note that we only need to pick one user for each episode uniformly and recompute its user embedding iteratively during sequential recommendation training with reinforcement learning. 
Within an episode, the action space of $\epsilon$-greedy influences the number of time steps, whereas the requirement to converge is correlated to the action space. 
Thus, HinCRec-RL is highly computationally efficient, and the total complexity is linearly proportional to the number of nodes and meta-path-based node pairs.

\begin{algorithm}[t!]
	\SetKwInOut{KwIn}{Input}
	\SetKwInOut{KwOut}{Output}
	
	\KwIn{Training set $\mathcal{U}_{train}$, clicking data $R$, number of time steps $T$, number of episodes $E$, discount factor $\gamma$, and $\epsilon$-greedy parameter $\epsilon$.}
	\KwOut{The learned recommendation policy $\pi_{\theta}$.}
	Meta-path sampling
	
	Initialize recommendation policy $\pi_{\theta}$ with random weights
	
	\For{episode $\gets 1$ \textbf{to} $E$}{
        Initialize the parameters of HIN $\mathcal{G}$ and user embeddings $\mathcal{U}_{train}$		
  
        Uniformly pick a user $u_0 \in \mathcal{U}_{train}$ 
		
		Learn the user embedding $\mathbf{u}_1$ by Eq.~\ref{uembd}
		
        
        ${t} \leftarrow {1}$
        
		\While{$c_t$ is correct and $t \leq T$} {
			Select the concept $c_{t}$ using $\epsilon$-greedy policy w.r.t $\pi_{\theta}$, observe reward $R(c_{t}|\mathbf{u}_{t})$
            
            \If{$c_t$ is correct} {
                Add a new link between the user $u_{t}$ and concept $c_{t}$ on the HIN $\mathcal{G}$
    			
    			Recompute the user embedding $\mathbf{u}_{t+1}$ by Eq.~\ref{uembd}
            }
            ${t} \leftarrow {t + 1}$
		}
		Update $\pi_{\theta}$'s weights $\theta$ according to Eq.~\ref{pg}
	}
	\caption{HinCRec-RL}
	\label{p_gradient}
\end{algorithm}

\section{Experimental Setup}\label{sec_experiments}
We carry out experiments by addressing the following research issues to assess the performance of our proposed model HinCRec-RL.
\begin{itemize}
	\item \textbf{RQ1:} How effective is the proposed model HinCRec-RL for concept recommendation when compared with baseline methods?
	\item \textbf{RQ2:} How do various involved meta-paths and their combinations affect user embedding?
	\item \textbf{RQ3:} How dose the proposed model HinCRec-RL perform when the hyper-parameters are changed?
\end{itemize}
\subsection{Experimental Dataset}
To conduct our experiments, we construct training and test datasets using the data from \textit{XuetangX} MOOC platform,
where the training set spans from 1st October, 2016 to 30th December, 2017, whereas the test set spans from 1st January, 2018 to 31st March, 2018. We pair a target course with all historical enrolled courses to form a sequence which is further considered as an instance of our datasets. For each instance when conducting the training phase, we consider the final course in the sequence to be the target course and the remaining to be the historical courses.
When conducting the test process, for a user, a target concept in the test set indicates a concept that must be involved in the enrolled courses, and historical concepts indicate the corresponding concepts to the same user in the training set. On the other hand, the same concept label is usually clicked by a user, and thus, multiple clicked concept records are generated. 
In our experiments, we serialize these records into a single one. 
We then paired each positive instance with a random sample of 99 negative ones in the test set.

\begin{table}[!t]
	\caption{The description of \textit{XuetangX} datasets.} 
	\centering
	\label{table_dataset}
	\begin{tabular}{m{4em}|m{4em}|m{8em}|m{6em}}
		\hline
		\textbf{Nodes} &\textbf{Count} &\textbf{Links} &\textbf{Count}\\
		\hline
		\multirow{2}{*}{\textbf{concept}}  &\multirow{2}{*}{2,527} &concept-course & 21,507 \\
		\cline{3-4}
		~ &~ & concept-video & 11,732 \\
		\hline
		\multirow{2}{*}{\textbf{user}}  &\multirow{2}{*}{3,111,637} &user-course & 15,045,219 \\
		\cline{3-4}
		~ &~ &user-video & 53,481,869 \\
		\hline
		\multirow{3}{*}{\textbf{course}}  &\multirow{3}{*}{7,327} &course-concept & 69,012 \\
		\cline{3-4}
		~ &~ &course-video & 811,841 \\
		\cline{3-4}
		~ &~ &course-user & 16,724,852 \\
		\hline
		\multirow{3}{*}{\textbf{video}}  &\multirow{3}{*}{62,191} &video-course & 247,433 \\
		\cline{3-4}
		~ &~ &video-concept & 11,732 \\
		\cline{3-4}
		~ &~ &video-user & 53,971,707 \\
		\hline
		\textbf{Total} &\textbf{3,183,682} & \textbf{Total} &\textbf{140,096,904} \\
		\hline
	\end{tabular}
\end{table}


Overall, the obtained dataset contains 62,191 videos, 7,327 courses, 3,111,637 users, 2,527 concepts and 140,096,904 relations between these entities. The description of this dataset is presented in Table \ref{table_dataset}.

\subsection{Evaluation Metrics} 
We pick several metrics commonly used in recommender systems to evaluate the proposed model. They are Mean Reciprocal Rank ($MRR$), Normalized Discounted Cumulative Gain of top-K items ($NDCG@K$) and Hit Ratio of top-K items ($HR@K$). We choose 5, 10 and 20 for $K$ in our experiments.

$HR@K$ is used to measure the proportion of ground-truth instances which are correctly recommended among the top-$K$ items. This is a recall-based metric and can be formulated as
\begin{equation}
	HR@K=\frac{{\#Hits}@K}{|GT|}\,,
\end{equation}
where $|GT|$ is the size of the test dataset.
$HR@K$ measures the percentage of successes in prediction to the number of attempts. $NDCG$ is usually applied to assess a retrieval syetem's performance when considering the graded relevance of retrieved entities. Thus, it is also used as another ranking metric in our experiments and can be defined as
\begin{equation}
	NDCG@K=\frac{1}{\left | Q \right |}\sum_{q=1}^{\left | Q \right |}Z_{kq}\sum_{j=1}^{k}\frac{2^{r(j)}-1}{\log(1+j)}\,,
\end{equation}
where $r(j)$ denotes a relevance score calculated for a retrieved entity for query $q$. $Z_{kq}$ indicates a normalization factor and is usually used to ensure that the $NDCG$ for a perfect ranking at $k$ for query $q$ is $1$. $MRR$ is used to measure the ranking of our searched results and can be defined as
\begin{equation}
	MRR=\frac{1}{|Q|}\sum_{i=1}^{|Q|}\frac{1}{rank_i}\,,
\end{equation}
where $|Q|$ indicates the size of a query set. 

Moreover, we leverage Area Under the ROC curve ($AUC$) to measure the quality of recommendation ranking.
\begin{table*}[!t]
	\centering
	\caption{Quantitative results (\%) of experiments on \textit{XuetangX} MOOCs datasets. (Best scores are in boldface.)}
	\resizebox{1\linewidth}{!}{	
		\begin{tabular}{lllllllll}
			\hline
			\textbf{Model} &\textbf{HR@5} &\textbf{HR@10} &\textbf{HR@20} &\textbf{NDCG@5} &\textbf{NDCG@10} &\textbf{NDCG@20} &\textbf{MRR} &\textbf{AUC}\\
			\hline\hline
			\textbf{BPR}  &41.50 &56.28 &72.12 &28.95  &33.74 &37.76  &28.68   &84.34\\
			\textbf{MLP}  &32.75   &52.66 &69.88 &17.67 &25.32 &30.71 &18.34   &82.73\\
			\textbf{FM}  &36.03   &54.40 &72.39  &23.99   &29.89 &34.44  &24.46   &83.20\\
			\textbf{FISM}  &36.41    &54.31 &70.79  &20.15  &27.73 &33.83  &21.65  &84.24\\
			\textbf{NAIS}  &36.47   &57.92 &77.99 &21.93    &28.10 &34.12  &22.15  &86.76\\
			\textbf{NASR}  &38.67   &53.21 &67.13  &21.19   &27.84 &32.93  &19.69   &79.90\\
			\hline
			\textbf{HinCRec-SL}  &60.73	&78.41	&\textbf{90.04}	&40.89	&46.62	&49.60	&37.85	&\textbf{92.10}\\
			\textbf{HinCRec-RL}  &\textbf{66.21}	&\textbf{81.53}	&88.74	&\textbf{46.21}	&\textbf{51.35}	&\textbf{53.22}	&\textbf{42.82}	&89.64\\
			\hline 
		\end{tabular}
	}
	\label{table_performance2}
\end{table*}
\begin{table*}[!t]
	\centering
	\caption{Different results  (\%) from different combinations of meta-paths. (Best scores are in boldface.)}
	\resizebox{1\linewidth}{!}{	
	\begin{tabular}{lllllllll}
		\hline
		\textbf{Meta-Path} &\textbf{HR@5} &\textbf{HR@10} &\textbf{HR@20} &\textbf{NDCG@5} &\textbf{NDCG@10} &\textbf{NDCG@20} &\textbf{MRR} &\textbf{AUC}\\
		\hline\hline
		\textbf{MP1}  &63.51  &75.23	&\textbf{90.99}	&43.08	&46.89	&50.90	&39.24	&91.89\\
		\textbf{MP2}  &45.33  &61.68	&79.44	&29.59	&34.90	&39.43	&28.44	&86.92\\
		\textbf{MP3}  &46.15  &67.42	&84.62	&30.37	&37.27	&41.66	&29.76	&88.69\\
		\textbf{MP4}  &50.68  &69.86	&86.76	&34.32	&40.53	&44.78	&33.01	&91.32\\
		\textbf{MP1 \& MP2}  &63.47  &77.17	&89.04	&44.33	&48.83	&51.81	&41.04	&\textbf{93.15}\\
		\textbf{MP1 \& MP3} &60.55	&76.15	&87.61	&43.37	&48.42	&51.39	&40.96	&93.12\\
		\textbf{MP1 \& MP4} &65.02	&78.48	&88.34	&45.69	&50.02	&52.48	&42.15	&89.69\\
		\textbf{MP2 \& MP3}  &46.98	&68.37	&81.86	&29.91	&36.78	&40.23	&28.64	&89.30\\
		\textbf{MP2 \& MP4}  &53.02	&73.02	&89.30	&35.68	&42.09	&46.20	&33.94	&92.09\\
		\textbf{MP3 \& MP4}  &47.44	&65.58	&84.65	&32.30	&38.16	&42.97	&31.56	&86.51\\
		\textbf{MP1 \& MP2 \& MP3}  &57.73	&75.45	&89.55	&41.06	&46.86	&50.47	&39.26	&91.82\\
		\textbf{MP1 \& MP2 \& MP4}  &59.35	&77.23	&89.76	&43.24	&49.12	&51.38	&41.36	&89.12\\
		\textbf{MP1 \& MP3 \& MP4}  &61.26	&80.82	&88.58	&44.73	&50.90	&52.84	&42.37	&89.04\\
		\textbf{MP2 \& MP3 \& MP4}  &38.01	&60.18	&80.09	&23.87	&30.93	&35.85	&24.00	&90.04\\
		\textbf{MP1 \& MP2 \& MP3 \& MP4} &\textbf{66.21}	&\textbf{81.53}	&88.74	&\textbf{46.21}	&\textbf{51.35}	&\textbf{53.22}	&\textbf{42.82}	&89.64\\
		\hline 
	\end{tabular}
	}
	\label{table_performance1}
\end{table*}

\subsection{Implementation Details}

The proposed model is trained on an Nvidia GeForce GTX1080Ti GPU with 11 GB of RAM. The dataset is split into two portions for our evaluation tasks. The first portion contains 80\% instances for training and the second portion contains 20\% instances for testing. 
We choose 64 as the dimension size of the final user embedding.
First, a cross-entropy loss function with 10,000 episodes is leveraged to pre-train the model.
The learning rate is set to 0.001, and the mini-batch size is set to 8.
Subsequently, the learning rate is reduced to 0.0001 and the loss function is adjusted to the policy gradient. 
Furthermore, the weight of the regularization term $\lambda$ is set to 0.08 to prevent the model from overfitting. 
The sequence length of recommended concepts is automatically maintained consistent with the mini-batch size. Through the entire training of the HinCRec-RL model, we update the parameter using the Adam optimizer~\cite{DBLP:journals/corr/KingmaB14}.

\section{Experimental Results} \label{sec_experimental_result}
In this section, we respond to the aforementioned research issues through conducting experiments and analyzing the results.
\subsection{Comparison with Baselines (RQ1)}
We compare the proposed model with the following baselines in order to evaluate its performance.
\begin{itemize}
    \item \textbf{BPR} \cite{steffen2012bpr} obeys a Bayesian method to minimize the pairwise ranking loss for recommendation tasks.
    \item \textbf{MLP} \cite{he2017neural} utilizes a multi-layer perceptron to compute the recommendation probability by leveraging embedding pairs of user and concept.
    \item \textbf{FM} \cite{steffen2012fm} is a typical matrix decomposition method that can easily address the problem of feature combination. 
    For fairness, we use only the embeddings of users and concepts in our experiments.
	\item \textbf{FISM} \cite{santosh2013fism} is a type of CF algorithm. It can utilize the average historical concept embedding and the target one to perform recommendation tasks.
	\item \textbf{NAIS} \cite{he2018nais} is another type of CF algorithm. It can differentiate among the importance of all historical concepts by utilizing the attention mechanism.
	\item \textbf{NASR} \cite{li2017nasr} is considered an improved GRU model, and can calculate an attention coefficient for each historical concept by using the corresponding hidden vectors produced by GRU.
	\item \textbf{HinCRec-SL} represents the proposed model that only applies an HIN for user embedding, without reinforcement learning.
	\item \textbf{HinCRec-RL} represents the proposed model that combines the HIN for user embedding and the reinforcement learning for better recommendation.
\end{itemize}

We conduct comparative experiments between our proposed HinCRec-RL model and several baseline ones using offline datasets. According to the comparative experiments results reported in Table~\ref{table_performance2}, HinCRec-RL largely outperforms all other baselines in terms of $HR@K$, $NDCG@K$, $MRR$ and $AUC$. We also compare the HinCRec-SL/RL model with these baselines and observe that the proposed HIN contains more useful semantic information of users. When comparing HinCRec-SL with HinCRec-RL, we can observe that the proposed reinforcement learning also has a positive effect on the performance based on supervised learning. For example, in terms of $HR@5$ and $NDCG@5$, HinCRec-RL is more than 10\% better than HinCRec-SL. This can be illustrated by the fact that the exploration mechanism in reinforcement learning can yield a better optimized result with our model. More specifically, HinCRec\-RL achieves an $HR@5$ of 66.21\%, an $HR@10$ of 81.53\%, an $HR@20$ of 88.74\%, an $NDCG@5$ of 46.21\%, an $NDCG@10$ of 51.35\%, an $NDCG@20$ of 53.22, and an $MRR$ of 42.82\% on the the dataset of $\textit{XuetangX}$ MOOCs. Although HinCRec-RL performs slightly worse than HinCRec-SL in terms of $HR@20$, it shows significant improvements when compared with the baselines (i.e., 9.74\%-27.54\% increment in terms of $HR@5$, $NDCG@5$, $MRR$ and $AUC$).
We attribute such performance enhancements to the fantastic HinCRec-RL modeling:
(1) The technique based on the hierarchical graph attention aggregation guided by the user's meta-paths is capable of encoding the interrelationship between concepts and users, hence providing a richer and more comprehensive representation learning of concepts and users.
(2) Benefiting from the representation learning, HinCRec-RL can preserve the user's state information and collect more informative rewards from the reinforcement learning process. 
(3) HinCRec-RL first combines the HIN and reinforcement learning to recommend knowledge concepts for users' course learning in MOOC platforms. In contrast, all the baselines ignore this strategy.


\subsection{Effect of Different Meta-Paths (RQ2)}\label{sec_meta_analysis}
We examine various meta-paths and their combinations to demonstrate the effectiveness of these heterogeneous graph patterns in improving the performance of HinCRec-RL.
More specifically, to describe the relatedness among pairs of users, four kinds of meta-paths are chosen for our study. These meta-paths are $MP_1\text{: } U_1\text{-}K_1\text{-}U_1$, $MP_2\text{: }U_1\text{-}K\text{-}U_2\text{-}K\text{-}U_3$, $MP_3\text{: } U_1\text{-}C_1\text{-}K\text{-}C_2\text{-}U_2$, and $MP_4\text{: }U_1\text{-}C_1\text{-}V\text{-}C_2\text{-}U_2$.
The combination of some single or simple meta-paths (e.g., $MP_1$) may affect our model's performance. To illustrate this, we conduct related experiments and show the results in Table~\ref{table_performance1}. Each single meta-path (i.e., $MP_1$, $MP_2$, $MP_3$, $MP_4$) and the combinations thereof contribute different performances to the proposed model. Thus, we can conclude that different single meta-paths can result in nonnegligible effects on HinCRec-RL's performance.
It is apparent that the adjacent combination of single or simple meta-paths has a positive correlation property. Among the four meta-paths, $MP1$ can achieve a high score in several metrics; thus, it can be considered the best designed meta-path for user representation. This is appropriate since the interests of users are usually affected by their neighbors.
Moreover, the more meta-paths the combinations include, the better is the model performance. The most convincing evidence to illustrate this aspect is that the combination of all the four meta-paths causes the best performance of the proposed model.

\subsection{Parameter Analysis (RQ3)}
\begin{figure*}[!t]
	\centering
	\begin{subfigure}[b]{0.24\textwidth}
		\includegraphics[width=\textwidth]{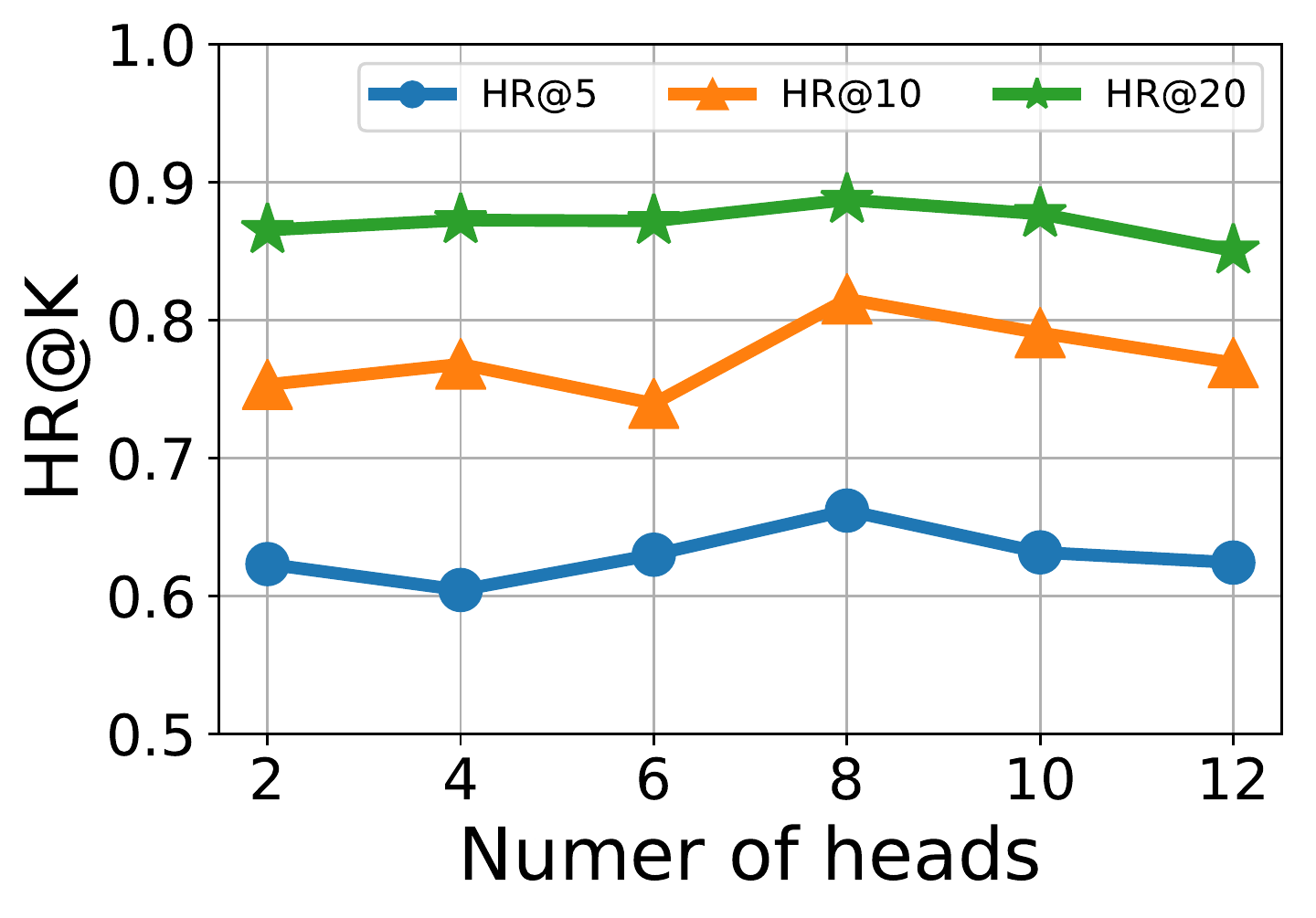}
		\caption{$L$ on $HR@K$}
		\label{fig_Heads_analysis_1}
	\end{subfigure}
	\begin{subfigure}[b]{0.24\textwidth}
		\includegraphics[width=\textwidth]{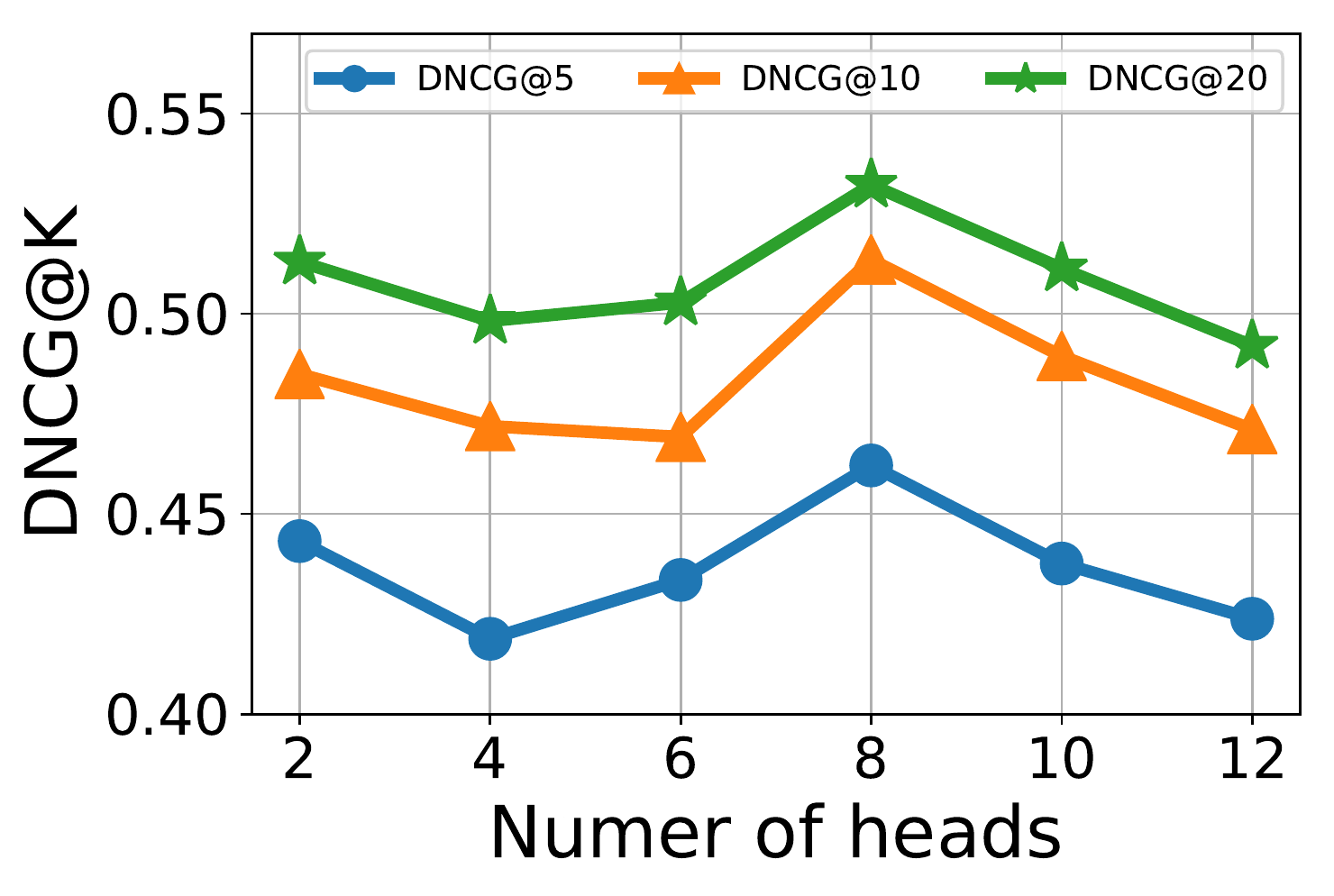}
		\caption{$L$ on $NDCG@K$}
		\label{fig_Heads_analysis_2}
	\end{subfigure}
	\begin{subfigure}[b]{0.24\textwidth}
		\includegraphics[width=\textwidth]{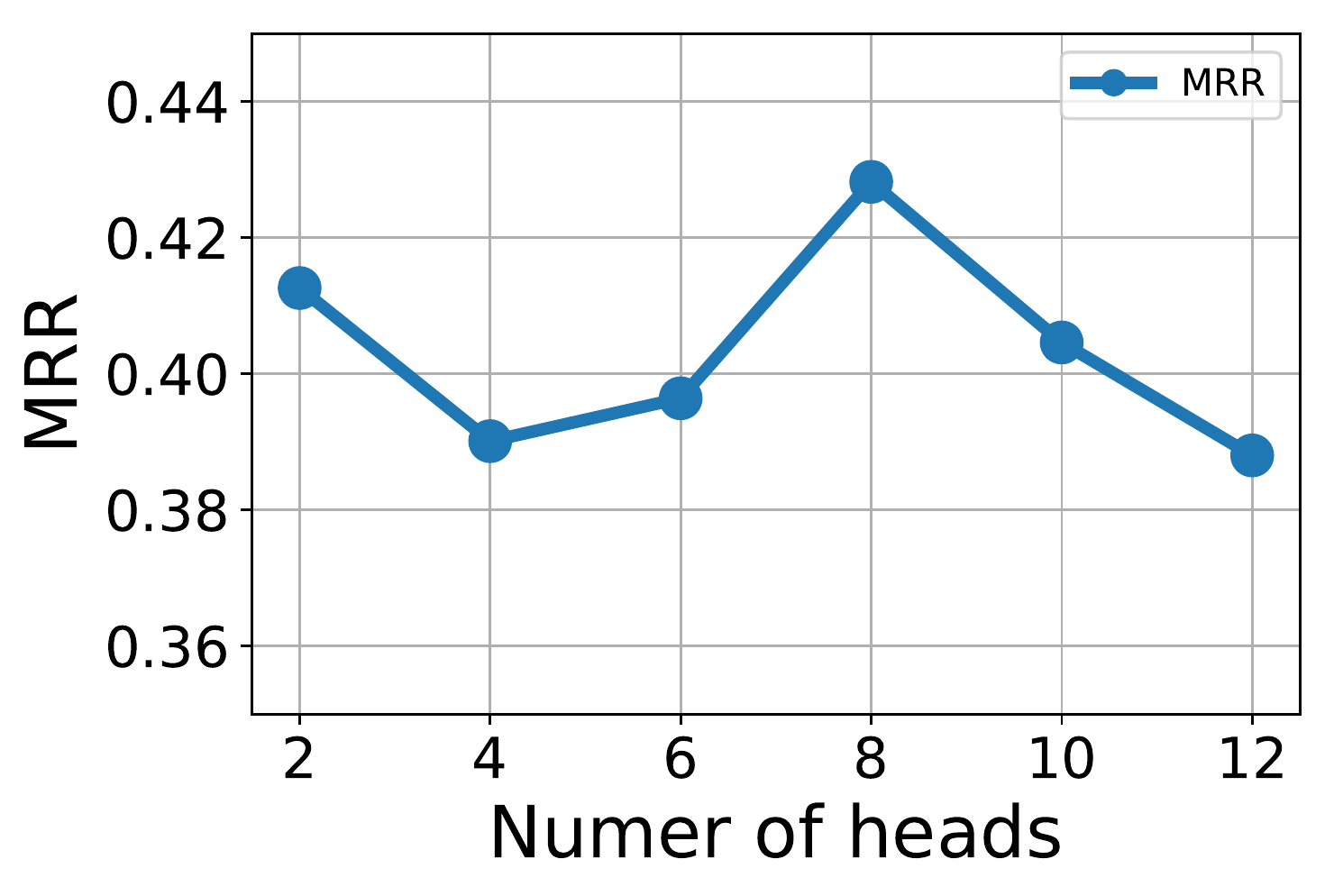}
		\caption{$L$ on $MRR$}
		\label{fig_Heads_analysis_3}
	\end{subfigure}
	\begin{subfigure}[b]{0.24\textwidth}
		\includegraphics[width=\textwidth]{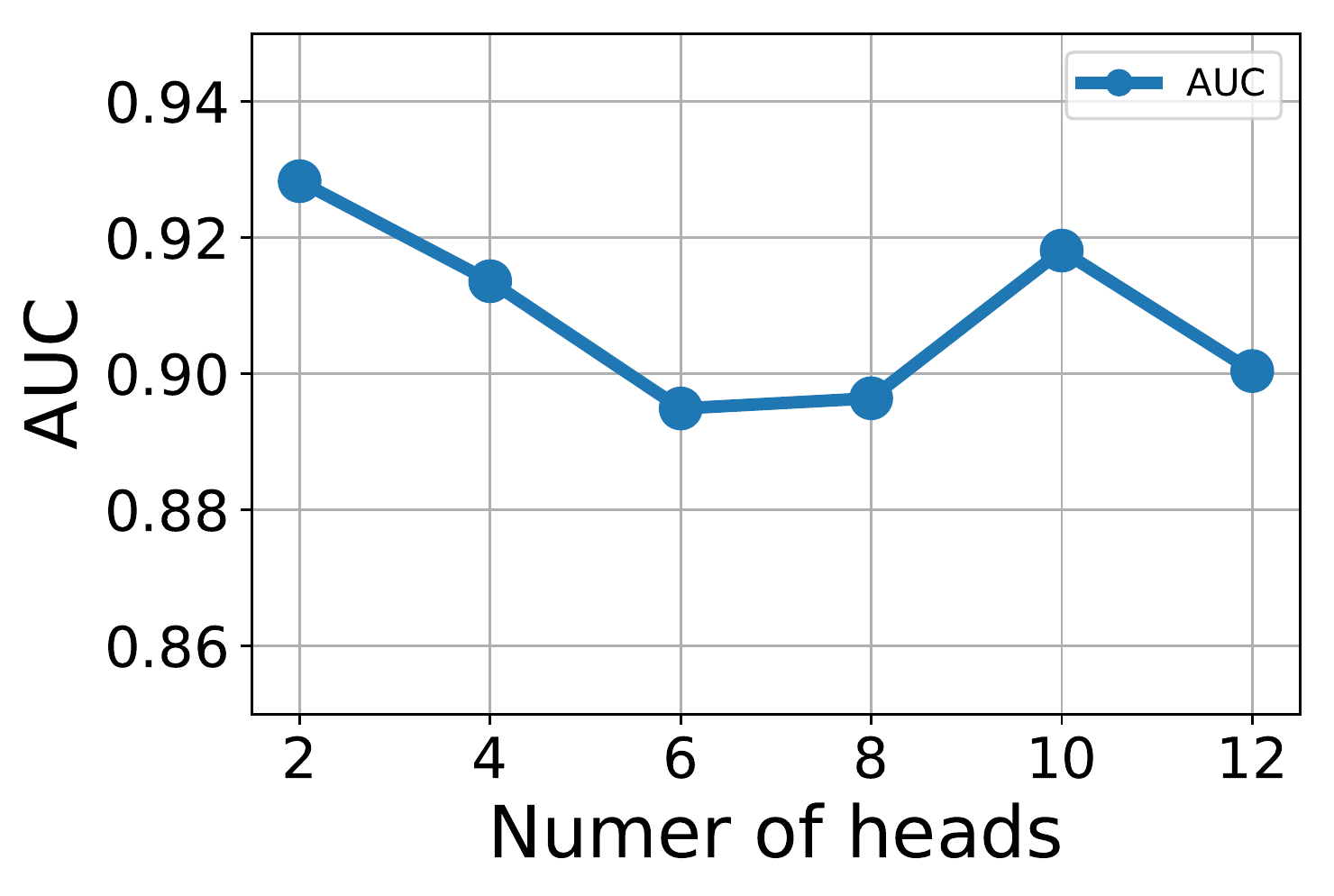}
		\caption{$L$ on $AUC$}
		\label{fig_Heads_analysis_4}
	\end{subfigure}
	\caption{Parameter sensitivity of HinCRec-RL over attention head $L$ in HIN embedding.}
	\label{fig_heads}
\end{figure*}
\begin{figure*}[!t]
	\centering
	\begin{subfigure}[b]{0.24\textwidth}
		\includegraphics[width=\textwidth]{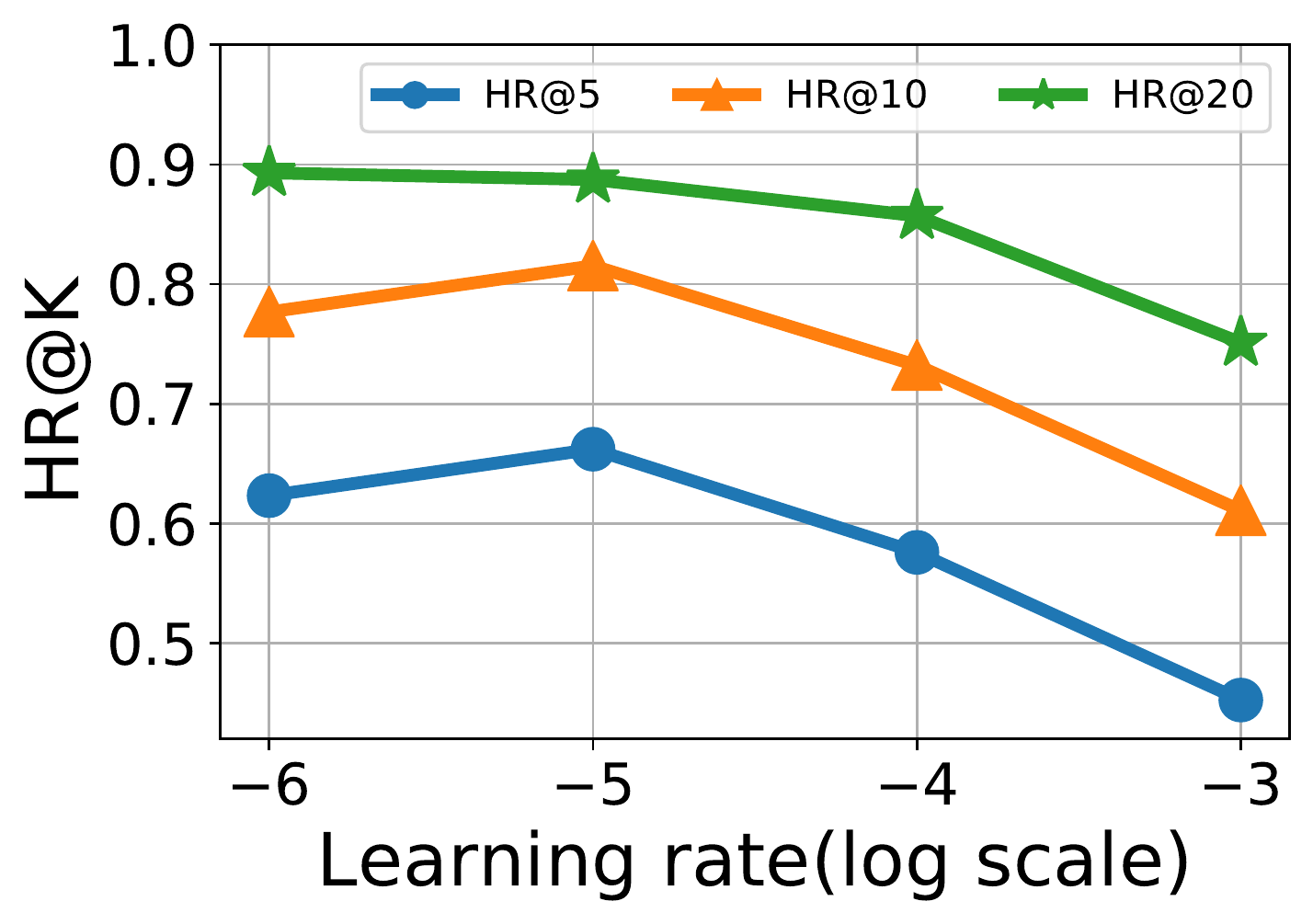}
		\caption{$lr$ on $HR@K$}
		\label{fig_lr_analysis_1}
	\end{subfigure}
	\begin{subfigure}[b]{0.24\textwidth}
		\includegraphics[width=\textwidth]{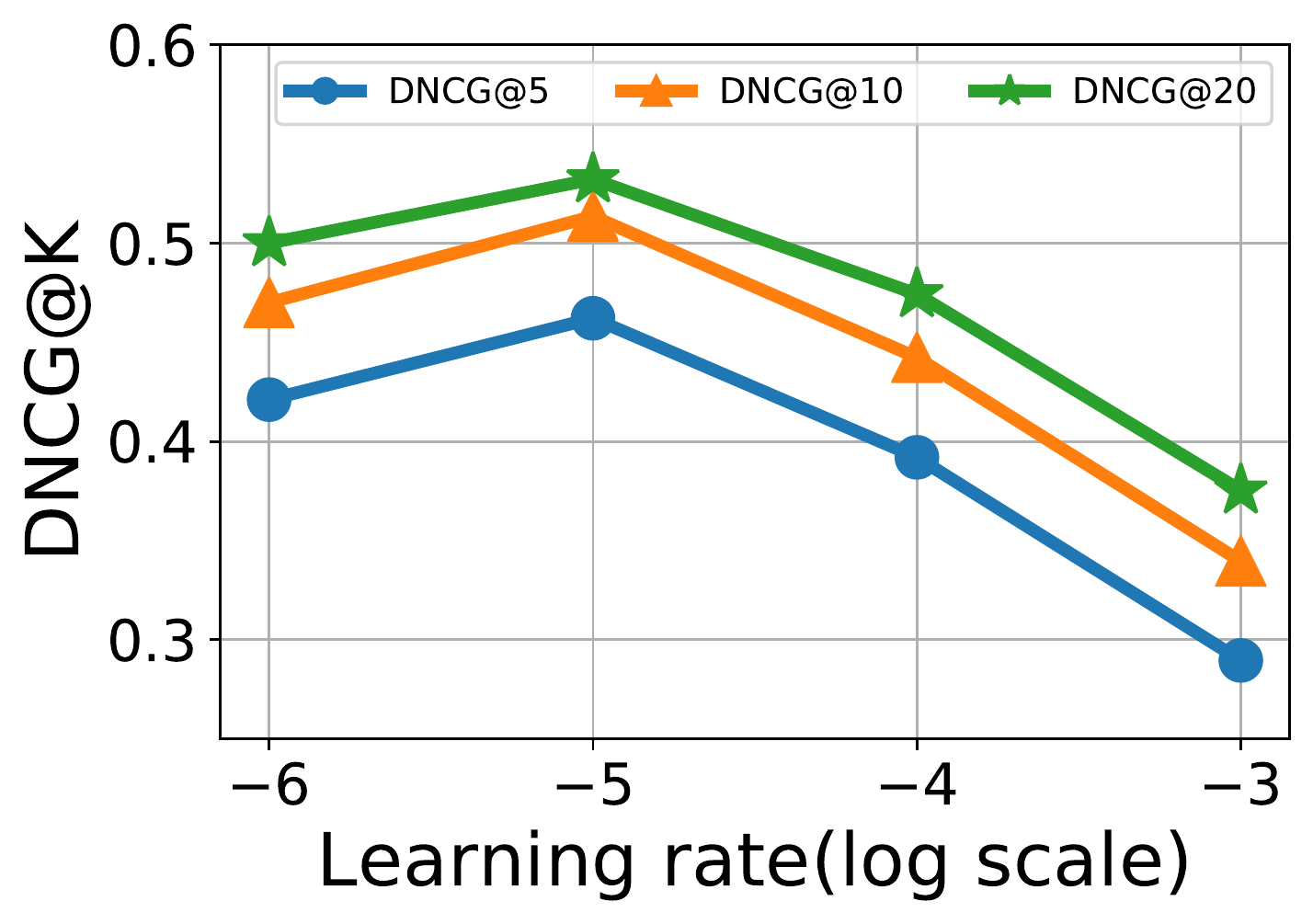}
		\caption{$lr$ on $NDCG@K$}
		\label{fig_lr_analysis_2}
	\end{subfigure}
	\begin{subfigure}[b]{0.24\textwidth}
		\includegraphics[width=\textwidth]{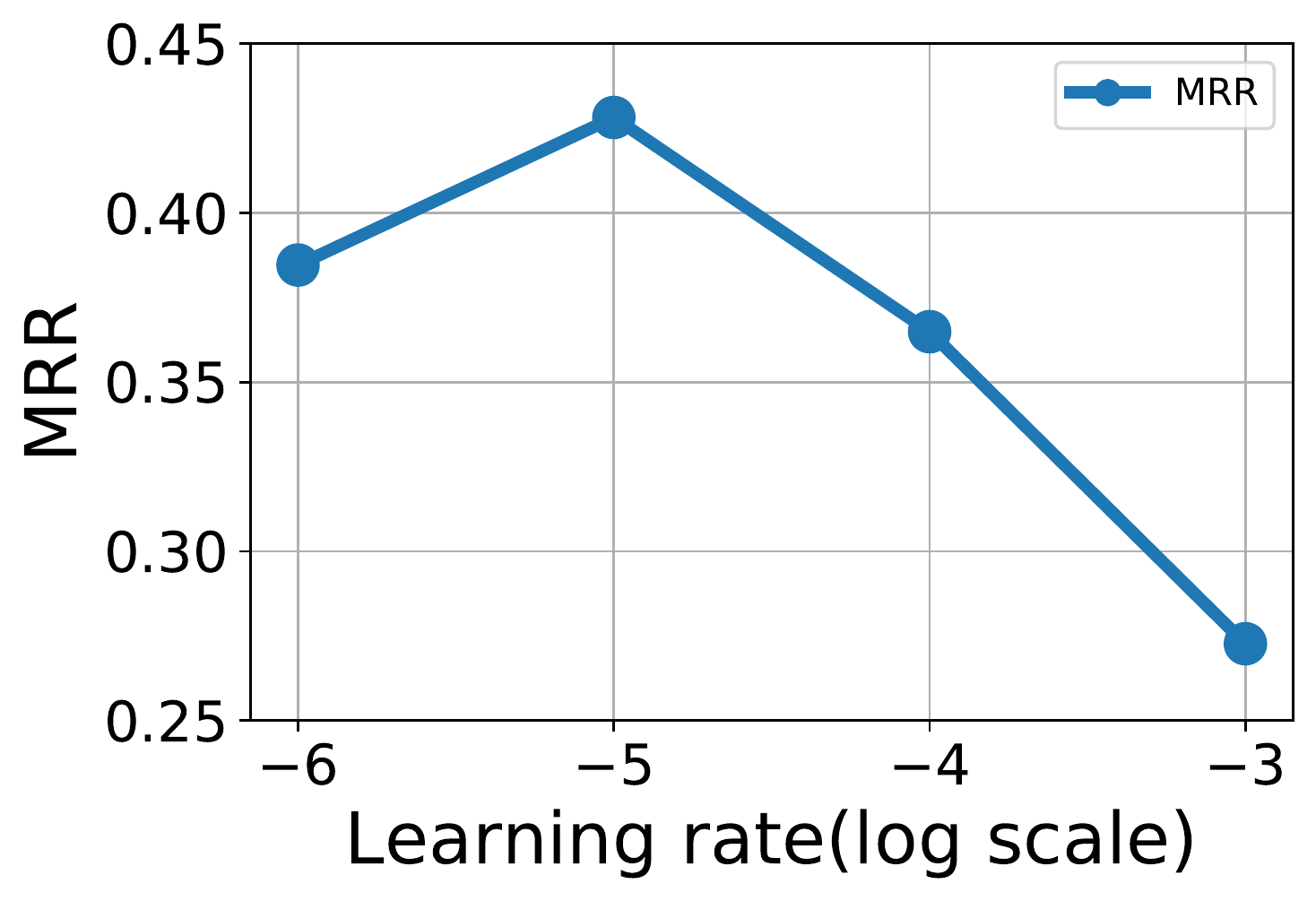}
		\caption{$lr$ on $MRR$}
		\label{fig_lr_analysis_3}
	\end{subfigure}
	\begin{subfigure}[b]{0.24\textwidth}
		\includegraphics[width=\textwidth]{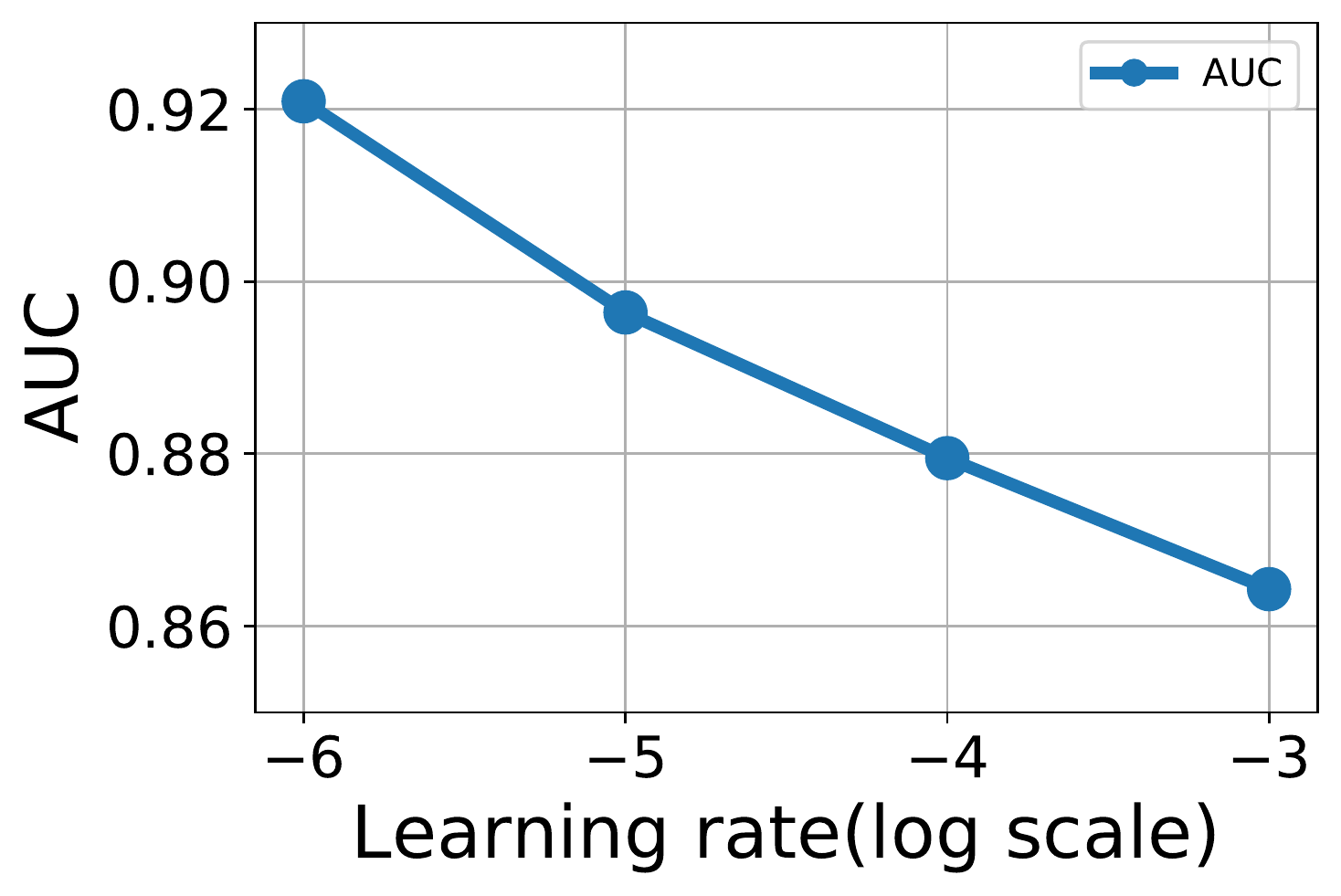}
		\caption{$lr$ on $AUC$}
		\label{fig_lr_analysis_4}
	\end{subfigure}
	\caption{Parameter sensitivity of HinCRec-RL over learning rate $lr$.}
	\label{fig_lr}
\end{figure*}

\begin{figure*}[!t]
	\centering
	\begin{subfigure}[b]{0.24\textwidth}
		\includegraphics[width=\textwidth]{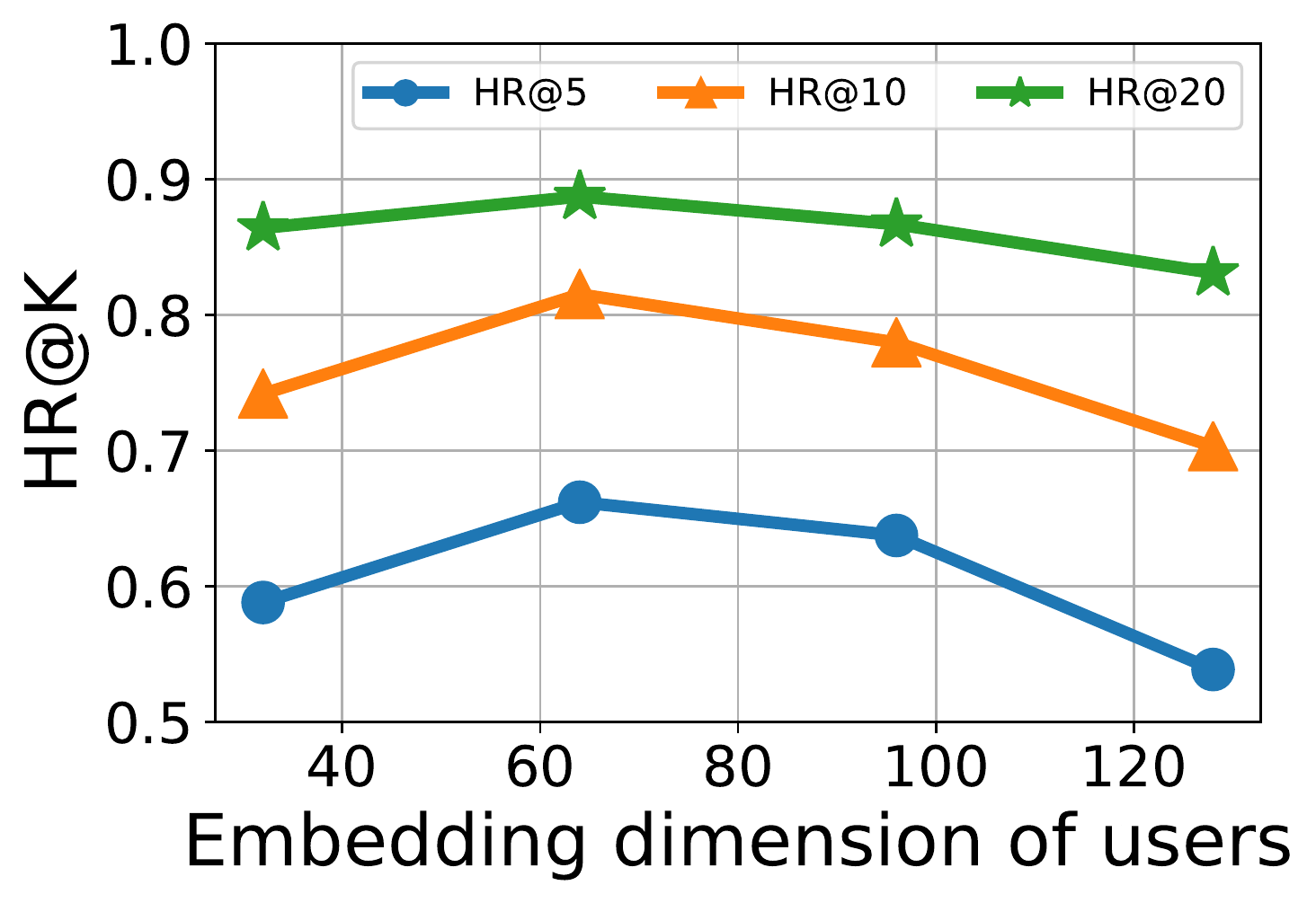}
		\caption{$d$ on $HR@K$}
		\label{fig_d_analysis_1}
	\end{subfigure}
	\begin{subfigure}[b]{0.24\textwidth}
		\includegraphics[width=\textwidth]{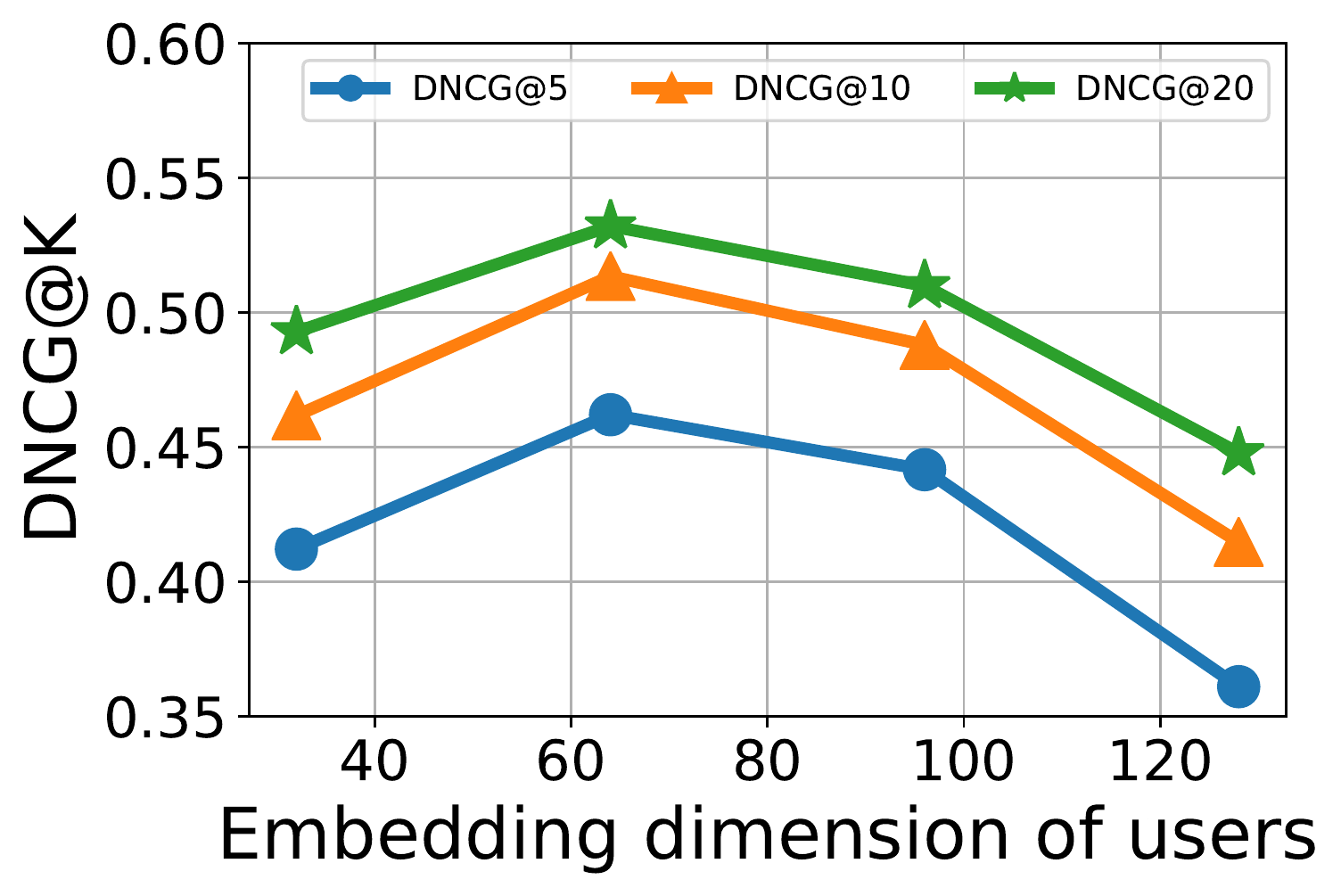}
		\caption{$d$ on $NDCG@K$}
		\label{fig_d_analysis_2}
	\end{subfigure}
	\begin{subfigure}[b]{0.24\textwidth}
		\includegraphics[width=\textwidth]{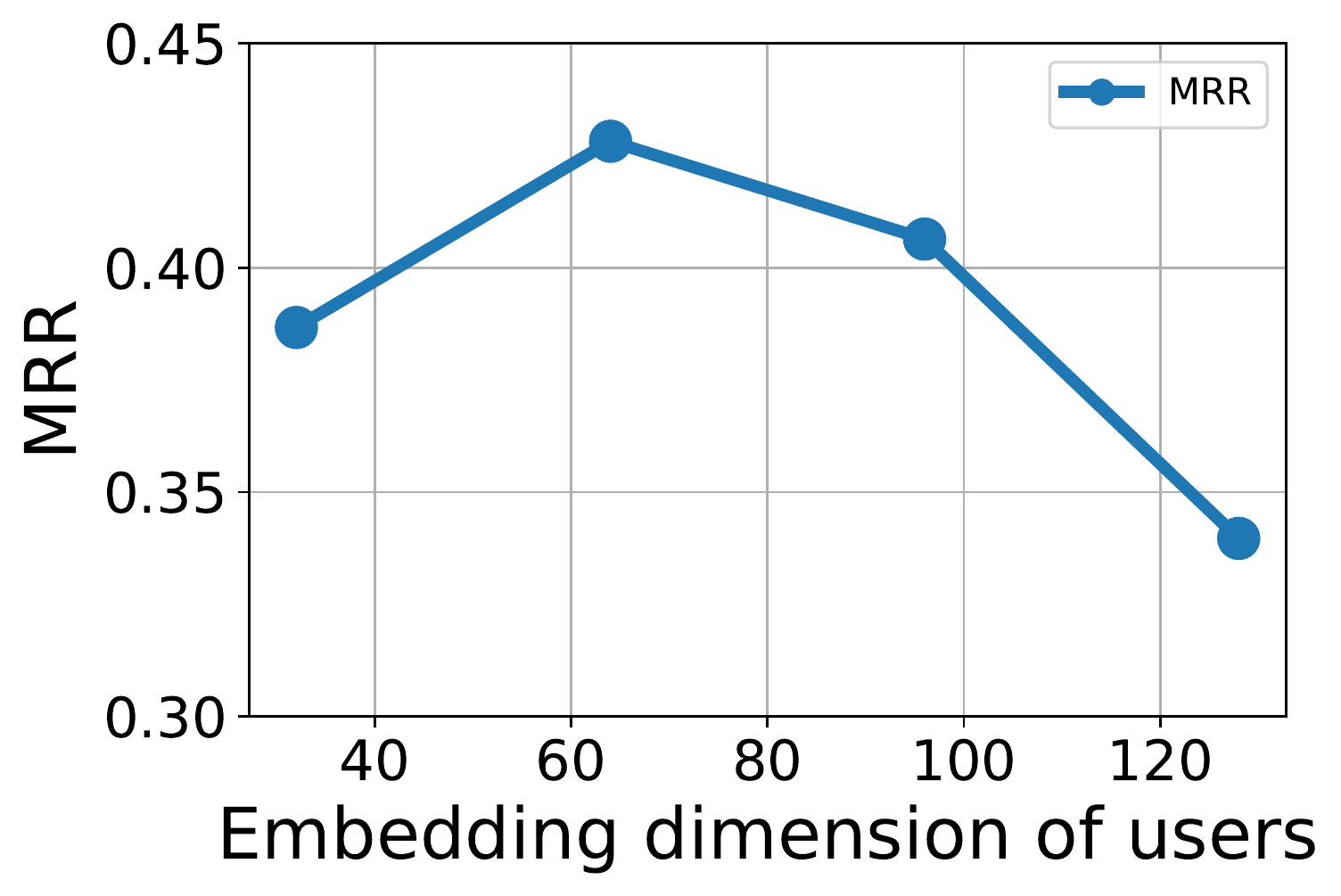}
		\caption{$d$ on $MRR$}
		\label{fig_d_analysis_3}
	\end{subfigure}
	\begin{subfigure}[b]{0.24\textwidth}
		\includegraphics[width=\textwidth]{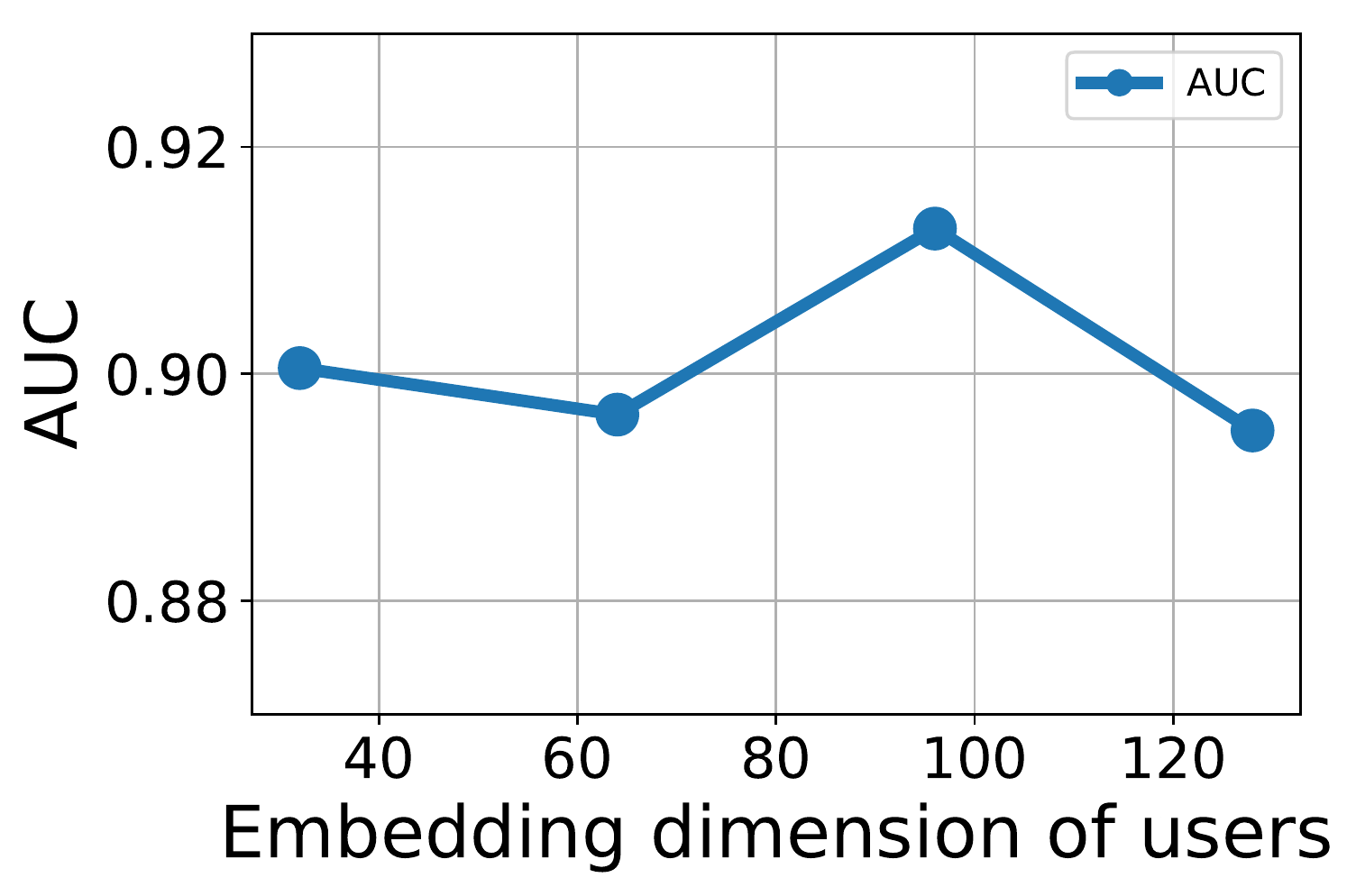}
		\caption{$d$ on $AUC$}
		\label{fig_d_analysis_4}
	\end{subfigure}
	\caption{Parameter sensitivity of HinCRec-RL over embedding dimension $d$ in HIN embedding.}
	\label{fig_d}
\end{figure*}
\begin{figure*}[!t]
	\centering
	\begin{subfigure}[b]{0.24\textwidth}
		\includegraphics[width=\textwidth]{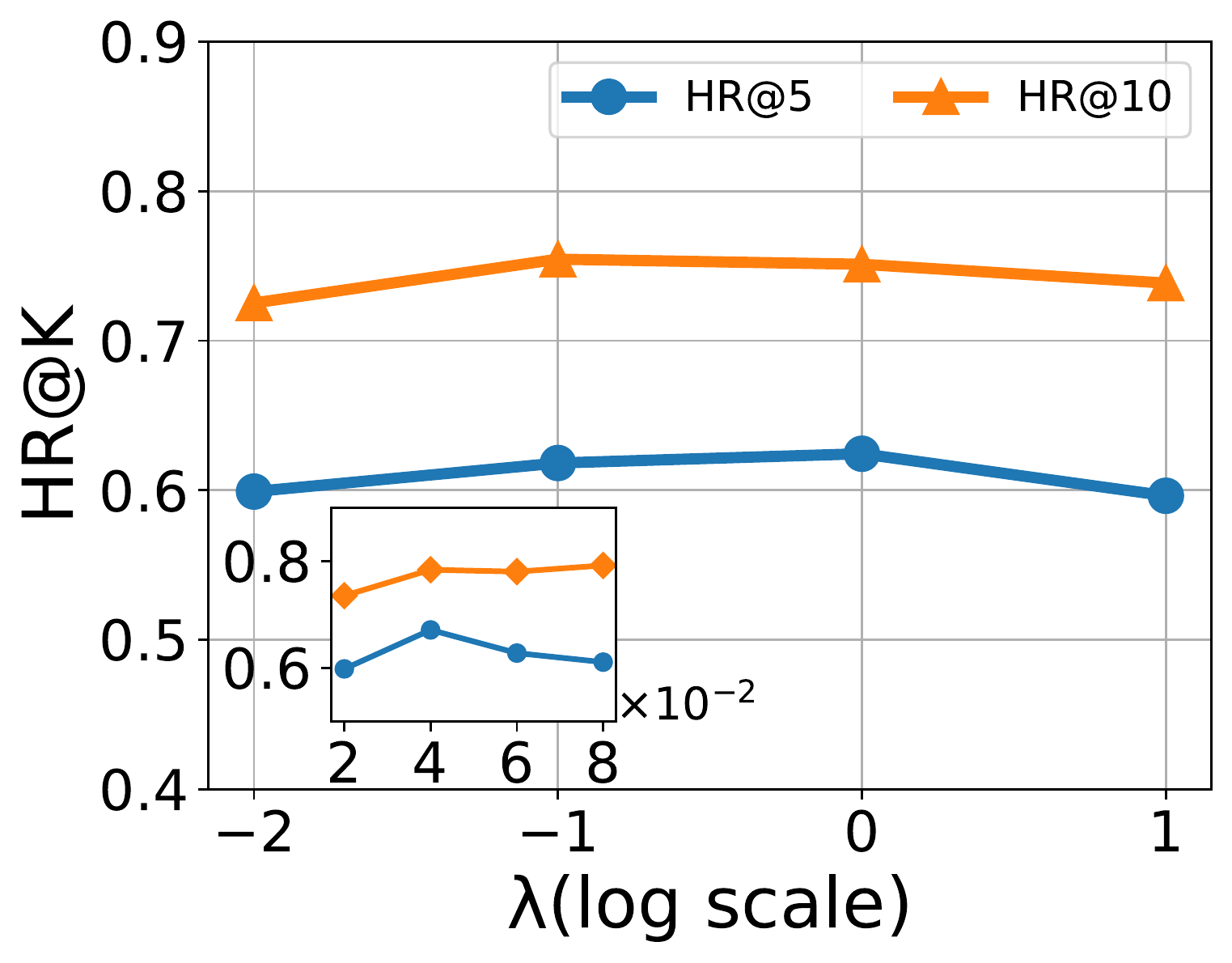}
		\caption{$\lambda$ on $HR@K$}
		\label{fig_lambda_analysis_1}
	\end{subfigure}
	\begin{subfigure}[b]{0.24\textwidth}
		\includegraphics[width=\textwidth]{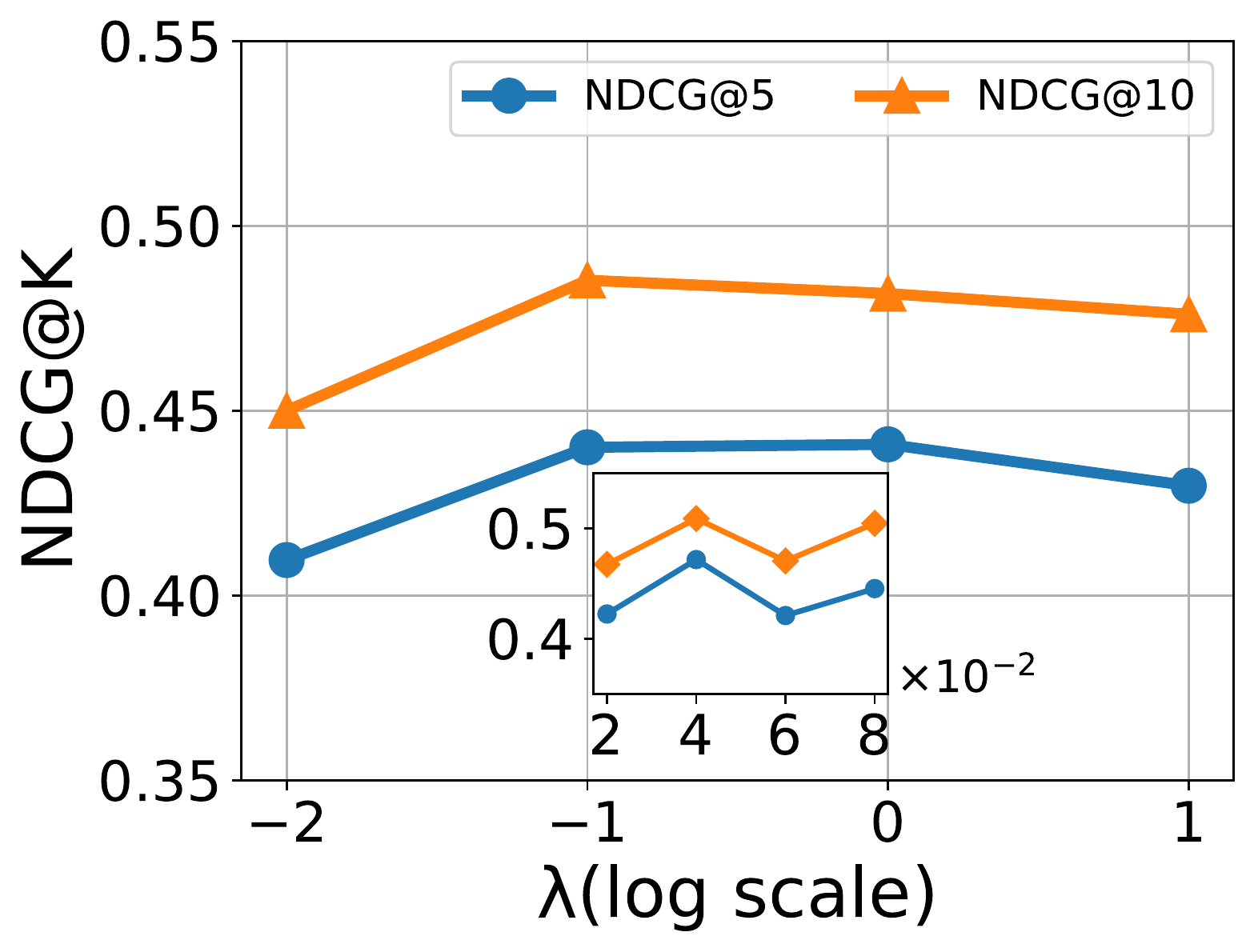}
		\caption{$\lambda$ on $NDCG@K$}
		\label{fig_lambda_analysis_2}
	\end{subfigure}
	\begin{subfigure}[b]{0.24\textwidth}
		\includegraphics[width=\textwidth]{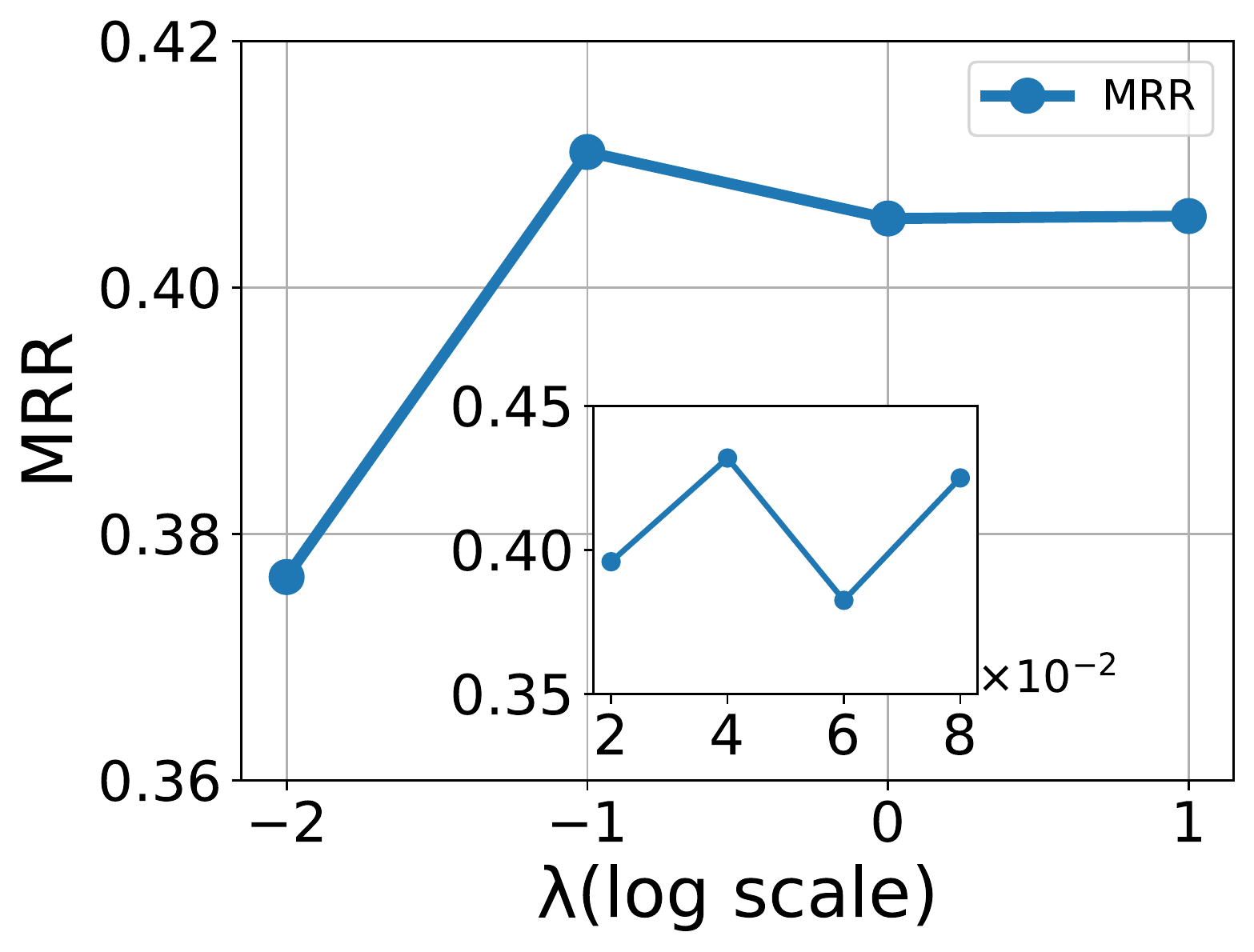}
		\caption{$\lambda$ on $MRR$}
		\label{fig_lambda_analysis_3}
	\end{subfigure}
	\begin{subfigure}[b]{0.24\textwidth}
		\includegraphics[width=\textwidth]{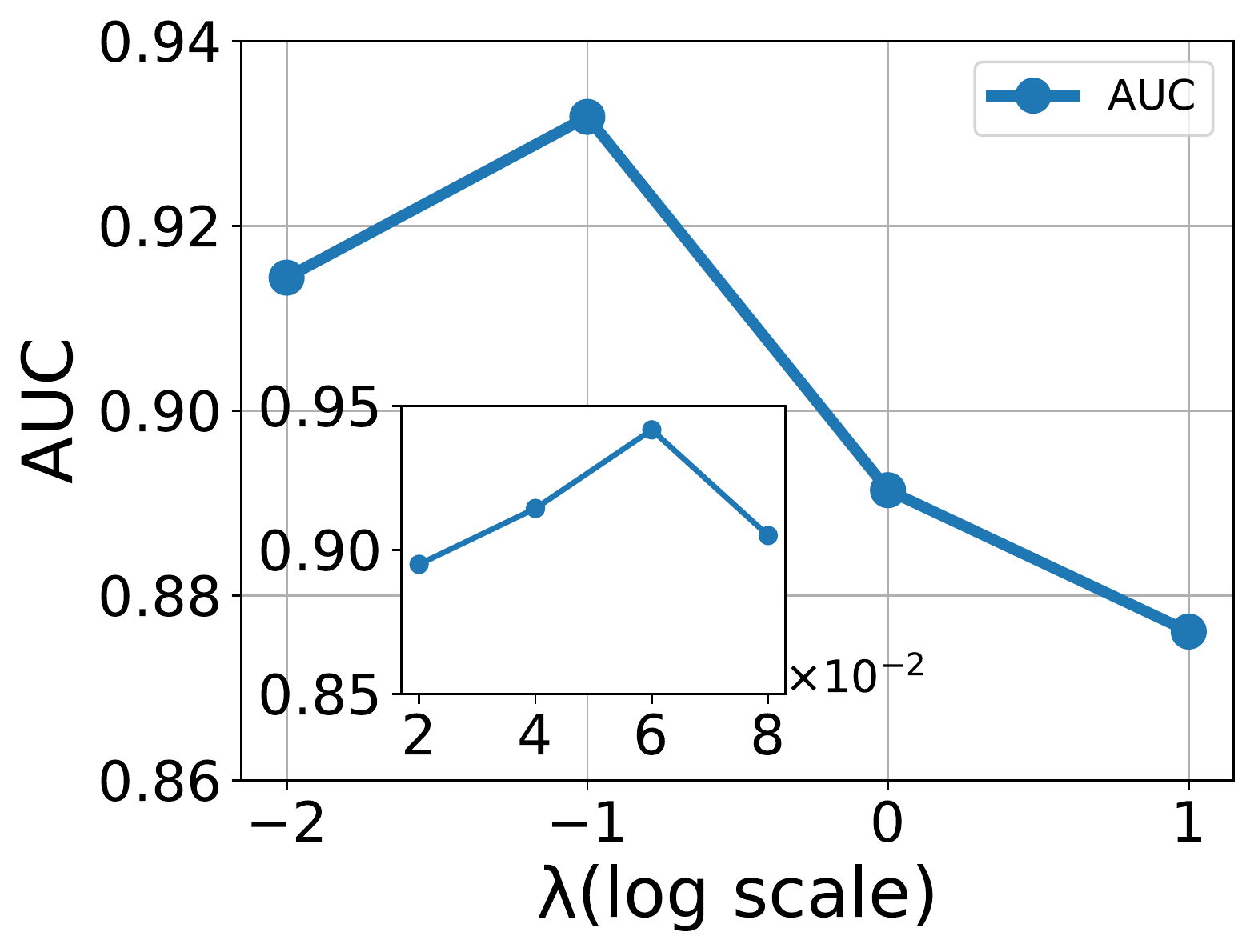}
		\caption{$\lambda$ on $AUC$}
		\label{fig_lambda_analysis_4}
	\end{subfigure}
\caption{Parameter sensitivity of HinCRec-RL over $\lambda$.}
	\label{fig_lambda}
\end{figure*}

\begin{figure*}[!t]
	\centering
	\begin{subfigure}[b]{0.24\textwidth}
		\includegraphics[width=\textwidth]{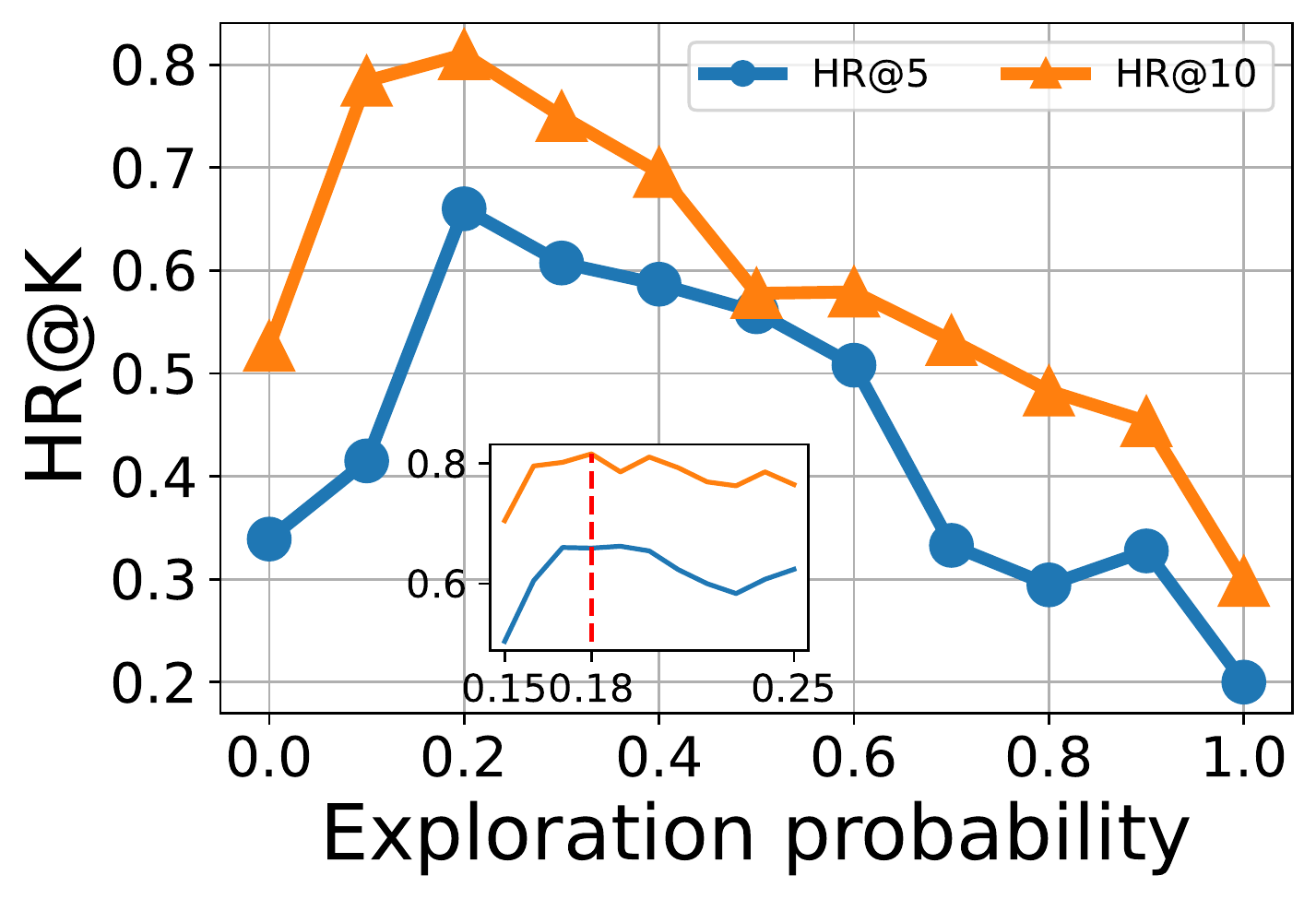}
		\caption{$\epsilon$ on $HR@K$}
		\label{fig_epsilon_analysis_1}
	\end{subfigure}
	\begin{subfigure}[b]{0.24\textwidth}
		\includegraphics[width=\textwidth]{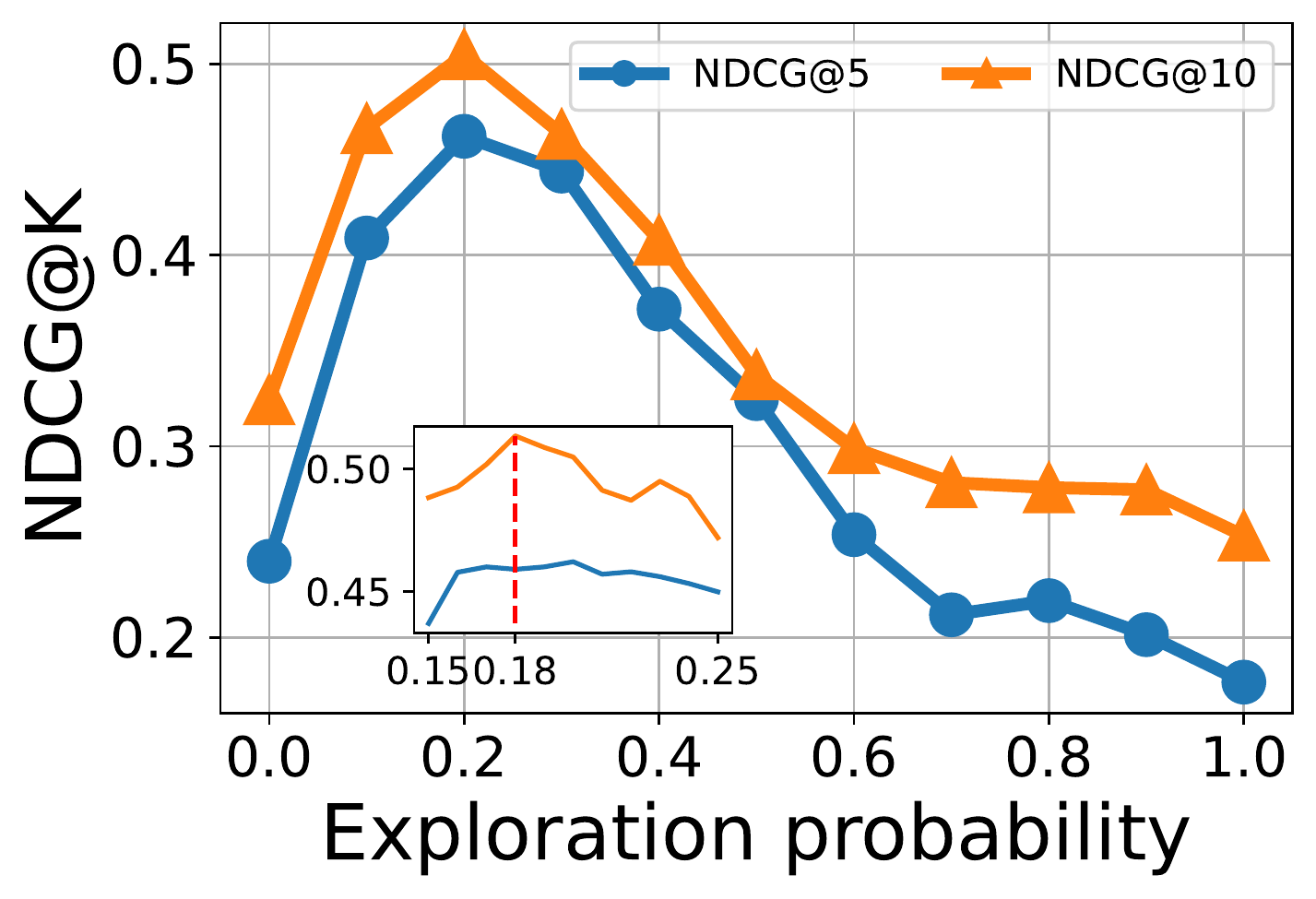}
		\caption{$\epsilon$ on $NDCG@K$}
		\label{fig_epsilon_analysis_2}
	\end{subfigure}
	\begin{subfigure}[b]{0.24\textwidth}
		\includegraphics[width=\textwidth]{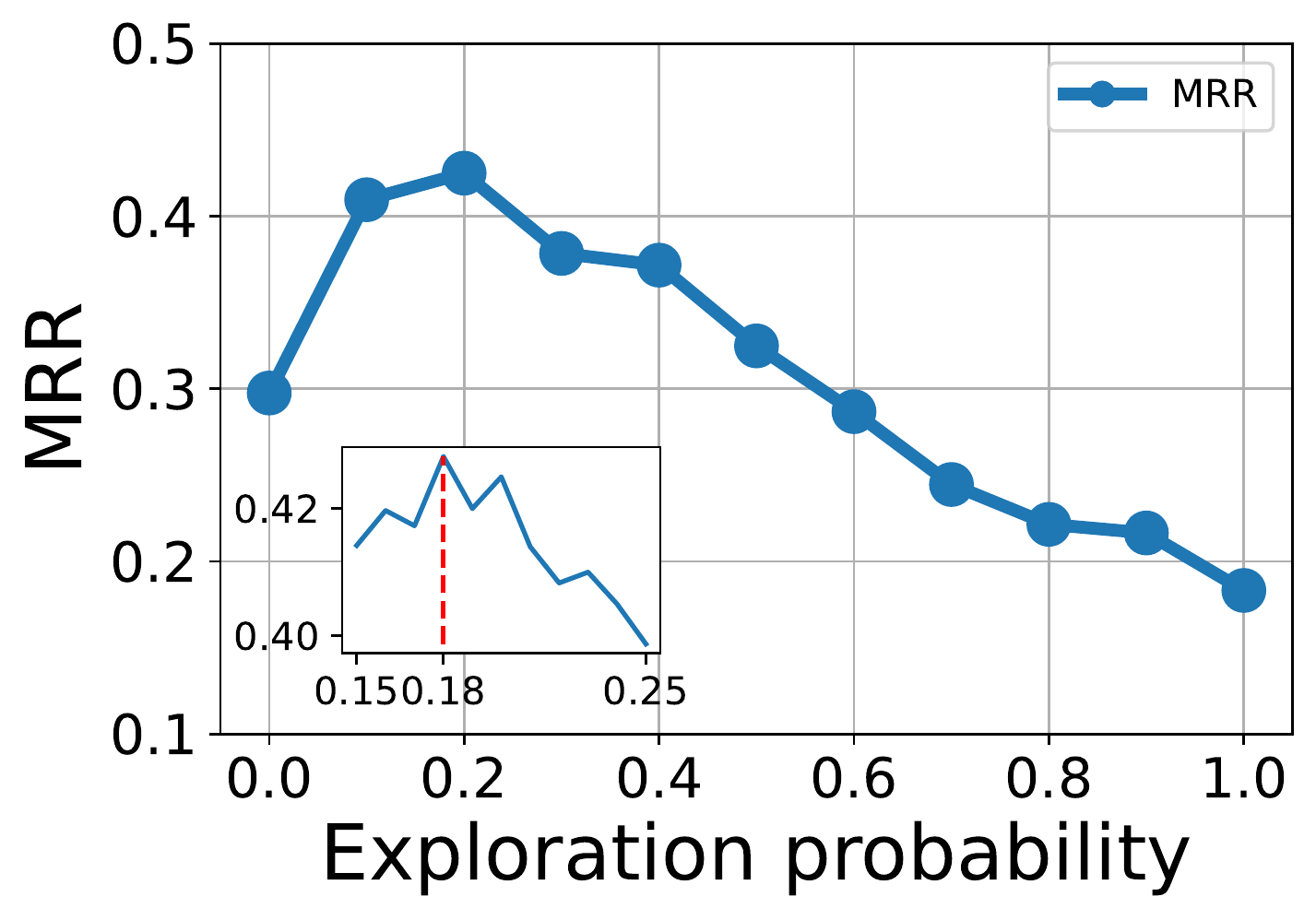}
		\caption{$\epsilon$ on $MRR$}
		\label{fig_epsilon_analysis_3}
	\end{subfigure}
	\begin{subfigure}[b]{0.24\textwidth}
		\includegraphics[width=\textwidth]{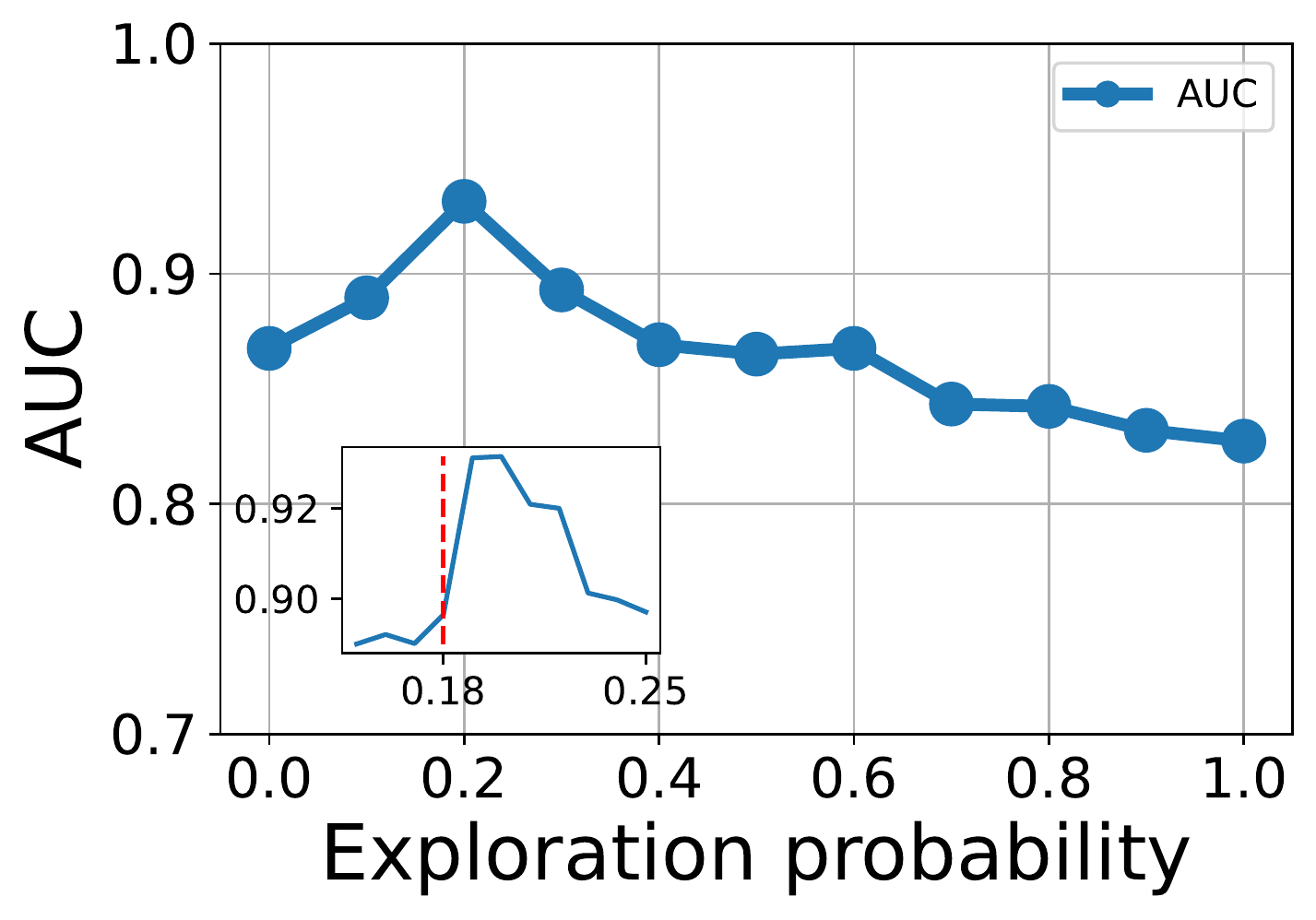}
		\caption{$\epsilon$ on $AUC$}
		\label{fig_epsilon_analysis_4}
	\end{subfigure}
\caption{Parameter sensitivity of HinCRec-RL over the probability of exploration $\epsilon$.}
	\label{fig_Epsilon}
\end{figure*}

We study parameter sensitivity and present the results of HinCRec-RL for diverse parameters in Figures~\ref{fig_heads}-\ref{fig_lambda}.

\subsubsection{Impact of Attention Head $L$ in HIN Embedding}
We analyze the performance of HinCRec-RL with varying numbers of attention heads, proving the validity of multi-head attention involved in HIN embeddings.
In sequence, there are 2, 4, 6, 8, 10 and 12 attention heads in our analytical design. The experimental results are given in Figure~\ref{fig_heads}.
We can see that the increase of the number of heads slightly elevates the model's performance. 
Although the attention mechanism included in our model requires slightly longer training time, the results prove that it can make our model more reliable and robust.

\begin{figure*}[!t]
	\centering
	\begin{subfigure}[b]{0.8\textwidth}
		\includegraphics[width=\textwidth]{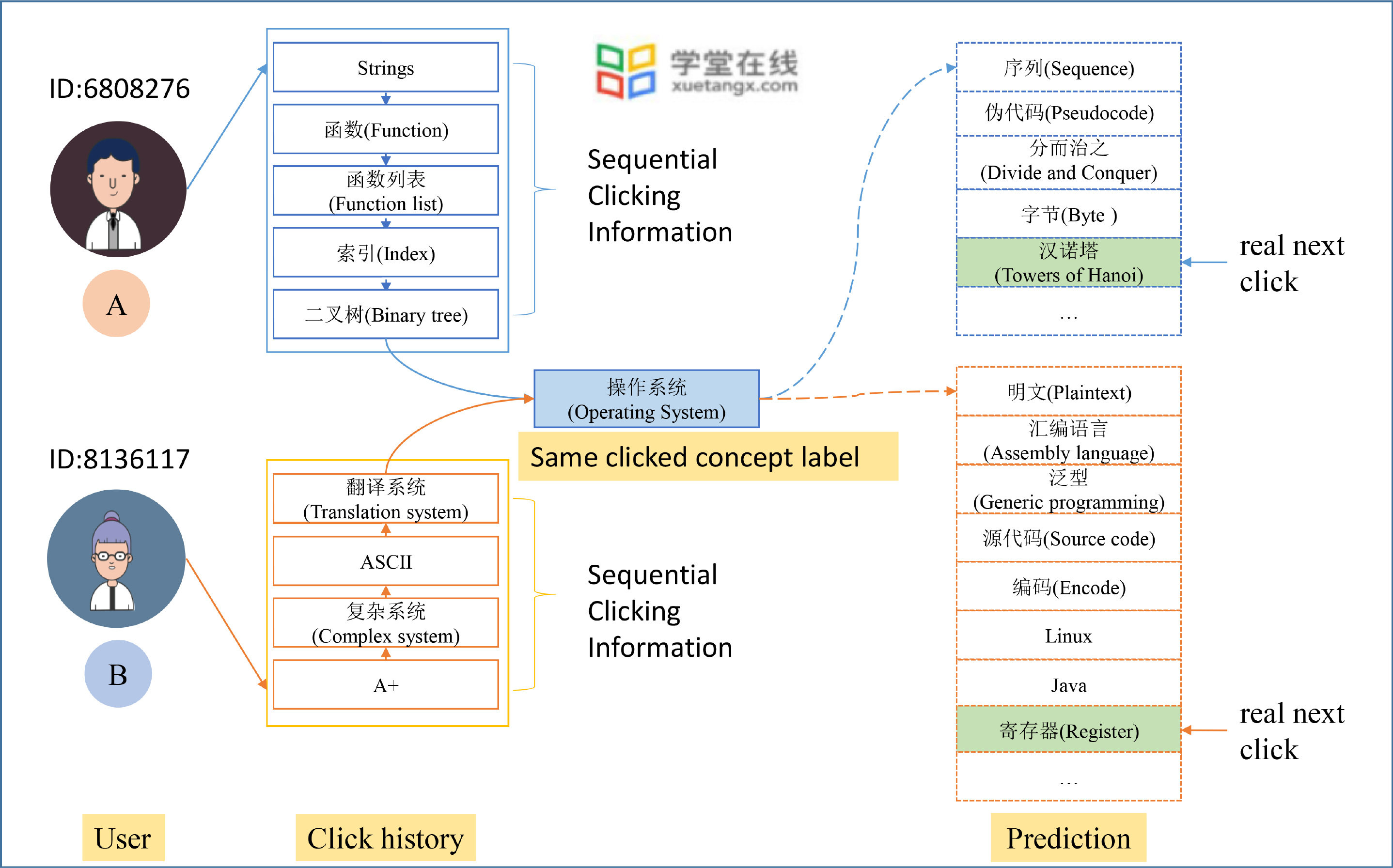}
		\caption{Case 1}
		\label{fig_data_analysis_1}
	\end{subfigure}
	\begin{subfigure}[b]{0.8\textwidth}
		\includegraphics[width=\textwidth]{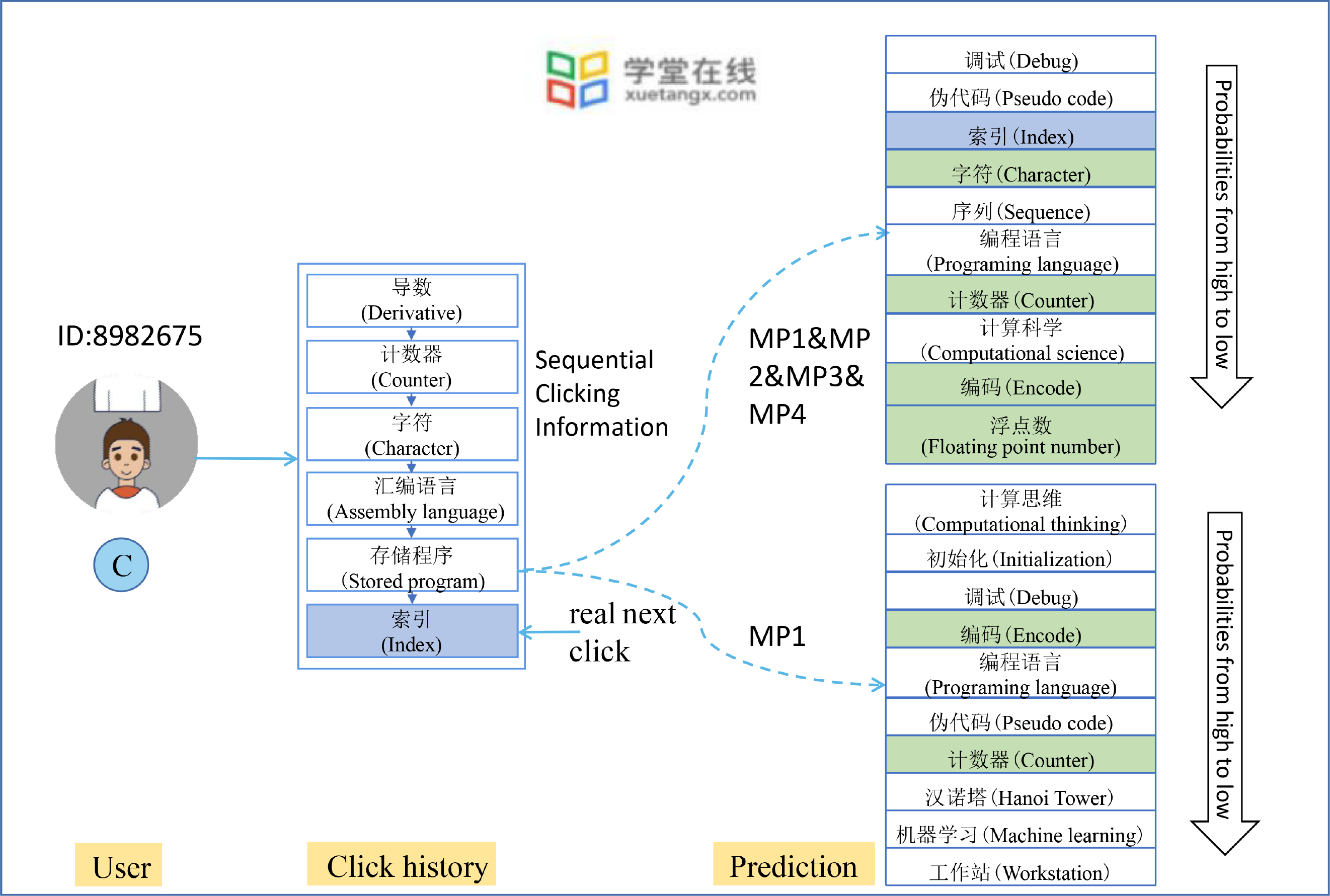}
		\caption{Case 2}
		\label{fig_data_analysis_4}
	\end{subfigure}
    \caption{A case of the diversity of  concept recommendation, which contains a sequence of clicked concepts within $\textit{XuetangX}$'s dynamic learning environment.}
	\label{fig_case_study}
	
\end{figure*}
\subsubsection{Impact of Learning Rate $lr$}
We tune $lr$ to learn a recommendation policy better. The impact of $lr$ on HinCRec-RL is described in Figure~\ref{fig_lr}. We can see that, when increasing the learning rate from $10e\text{-}6$ to $10e\text{-}5$, the performance is improved in terms of the $HR@K$, $NDCG@K$, and $MRR$, whereas the performance is degraded when increasing $lr$ further. Therefore, we can conclude that the learning rate affects HinCRec-RL. The best $lr$ is approximately $10e\text{-}5$.

\subsubsection{Impact of Embedding Dimension $d$}
Aiming to investigate the embedding dimension settings of entities, we conduct experiments with various numbers of dimensions (i.e., 32, 64, 96 and 128). The experiment results are shown in Figure~\ref{fig_d}. 
When increasing $d$ from 32 to 64, the performance is improved in terms of $HR@K$, $NDCG@K$, and $MRR$, whereas the performance is degraded when increasing $d$ further. 
Our final settings have the user and knowledge concept vector dimension set to 64.
In light of the experiment findings, we can conclude that a reasonable embedding dimension is crucial to enhance recommendation performance in our task.

\subsubsection{Impact of Regularization Rate $\lambda$}
We conduct experiments on entropy regularization to investigate the efficacy of the parameter $\lambda$. We set $\lambda$ to 0.01, 0.1, 1 and 10, and show the corresponding experimental results in Figure~\ref{fig_lambda}. In addition, for more fine-grained analysis, we set $\lambda$ to 0.02, 0.04, 0.06 and 0.08, and present the results in a zoomed view. Figure~\ref{fig_lambda} shows that, 
when increasing $\lambda$ from $10e\text{-}2$ to $10e\text{-}1$, the performance is improved in terms of the $HR@K$, $NDCG@K$, $MRR$ and $AUC$, whereas the performance is degraded when increasing $\lambda$ further. 
Therefore, we can conclude that $\lambda$ affects the performance of HinCRec-RL, and the best $\lambda$ is approximately $10e\text{-}1$.

\subsubsection{Impact of Exploration Probability $\epsilon$}\label{parameters_exploration_probability}
We investigate the random action probability in an $\epsilon$-greedy policy. We set the range of $\epsilon$ range from 0 to 1 in steps of 0.1, and present the corresponding results in Figure~\ref{fig_Epsilon}. 
When increasing $\epsilon$ from 0 to 0.2, the performance is improved in terms of the $HR@K$, $NDCG@K$, $MRR$ and $AUC$, whereas the performance is degraded when increasing $\epsilon$ from 0.2 to 1 further.
The thumbnail of Figure~\ref{fig_Epsilon} shows that HinCRec-RL performs optimally when $\epsilon$ is approximately 0.18. 
Therefore, the exploration probability affects the performance of HinCRec-RL. 

\subsection{Case Study}
Figure~\ref{fig_case_study} shows a case study to examine the availability of HinCRec-RL, particularly its ability to adapt to dynamic environments when providing personalized concept recommendations.
User A (ID: 6806276) and User B (ID: 8136117) are randomly chosen from the testing dataset. They have different educational backgrounds and consequently experience different learning paths in the $\textit{XuetangX}$ MOOC platform. Figure~\ref{fig_data_analysis_1} first shows the clicked history of the two users and then displays the concept lists recommended to them separately.
From this figure, we can observe that User A is learning a course named ``\textit{Introduction to Algorithms}'', whereas User B is learning a course named ``\textit{Natural Language Processing}''.
Both users selected distinct knowledge concepts pertaining to the same target course ``\textit{Operating System}''. In this case, our model can still adaptively consider the currently selected concepts, and creatively recommend the next one depending on the past.
We attribute it to the reinforcement learning mechanism we have adopted, i.e., the proposed approach can dynamically interact with the environment.

On the other hand, we also show a case of recommending different concept lists for a given user according to different meta-paths, as shown in Figure~\ref{fig_data_analysis_4}. For User C with ID 8982675, given his click history, we show the recommended results by meta-paths $MP1\&MP2\&MP3\&MP4$ and $MP1$, respectively. The result recommended by $MP1\&MP2\&MP3\&MP4$ is more precise than that recommended by $MP1$.  We attribute this to the fact that the meta-paths can aggregate the information related to users from neighbors, as data augmentation. This also demonstrates that introducing heterogeneous graph information into user embedding is effective.

\section{Related Work}\label{sec_related_work}
We will present the relevant studies from three aspects, namely, mining in MOOCs, HIN learning, and recommender systems.



\subsection{Mining in MOOCs}
MOOCs can be viewed as interactive systems containing large-scale multimedia objects, user interaction records, and HINs. Several data mining techniques have been employed to mine these plentiful resources, to improve the user experience on online learning platforms. 
These data can also be exploited to reveal, in detail, how students behave and how learning takes place. Existing studies on MOOCs fall under two major categories. One line of work focused on the course content, such as course concept extraction~\cite{pan2017course,lu2019concept}, course concept expansion~\cite{JifanACL19}, and prerequisite relation discovery~\cite{pan2017prerequisite,xz1}.
It is common to learn how the representations of entities leverage both context information and content. In addition to the title and description of each course, other types of entities, such as teachers, videos, and concepts, can be constructed into heterogeneous networks to explore more corresponding beneficial semantic relationships. 
\citet{pan2017prerequisite} explored the prerequisite relations among concepts by leveraging the concept semantics of course , the video context of course and the structure of course. The other line of work pays attention to the learning behavior patterns of students~\cite{qiu2016modeling}. For example, some studies~\cite{jing2017guess,zhang2019hierarchical} recommended courses to students. 
Additionally, some studies~\cite{feng2019understanding,xz2} predicted dropouts, whereas other studies~\cite{xz3} predicted the performances of students.
Diverse students' behaviors are essential for MOOC research because they are meaningful to mine the motivations~\cite{feng2019understanding}, cognitive abilities~\cite{xz4}, and social activities~\cite{ke2021cross} of students static course research and contents. 
\citet{xz5} collected student behaviors data from MOOCs, which indicates the relationships between courses and concepts. These records include student profiles, video watching behaviors, exercise behaviors, comments and replies.

Among the related studies, course recommendation is the task most relevant to ours. Our study not only deploys concept recommendation in the dynamic interactive environment, but also puts more focus on modeling the dynamic interaction among users and concepts, and enriching the representation of users based on heterogeneous information networks. Another related work to ours is~\cite{gong2020attentional}, which proposed an attentional graph convolutional network for concept recommendation. 
Different to ~\cite{gong2020attentional}, we resort to modeling the dynamic interactions among users and concepts by reinforcement learning.

\subsection{Learning with Heterogeneous Information Networks}
Many applications, e.g recommender systems~\cite{hu2018local}, social event detection systems~\cite{10.1145/3447585,10.1145/3442381.3449834}, etc, capture the meaningful semantic and structural information by constructing an HIN in auxiliary data. Specially, lots of entity types (e.g., users, courses, and knowledge concepts) and rich edges among them which naturally constitute an HIN, can be easily found in E-commerce applications or online platforms (e.g., Amazon, eBay and \textit{XuetangX} MOOC).

HIN-based methods have outperformed the other previous methods in mining more meaningful information owing to their excellent ability.  
\citet{sun2011pathsim} suggested using meta-paths to explore the information network's meta-structure which can systematically explore fruitful semantic relationships across several kinds of objects. Metapath2vec~\cite{dong2017metapath2vec} generated the meta-path by utilizing skip-gram and random walk to represent the HIN. 
\citet{cai2018generative} proposed GAN-HBNR which is based on generative adversarial bibliographic network module for citation recommendation.  \citet{DBLP:conf/kdd/CenZZYZ019} introduced an embedding method for the attributed multiplex heterogeneous network to boost the behavior of the recommender system. 
\citet{chen2018pme} proposed a novel metric learning based heterogeneous information network embedding model named PME which uses a unified technique to capture 1st-order and 2nd-order proximities. 
HAN~\cite{DBLP:conf/www/WangJSWYCY19} introduced the attention mechanism into an HIN and achieved a comparable performance in node classification and clustering. 
\citet{sun2021heterogeneous} presented a heterogeneous hypergraph learning model based on graph neural networks to characterize multiple non-pairwise relations. Furthermore, the attention mechanism has been adopted in several approaches, such as HERec~\cite{shi2019heterogeneous} and MCRec~\cite{hu2018leveraging}. 

MOOCs are more typical for utilizing the emerging HINs because they contain richer links and semantics in objects, and are thus a fresh direction for data mining. 
We construct a framework by combining extended deep reinforcement learning with the representations of user status to provide recommendations of relevant knowledge concepts in this study.

\subsection{Recommender System}
Recent recommendation works can be grouped approximately into four  categories~\cite{13,15,liu2022re}: collaborative-filtering (CF)  based recommendation, content-based (CB) recommendation, conventional hybrid recommendation, and deep-learning-based recommendation.

\subsubsection{CF-based Recommendation.}
CF has become one of the most efficient recommendation algorithms because of its simplicity and efficiency. It aims to recommend items on the basis of historical data of user–item interactions~\cite{15}. Recommender systems rely on either explicit or implicit feedback, which contains abundant information about the interests of users~\cite{17}. There are two kinds of algorithms deriving from CF, namely user-based CF (UserCF) and item-based CF (ItemCF)~\cite{18}. The major challenges of CF approaches are data sparsity issues~\cite{23} and the cold-start problem~\cite{24}.
In order to alleviate these problems in MOOC course recommendation, \citet{25} proposed MLBR, which uses learner-course vectors rather than a huge and sparse learner-course matrix. 
\citet{26} combined ontology and CF to recommend personalized courses to online students, where ontology can provide a semantic description of students to mitigate the cold-start problem.

\subsubsection{Content-based Recommendation.}
Compared with CF, CB prioritizes the ratings of the target users and the attributes of items liked by them in the past. Currently, several CB approaches have been proposed for course recommendations on MOOCs. \citet{29}  utilized feature modeling for items and user profiles constructing and matching the item features with the user's preferences. In comparison to CF, the CB suggestion does not require a substantial rating matrix and offers the benefits of user independence and the capacity to suggest new goods.
Although CB approaches are significantly better at proposing new things, they cannot yet deliver services to new system users. 

\subsubsection{Hybrid Recommendation.}
Hybrid models seek to get the best of both worlds by combining both the basic and the enhanced methods~\cite{15}. They can incorporate the advantages of different types of recommendation methods to create a unified method that is more robust in a wide range of contexts. 
Typically, basic models contain CF and CB methods, whereas enhanced ones include factorization machine (FM)~\cite{31,meng2022privacy}, GBDT~\cite{32}, and other combined variants. Classical methods include GBDT+LR~\cite{33} and LS-PLM~\cite{34}. 
The GBDT+LR model has the ability to perform higher-order feature combination. It promotes the trend of feature engineering modeling. The LS-PLM model is a three-layer neural network with an attention mechanism, which has a stronger expression. Several related studies leverage hybrid models for online course recommendation~\cite{21,37}.
\citet{36} proposed an approach called SI-IFL for recommending courses to learners, which integrates sequential pattern mining, the learner influence model, and self-organization-based recommendation strategy. Although deep learning has demonstrated considerable potential for learning effective representations, traditional recommendation models still retain irreplaceable advantages owing to their strong interpretability and fast training speed. In general, there are complicated relationships between traditional recommendation models and deep learning models.

\subsubsection{Deep-learning-based Recommendation.}
Compared with traditional recommendation methods, deep learning recommendation methods have deeper neural network layers, and hence, they can explore more complex connections between users and items.

Graph neural networks (GNNs), particularly dynamic GNNs~\cite{PENG2020277}, graph convolutional networks (GCNs)~\cite{welling2016semi}, graph inductive representation learning (GraphSAGE)~\cite{hamilton2017inductive}, graph attention networks~\cite{velickovic2017graph}, and graph isomorphism networks~\cite{xu2018powerful}, have gained considerable attention. The main idea behind GNN is to aggregate the neighbor features of nodes by introducing convolution. 
For example, MixGCF~\cite{huang2021mixgcf} introduces negative samples by using 
both positive mixing and hop mixing methods, which leverages the information of several local graphs to train GNN-based CF recommendation models. 
SHCF~\cite{41-zy} is a novel sequence-aware graph-based CF method for recommendation that combines both high-order collaborative messages and sequential patterns. \citet{DBLP:conf/coling/DengPXLHY20}  predicted paper review rating via a hierarchical network (HabNet) that leverages a bidirectional self-attention network. 
\citet{fan2019metapath} proposed a heterogeneous graph network through a meta-path-guided method to model complex entities and rich edges for intent recommendation. 
\citet{jin2020efficient} proposed an interaction model considering the information of neighborhoods for recommendation in heterogeneous views. 
Another study~\cite{zhang2021we} presented a comprehensive analysis on network embedding methods for recommendation. 
Further, the interpretability~\cite{chen2021temporal} and privacy of recommender systems~\cite{bosri2020integrating,omar2019towards} have been investigated.

Recently, reinforcement learning has been widely utilized in several fields; for example, OpenAI’s AlphaStar~\cite{vinyals2019grandmaster} defeated professional players in StarCraft II. 
\citet{silver2013concurrent}  proposed a reinforcement learning method to build a system for customer interaction.  
\citet{peng2022reinforced} employed a multi-agent RL algorithm to select optimal aggregation strategies in a multi-relational GNN for social event detection tasks. Especially in recommender systems, as a method that focuses on how the agents act in the environment, reinforcement learning enables the recommender to capture the real-time feedback of users effectively for learning optimal recommendation strategies. In practice, several studies have begun to regard recommendation as a deep reinforcement learning problem, and have achieved comparable performances in movie recommendation~\cite{zhao2019leveraging,liu2018deep}, news recommendation~\cite{zheng2018drn,shen2018interactive}, music recommendation~\cite{pan2019policy}, e-commerce~\cite{zhao2018deep}, and health care~\cite{zhang2017leap}. 

More specifically, \citet{zhou2020interactive} proposed a knowledge-graph-enhanced RL method on an interactive recommender system to deal with the sample efficiency issues. 
\citet{chen2020knowledge} introduced the attention method to aggregate the state vector of knowledge graphs and then calculated the Q value in the user–item interest graph through graph convolution. 
\citet{zou2019reinforcement}  introduced a novel Q network to explore user's long-term interest. On this basis, they also proposed a RL framework~\cite{zou2020pseudo} to guarantee stable convergence and low time complexity. 
\citet{huang2021deep} proposed a method using RNN and RL to improve the accuracy of long-term recommendations. 
\citet{yu2019vision} designed VL-Rec, a vision-language recommendation framework with an attribute-augmented RL, which provides natural language feedback to perform effective interactions with the user. 
\citet{zhao2020mahrl} proposed MaHRL based on hierarchical RL for recommendation. The low-level agents learn the short-term preferences of users while the high-level agents focus on learning the uses' long-term preferences and guide the low-level agents. 
\citet{xie2021hierarchical} addressed the integrated recommendation task which also employed a hierarchical model. 
\citet{zhang2019hierarchical} removed noisy courses from historical course sequences and then performed hierarchical reinforcement learning.  
\citet{feng2018learning} and \citet{zhao2020whole}  combined two recommendation scenarios which contain the homepage and product detail page into one recommendation target on an e-commerce platform.  
\citet{peng2021reinforced} proposed to address the complexity of networks by designing a novel neighborhood-selection-guided multi-relational GNN, which maintains relation-dependent representations simultaneously.

Different from existing recommendation methods based on reinforcement learning~\cite{pan2019policy,chen2019neural}, our model leverages rich semantic meta-path-based context, which can learn better representations by specific interactions for students, courses, concepts, items, etc.

\section{Discussion}\label{sec_discussions}
In this section, we discuss our study from three aspects: comparison with our previous model, verifying the effectiveness our model, and presenting the threats to validity.

\subsection{Comparison with Our Previous Work}
We revisit our previous model ACKRec~\cite{gong2020attentional} for knowledge concept recommendation, and compare it with HinCRec-RL.

\paragraph{Attention-based GCNs versus Hierarchical Attention Networks.}
By combining information from neighbors based on meta-paths,
our models can learn the node representation in heterogeneous graphs.  In node-level aggregation, we use information propagation in GCNs within meta-path-based neighbors in ACKRec, which sets the symmetric normalization term between two nodes as ${1/\sqrt{|\mathcal{N}_{i}||\mathcal{N}_{j}|}}$. 
This encodes the prior that the intermediate representation of node $j\in{\mathcal{N}_{i}^{{\Phi}_{k}}}$ to node $i$ under a single meta-path ${\Phi}_{k}$ exploits the proximity structure of the graph. 
HinCRec-RL, in contrast to GCN, uses self-attention to learn the importance of ${\alpha}_{i j}^{{\Phi}_{k}}$ among meta-path-based neighbors. 
By assigning weights to the meta-path-based neighbors, HinCRec-RL may be applied to graph nodes with varied degrees.  Both of our models use the attention mechanism during path-level aggregation to combine the representations of entities learnt while following various meta-paths and produce the final attentional joint representation.  Experimental results show that, in the representation learning stage, both ACKRec and HinCRec-RL can both fully utilize the structure of HINs to learn excellent node embeddings. This is useful for the further downstream recommendation task.

\paragraph{Extended Matrix Factorization versus REINFORCE}
We feed both content and context information into the basic matrix factorization model to form our previous, robust and explainable model, ACKRec. It adopts gradient descent to obtain a scalar product of user latent vectors ${p}_{k}\in{\mathbb{R}^{d}}$ and knowledge concept ones ${q}_{u}\in{\mathbb{R}^{d}}$ plus users representations ${e}_{u}$ and knowledge concepts ${e}_{k}$ approximate original ratings ${r}_{u k}$. Although ACKRec only slightly alleviates the classic data sparsity problem, it is no longer useful in incremental recommendation settings.
The fundamental problem with such a system is that it continuously welcomes new users and concepts.  The recommender system for  top-K concepts used at \textit{XuetangX} is implemented in practice by the HinCRec-RL model presented in this paper. When the system is running, it may concurrently learn the interests of the users and utilize those interests. This is distinct from the offline setting of ACKRec.  In this study, we apply REINFORCE to enable HinCRec-RL to adapt to the dynamic interaction environment between users and concepts, where simultaneous exploration and exploitation of the search space takes place. 
In the simplest case, the recommendation agent may have to decide the concept lists to recommend. The agent would either earn a reward for the success of the recommendation or receive a penalty when a user clicks the particular notion on the list. 
Thus, the online recommendation policy network changes can be performed in real time. 
Experimental results show that 
HinCRec-RL can perform well in dynamic environment. We will run  HinCRec-RL stably in a live production system in the future.

\subsection{Why does HinCRec-RL Work?}
Here we summarize three main features of HinCRec-RL that may explain its effectiveness in concept recommendation in MOOCs.

\paragraph{Mining Latent Information in Heterogeneous Graphs.} 
Traditional approaches aimed at solving the data sparsity issue by supplementing information about the context and its own attributes are not work effective in our scenario owing to sparse interactions between users and knowledge concepts coupled with a lack of attributes.
To enhance the relatedness between users and concepts, we introduce course and video entities and build a heterogeneous graph as a bridge for message passing from concepts to users. In addition to direct interactions between users and concepts, the heterogeneous graph contains rich latent semantic information which is not easy to mine but is crucial.

\paragraph{Attentional Multiple Meta-path-based Node Aggregation Representations.} 
HinCRec-RL learns user embedding representation by integrating the representations of other adjacent nodes, which are those nodes on a path selected based on a specific meta-path schema in the heterogeneous graph. In the model, node weights and meta-paths are included in addition to node properties.  Therefore, it can capture the fine-grained connection between users and concepts better and interpret  users' preferences by tracking and tracing the users' historical behaviors with the node relationships in these meta-paths.

\paragraph{Dynamic Interaction through Reinforcement Learning.} 
In several real-world scenarios, a user's interest always changes rapidly with time, which requires methods that can react quickly to such changes. Traditional deep-learning-based methods, owing to their training and learning in a known data context and static environment, have intuitive limitations in capturing dynamic changes.  We formulate the concept recommendation task as a sequential recommendation problem and model it as an MDP. Every action taken by a reinforcement learning agent relies on a constantly changing user embedding vector. Thus, HinCRec-RL can better adapt to the dynamic environment in MOOCs.

\subsection{Threats to Validity}
The validity of our assessment is mostly at risk from two factors.
\begin{itemize}
    \item 
    \textit{Generalizability of our framework.} The main objective of our study is to improve the performance of personalized recommendations in MOOCs through a fine-grained concept recommendation approach. One threat is the differences between education scenes and other scenes (e.g. e-commerce) in the real world. In our experiments, we used a large dataset, containing 3,111,637 users and 2,527 concepts, to account for as many user behaviors as possible. Moreover, we made the training and testing datasets as significantly different as possible. In our future work, we will utilize even larger-scale datasets and datasets of other scenarios to evaluate the effectiveness of HinCRec-RL. 
    \item 
    \textit{The returned recommendations may have low relevance scores.} We consider only the top 20 returned results based on the ranking score. Increasing the scale and diversity of datasets can avoid such low relevance threat. Moreover, we believe that it is reasonable to consider the top 20 results. Users typically pay attention to the top 20 or even fewer items in the recommended list in real-world recommender systems and disregard the other items.  This indicates that items starting from $k+1$ ($k$=20) are not significant from the perspective of users.
\end{itemize}

\section{Conclusion}\label{sec_conclusion}
This paper presented HinCRec-RL, a novel reinforcement learning architecture that incorporates HINs, for the task of concept recommendation in MOOCs.
HinCRec-RL has a better ability to learn user representations by meta-path sampling over an HIN, and can also accommodate the interactions between users and knowledge concepts in the dynamic context of MOOCs. From experimental results, we can have a conclusion that the proposed HinCRec-RL model can considerably improve both the precision and diversity of the  recommended list. Our model can be applied in not only MOOC platforms for knowledge concept recommendation, but also other practical scenarios, such as user recommendation and commodity recommendation.
In the future, we will deploy the concept recommendation algorithm on the real-world MOOC platform \textit{XuetangX} and provide a commercial concepts recommendation service with personalized and dynamic characteristics. In addition, we will explore how to apply inverse reinforcement learning into the recommender system and further unleash the power of HinCRec-RL.


\begin{acks}
The work is supported by National Key R\&D Program of China through grant 2022YFB3104700 and Hebei Natural Science Foundation of China through grant F2022203072.
\end{acks}
\bibliographystyle{ACM-Reference-Format}
\bibliography{ref}


\begin{thebibliography}{99}


\ifx \showCODEN    \undefined \def \showCODEN     #1{\unskip}     \fi
\ifx \showDOI      \undefined \def \showDOI       #1{#1}\fi
\ifx \showISBNx    \undefined \def \showISBNx     #1{\unskip}     \fi
\ifx \showISBNxiii \undefined \def \showISBNxiii  #1{\unskip}     \fi
\ifx \showISSN     \undefined \def \showISSN      #1{\unskip}     \fi
\ifx \showLCCN     \undefined \def \showLCCN      #1{\unskip}     \fi
\ifx \shownote     \undefined \def \shownote      #1{#1}          \fi
\ifx \showarticletitle \undefined \def \showarticletitle #1{#1}   \fi
\ifx \showURL      \undefined \def \showURL       {\relax}        \fi
\providecommand\bibfield[2]{#2}
\providecommand\bibinfo[2]{#2}
\providecommand\natexlab[1]{#1}
\providecommand\showeprint[2][]{arXiv:#2}

\bibitem[\protect\citeauthoryear{Aggarwal et~al\mbox{.}}{Aggarwal
  et~al\mbox{.}}{2016}]%
        {15}
\bibfield{author}{\bibinfo{person}{Charu~C Aggarwal} {et~al\mbox{.}}}
  \bibinfo{year}{2016}\natexlab{}.
\newblock \bibinfo{booktitle}{\emph{Recommender systems}}.
  Vol.~\bibinfo{volume}{1}.
\newblock \bibinfo{publisher}{Springer}.
\newblock


\bibitem[\protect\citeauthoryear{Bosri, Rahman, Bhuiyan, and Al~Omar}{Bosri
  et~al\mbox{.}}{2021}]%
        {bosri2020integrating}
\bibfield{author}{\bibinfo{person}{Rabeya Bosri},
  \bibinfo{person}{Mohammad~Shahriar Rahman}, \bibinfo{person}{Md~Zakirul~Alam
  Bhuiyan}, {and} \bibinfo{person}{Abdullah Al~Omar}.}
  \bibinfo{year}{2021}\natexlab{}.
\newblock \showarticletitle{Integrating blockchain with artificial intelligence
  for privacy-preserving recommender systems}.
\newblock \bibinfo{journal}{\emph{IEEE Transactions on Network Science and
  Engineering}} \bibinfo{volume}{8}, \bibinfo{number}{2}
  (\bibinfo{year}{2021}), \bibinfo{pages}{1009--1018}.
\newblock


\bibitem[\protect\citeauthoryear{Cai, Han, and Yang}{Cai et~al\mbox{.}}{2018}]%
        {cai2018generative}
\bibfield{author}{\bibinfo{person}{Xiaoyan Cai}, \bibinfo{person}{Junwei Han},
  {and} \bibinfo{person}{Libin Yang}.} \bibinfo{year}{2018}\natexlab{}.
\newblock \showarticletitle{Generative adversarial network based heterogeneous
  bibliographic network representation for personalized citation
  recommendation}. In \bibinfo{booktitle}{\emph{Thirty-Second AAAI Conference
  on Artificial Intelligence}}.
\newblock


\bibitem[\protect\citeauthoryear{Cao, Peng, Wu, Dou, Li, and Yu}{Cao
  et~al\mbox{.}}{2021}]%
        {10.1145/3442381.3449834}
\bibfield{author}{\bibinfo{person}{Yuwei Cao}, \bibinfo{person}{Hao Peng},
  \bibinfo{person}{Jia Wu}, \bibinfo{person}{Yingtong Dou},
  \bibinfo{person}{Jianxin Li}, {and} \bibinfo{person}{Philip~S. Yu}.}
  \bibinfo{year}{2021}\natexlab{}.
\newblock \showarticletitle{Knowledge-Preserving Incremental Social Event
  Detection via Heterogeneous GNNs}. In \bibinfo{booktitle}{\emph{Proceedings
  of the Web Conference 2021}} (Ljubljana, Slovenia)
  \emph{(\bibinfo{series}{WWW '21})}. \bibinfo{publisher}{Association for
  Computing Machinery}, \bibinfo{address}{New York, NY, USA},
  \bibinfo{pages}{3383–3395}.
\newblock
\showISBNx{9781450383127}


\bibitem[\protect\citeauthoryear{Cen, Zou, Zhang, Yang, Zhou, and Tang}{Cen
  et~al\mbox{.}}{2019}]%
        {DBLP:conf/kdd/CenZZYZ019}
\bibfield{author}{\bibinfo{person}{Yukuo Cen}, \bibinfo{person}{Xu Zou},
  \bibinfo{person}{Jianwei Zhang}, \bibinfo{person}{Hongxia Yang},
  \bibinfo{person}{Jingren Zhou}, {and} \bibinfo{person}{Jie Tang}.}
  \bibinfo{year}{2019}\natexlab{}.
\newblock \showarticletitle{Representation Learning for Attributed Multiplex
  Heterogeneous Network}. In \bibinfo{booktitle}{\emph{Proceedings of the 25th
  {ACM} {SIGKDD} International Conference on Knowledge Discovery {\&} Data
  Mining, {KDD} 2019, Anchorage, AK, USA, August 4-8, 2019}},
  \bibfield{editor}{\bibinfo{person}{Ankur Teredesai}, \bibinfo{person}{Vipin
  Kumar}, \bibinfo{person}{Ying Li}, \bibinfo{person}{R{\'{o}}mer Rosales},
  \bibinfo{person}{Evimaria Terzi}, {and} \bibinfo{person}{George Karypis}}
  (Eds.). \bibinfo{publisher}{{ACM}}, \bibinfo{pages}{1358--1368}.
\newblock


\bibitem[\protect\citeauthoryear{Chen, Li, Sun, Xu, and Yin}{Chen
  et~al\mbox{.}}{2021}]%
        {chen2021temporal}
\bibfield{author}{\bibinfo{person}{Hongxu Chen}, \bibinfo{person}{Yicong Li},
  \bibinfo{person}{Xiangguo Sun}, \bibinfo{person}{Guandong Xu}, {and}
  \bibinfo{person}{Hongzhi Yin}.} \bibinfo{year}{2021}\natexlab{}.
\newblock \showarticletitle{Temporal meta-path guided explainable
  recommendation}. In \bibinfo{booktitle}{\emph{Proceedings of the 14th ACM
  international conference on web search and data mining}}.
  \bibinfo{pages}{1056--1064}.
\newblock


\bibitem[\protect\citeauthoryear{Chen, Yin, Wang, Wang, Nguyen, and Li}{Chen
  et~al\mbox{.}}{2018b}]%
        {chen2018pme}
\bibfield{author}{\bibinfo{person}{Hongxu Chen}, \bibinfo{person}{Hongzhi Yin},
  \bibinfo{person}{Weiqing Wang}, \bibinfo{person}{Hao Wang},
  \bibinfo{person}{Quoc Viet~Hung Nguyen}, {and} \bibinfo{person}{Xue Li}.}
  \bibinfo{year}{2018}\natexlab{b}.
\newblock \showarticletitle{PME: projected metric embedding on heterogeneous
  networks for link prediction}. In \bibinfo{booktitle}{\emph{Proceedings of
  the 24th ACM SIGKDD international conference on knowledge discovery \& data
  mining}}. \bibinfo{pages}{1177--1186}.
\newblock


\bibitem[\protect\citeauthoryear{Chen, Lu, Zheng, and Pian}{Chen
  et~al\mbox{.}}{2018a}]%
        {xz4}
\bibfield{author}{\bibinfo{person}{Penghe Chen}, \bibinfo{person}{Yu Lu},
  \bibinfo{person}{Vincent~W Zheng}, {and} \bibinfo{person}{Yang Pian}.}
  \bibinfo{year}{2018}\natexlab{a}.
\newblock \showarticletitle{Prerequisite-driven deep knowledge tracing}. In
  \bibinfo{booktitle}{\emph{2018 IEEE International Conference on Data Mining
  (ICDM)}}. IEEE, \bibinfo{pages}{39--48}.
\newblock


\bibitem[\protect\citeauthoryear{Chen, Niu, Zhao, and Li}{Chen
  et~al\mbox{.}}{2014}]%
        {37}
\bibfield{author}{\bibinfo{person}{Wei Chen}, \bibinfo{person}{Zhendong Niu},
  \bibinfo{person}{Xiangyu Zhao}, {and} \bibinfo{person}{Yi Li}.}
  \bibinfo{year}{2014}\natexlab{}.
\newblock \showarticletitle{A hybrid recommendation algorithm adapted in
  e-learning environments}.
\newblock \bibinfo{journal}{\emph{World Wide Web}} \bibinfo{volume}{17},
  \bibinfo{number}{2} (\bibinfo{year}{2014}), \bibinfo{pages}{271--284}.
\newblock


\bibitem[\protect\citeauthoryear{Chen, Huang, Yao, Wang, Zhang,
  et~al\mbox{.}}{Chen et~al\mbox{.}}{2020}]%
        {chen2020knowledge}
\bibfield{author}{\bibinfo{person}{Xiaocong Chen}, \bibinfo{person}{Chaoran
  Huang}, \bibinfo{person}{Lina Yao}, \bibinfo{person}{Xianzhi Wang},
  \bibinfo{person}{Wenjie Zhang}, {et~al\mbox{.}}}
  \bibinfo{year}{2020}\natexlab{}.
\newblock \showarticletitle{Knowledge-guided deep reinforcement learning for
  interactive recommendation}. In \bibinfo{booktitle}{\emph{2020 International
  Joint Conference on Neural Networks (IJCNN)}}. IEEE, \bibinfo{pages}{1--8}.
\newblock


\bibitem[\protect\citeauthoryear{Chen, Li, Li, Jiang, and Song}{Chen
  et~al\mbox{.}}{2019}]%
        {chen2019neural}
\bibfield{author}{\bibinfo{person}{Xinshi Chen}, \bibinfo{person}{Shuang Li},
  \bibinfo{person}{Hui Li}, \bibinfo{person}{Shaohua Jiang}, {and}
  \bibinfo{person}{Le Song}.} \bibinfo{year}{2019}\natexlab{}.
\newblock \showarticletitle{Neural Model-Based Reinforcement Learning for
  Recommendation}. In \bibinfo{booktitle}{\emph{International Conference on
  Learning Representations}}.
\newblock


\bibitem[\protect\citeauthoryear{Deng, Peng, Xia, Li, He, and Yu}{Deng
  et~al\mbox{.}}{2020}]%
        {DBLP:conf/coling/DengPXLHY20}
\bibfield{author}{\bibinfo{person}{Zhongfen Deng}, \bibinfo{person}{Hao Peng},
  \bibinfo{person}{Congying Xia}, \bibinfo{person}{Jianxin Li},
  \bibinfo{person}{Lifang He}, {and} \bibinfo{person}{Philip~S. Yu}.}
  \bibinfo{year}{2020}\natexlab{}.
\newblock \showarticletitle{Hierarchical Bi-Directional Self-Attention Networks
  for Paper Review Rating Recommendation}. In
  \bibinfo{booktitle}{\emph{Proceedings of the 28th International Conference on
  Computational Linguistics, {COLING} 2020, Barcelona, Spain (Online), December
  8-13, 2020}}, \bibfield{editor}{\bibinfo{person}{Donia Scott},
  \bibinfo{person}{N{\'{u}}ria Bel}, {and} \bibinfo{person}{Chengqing Zong}}
  (Eds.). \bibinfo{publisher}{International Committee on Computational
  Linguistics}, \bibinfo{pages}{6302--6314}.
\newblock


\bibitem[\protect\citeauthoryear{Dong, Chawla, and Swami}{Dong
  et~al\mbox{.}}{2017}]%
        {dong2017metapath2vec}
\bibfield{author}{\bibinfo{person}{Yuxiao Dong}, \bibinfo{person}{Nitesh~V
  Chawla}, {and} \bibinfo{person}{Ananthram Swami}.}
  \bibinfo{year}{2017}\natexlab{}.
\newblock \showarticletitle{metapath2vec: Scalable representation learning for
  heterogeneous networks}. In \bibinfo{booktitle}{\emph{Proceedings of the 23rd
  ACM SIGKDD International Conference on Knowledge Discovery and Data Mining}}.
  ACM, \bibinfo{pages}{135--144}.
\newblock


\bibitem[\protect\citeauthoryear{Fan, Zhu, Han, Shi, Hu, Ma, and Li}{Fan
  et~al\mbox{.}}{2019}]%
        {fan2019metapath}
\bibfield{author}{\bibinfo{person}{Shaohua Fan}, \bibinfo{person}{Junxiong
  Zhu}, \bibinfo{person}{Xiaotian Han}, \bibinfo{person}{Chuan Shi},
  \bibinfo{person}{Linmei Hu}, \bibinfo{person}{Biyu Ma}, {and}
  \bibinfo{person}{Yongliang Li}.} \bibinfo{year}{2019}\natexlab{}.
\newblock \showarticletitle{Metapath-guided heterogeneous graph neural network
  for intent recommendation}. In \bibinfo{booktitle}{\emph{Proceedings of the
  25th ACM SIGKDD international conference on knowledge discovery \& data
  mining}}. \bibinfo{pages}{2478--2486}.
\newblock


\bibitem[\protect\citeauthoryear{Feng, Li, Huang, Liu, Ou, Wang, and Zhu}{Feng
  et~al\mbox{.}}{2018}]%
        {feng2018learning}
\bibfield{author}{\bibinfo{person}{Jun Feng}, \bibinfo{person}{Heng Li},
  \bibinfo{person}{Minlie Huang}, \bibinfo{person}{Shichen Liu},
  \bibinfo{person}{Wenwu Ou}, \bibinfo{person}{Zhirong Wang}, {and}
  \bibinfo{person}{Xiaoyan Zhu}.} \bibinfo{year}{2018}\natexlab{}.
\newblock \showarticletitle{Learning to collaborate: Multi-scenario ranking via
  multi-agent reinforcement learning}. In \bibinfo{booktitle}{\emph{Proceedings
  of the 2018 World Wide Web Conference}}. \bibinfo{pages}{1939--1948}.
\newblock


\bibitem[\protect\citeauthoryear{Feng, Tang, and Liu}{Feng
  et~al\mbox{.}}{2019}]%
        {feng2019understanding}
\bibfield{author}{\bibinfo{person}{Wenzheng Feng}, \bibinfo{person}{Jie Tang},
  {and} \bibinfo{person}{Tracy~Xiao Liu}.} \bibinfo{year}{2019}\natexlab{}.
\newblock \showarticletitle{Understanding dropouts in MOOCs}. In
  \bibinfo{booktitle}{\emph{Proceedings of the AAAI Conference on Artificial
  Intelligence}}, Vol.~\bibinfo{volume}{33}. \bibinfo{pages}{517--524}.
\newblock


\bibitem[\protect\citeauthoryear{Gai, Zhu, Li, Liu, and Wang}{Gai
  et~al\mbox{.}}{2017}]%
        {34}
\bibfield{author}{\bibinfo{person}{Kun Gai}, \bibinfo{person}{Xiaoqiang Zhu},
  \bibinfo{person}{Han Li}, \bibinfo{person}{Kai Liu}, {and}
  \bibinfo{person}{Zhe Wang}.} \bibinfo{year}{2017}\natexlab{}.
\newblock \showarticletitle{Learning piece-wise linear models from large scale
  data for ad click prediction}.
\newblock \bibinfo{journal}{\emph{arXiv preprint arXiv:1704.05194}}
  (\bibinfo{year}{2017}).
\newblock


\bibitem[\protect\citeauthoryear{Ghazarian and Nematbakhsh}{Ghazarian and
  Nematbakhsh}{2015}]%
        {18}
\bibfield{author}{\bibinfo{person}{Sarik Ghazarian} {and}
  \bibinfo{person}{Mohammad~Ali Nematbakhsh}.} \bibinfo{year}{2015}\natexlab{}.
\newblock \showarticletitle{Enhancing memory-based collaborative filtering for
  group recommender systems}.
\newblock \bibinfo{journal}{\emph{Expert systems with applications}}
  \bibinfo{volume}{42}, \bibinfo{number}{7} (\bibinfo{year}{2015}),
  \bibinfo{pages}{3801--3812}.
\newblock


\bibitem[\protect\citeauthoryear{Gong, Wang, Wang, Feng, Peng, Tang, and
  Yu}{Gong et~al\mbox{.}}{2020}]%
        {gong2020attentional}
\bibfield{author}{\bibinfo{person}{Jibing Gong}, \bibinfo{person}{Shen Wang},
  \bibinfo{person}{Jinlong Wang}, \bibinfo{person}{Wenzheng Feng},
  \bibinfo{person}{Hao Peng}, \bibinfo{person}{Jie Tang}, {and}
  \bibinfo{person}{Philip~S Yu}.} \bibinfo{year}{2020}\natexlab{}.
\newblock \showarticletitle{Attentional graph convolutional networks for
  knowledge concept recommendation in moocs in a heterogeneous view}. In
  \bibinfo{booktitle}{\emph{Proceedings of the 43rd International ACM SIGIR
  Conference on Research and Development in Information Retrieval}}.
  \bibinfo{pages}{79--88}.
\newblock


\bibitem[\protect\citeauthoryear{Hamilton, Ying, and Leskovec}{Hamilton
  et~al\mbox{.}}{2017}]%
        {hamilton2017inductive}
\bibfield{author}{\bibinfo{person}{William~L Hamilton}, \bibinfo{person}{Rex
  Ying}, {and} \bibinfo{person}{Jure Leskovec}.}
  \bibinfo{year}{2017}\natexlab{}.
\newblock \showarticletitle{Inductive representation learning on large graphs}.
  In \bibinfo{booktitle}{\emph{Proceedings of the 31st International Conference
  on Neural Information Processing Systems}}. \bibinfo{pages}{1025--1035}.
\newblock


\bibitem[\protect\citeauthoryear{He, He, Song, Liu, Jiang, and Chua}{He
  et~al\mbox{.}}{2018}]%
        {he2018nais}
\bibfield{author}{\bibinfo{person}{Xiangnan He}, \bibinfo{person}{Zhankui He},
  \bibinfo{person}{Jingkuan Song}, \bibinfo{person}{Zhenguang Liu},
  \bibinfo{person}{Yu{-}Gang Jiang}, {and} \bibinfo{person}{Tat{-}Seng Chua}.}
  \bibinfo{year}{2018}\natexlab{}.
\newblock \showarticletitle{{NAIS:} Neural Attentive Item Similarity Model for
  Recommendation}.
\newblock \bibinfo{journal}{\emph{{IEEE} Trans. Knowl. Data Eng.}}
  \bibinfo{volume}{30}, \bibinfo{number}{12} (\bibinfo{year}{2018}),
  \bibinfo{pages}{2354--2366}.
\newblock


\bibitem[\protect\citeauthoryear{He, Liao, Zhang, Nie, Hu, and Chua}{He
  et~al\mbox{.}}{2017}]%
        {he2017neural}
\bibfield{author}{\bibinfo{person}{Xiangnan He}, \bibinfo{person}{Lizi Liao},
  \bibinfo{person}{Hanwang Zhang}, \bibinfo{person}{Liqiang Nie},
  \bibinfo{person}{Xia Hu}, {and} \bibinfo{person}{Tat-Seng Chua}.}
  \bibinfo{year}{2017}\natexlab{}.
\newblock \showarticletitle{Neural collaborative filtering}. In
  \bibinfo{booktitle}{\emph{Proceedings of the 26th International Conference on
  World Wide Web}}. International World Wide Web Conferences Steering
  Committee, \bibinfo{pages}{173--182}.
\newblock


\bibitem[\protect\citeauthoryear{He, Pan, Jin, Xu, Liu, Xu, Shi, Atallah,
  Herbrich, Bowers, et~al\mbox{.}}{He et~al\mbox{.}}{2014}]%
        {33}
\bibfield{author}{\bibinfo{person}{Xinran He}, \bibinfo{person}{Junfeng Pan},
  \bibinfo{person}{Ou Jin}, \bibinfo{person}{Tianbing Xu}, \bibinfo{person}{Bo
  Liu}, \bibinfo{person}{Tao Xu}, \bibinfo{person}{Yanxin Shi},
  \bibinfo{person}{Antoine Atallah}, \bibinfo{person}{Ralf Herbrich},
  \bibinfo{person}{Stuart Bowers}, {et~al\mbox{.}}}
  \bibinfo{year}{2014}\natexlab{}.
\newblock \showarticletitle{Practical lessons from predicting clicks on ads at
  facebook}. In \bibinfo{booktitle}{\emph{Proceedings of the Eighth
  International Workshop on Data Mining for Online Advertising}}.
  \bibinfo{pages}{1--9}.
\newblock


\bibitem[\protect\citeauthoryear{Hu, Shi, Zhao, and Yang}{Hu
  et~al\mbox{.}}{2018a}]%
        {hu2018local}
\bibfield{author}{\bibinfo{person}{Binbin Hu}, \bibinfo{person}{Chuan Shi},
  \bibinfo{person}{Wayne~Xin Zhao}, {and} \bibinfo{person}{Tianchi Yang}.}
  \bibinfo{year}{2018}\natexlab{a}.
\newblock \showarticletitle{Local and Global Information Fusion for Top-N
  Recommendation in Heterogeneous Information Network}. In
  \bibinfo{booktitle}{\emph{Proceedings of the 27th ACM International
  Conference on Information and Knowledge Management}}. ACM,
  \bibinfo{pages}{1683--1686}.
\newblock


\bibitem[\protect\citeauthoryear{Hu, Shi, Zhao, and Yu}{Hu
  et~al\mbox{.}}{2018b}]%
        {hu2018leveraging}
\bibfield{author}{\bibinfo{person}{Binbin Hu}, \bibinfo{person}{Chuan Shi},
  \bibinfo{person}{Wayne~Xin Zhao}, {and} \bibinfo{person}{Philip~S. Yu}.}
  \bibinfo{year}{2018}\natexlab{b}.
\newblock \showarticletitle{Leveraging meta-path based context for top-n
  recommendation with a neural co-attention model}. In
  \bibinfo{booktitle}{\emph{Proceedings of the 24th ACM SIGKDD International
  Conference on Knowledge Discovery \& Data Mining}}. ACM,
  \bibinfo{pages}{1531--1540}.
\newblock


\bibitem[\protect\citeauthoryear{Huang, Fu, Li, Qu, Liu, and Chen}{Huang
  et~al\mbox{.}}{2021b}]%
        {huang2021deep}
\bibfield{author}{\bibinfo{person}{Liwei Huang}, \bibinfo{person}{Mingsheng
  Fu}, \bibinfo{person}{Fan Li}, \bibinfo{person}{Hong Qu},
  \bibinfo{person}{Yangjun Liu}, {and} \bibinfo{person}{Wenyu Chen}.}
  \bibinfo{year}{2021}\natexlab{b}.
\newblock \showarticletitle{A deep reinforcement learning based long-term
  recommender system}.
\newblock \bibinfo{journal}{\emph{Knowledge-Based Systems}}
  \bibinfo{volume}{213} (\bibinfo{year}{2021}), \bibinfo{pages}{106706}.
\newblock


\bibitem[\protect\citeauthoryear{Huang and Lu}{Huang and Lu}{2018}]%
        {29}
\bibfield{author}{\bibinfo{person}{Ran Huang} {and} \bibinfo{person}{Ran Lu}.}
  \bibinfo{year}{2018}\natexlab{}.
\newblock \showarticletitle{Research on Content-based MOOC Recommender Model}.
  In \bibinfo{booktitle}{\emph{2018 5th International Conference on Systems and
  Informatics (ICSAI)}}. IEEE, \bibinfo{pages}{676--681}.
\newblock


\bibitem[\protect\citeauthoryear{Huang, Dong, Ding, Yang, Feng, Wang, and
  Tang}{Huang et~al\mbox{.}}{2021a}]%
        {huang2021mixgcf}
\bibfield{author}{\bibinfo{person}{Tinglin Huang}, \bibinfo{person}{Yuxiao
  Dong}, \bibinfo{person}{Ming Ding}, \bibinfo{person}{Zhen Yang},
  \bibinfo{person}{Wenzheng Feng}, \bibinfo{person}{Xinyu Wang}, {and}
  \bibinfo{person}{Jie Tang}.} \bibinfo{year}{2021}\natexlab{a}.
\newblock \showarticletitle{Mixgcf: An improved training method for graph
  neural network-based recommender systems}. In
  \bibinfo{booktitle}{\emph{Proceedings of the 27th ACM SIGKDD Conference on
  Knowledge Discovery \& Data Mining}}. \bibinfo{pages}{665--674}.
\newblock


\bibitem[\protect\citeauthoryear{Intayoad, Kamyod, and Temdee}{Intayoad
  et~al\mbox{.}}{2020}]%
        {intayoad2020reinforcement}
\bibfield{author}{\bibinfo{person}{Wacharawan Intayoad},
  \bibinfo{person}{Chayapol Kamyod}, {and} \bibinfo{person}{Punnarumol
  Temdee}.} \bibinfo{year}{2020}\natexlab{}.
\newblock \showarticletitle{Reinforcement Learning Based on Contextual Bandits
  for Personalized Online Learning Recommendation Systems}.
\newblock \bibinfo{journal}{\emph{Wireless Personal Communications}}
  (\bibinfo{year}{2020}), \bibinfo{pages}{1--16}.
\newblock


\bibitem[\protect\citeauthoryear{Jin, Qin, Fang, Du, Zhang, Yu, Zhang, and
  Smola}{Jin et~al\mbox{.}}{2020}]%
        {jin2020efficient}
\bibfield{author}{\bibinfo{person}{Jiarui Jin}, \bibinfo{person}{Jiarui Qin},
  \bibinfo{person}{Yuchen Fang}, \bibinfo{person}{Kounianhua Du},
  \bibinfo{person}{Weinan Zhang}, \bibinfo{person}{Yong Yu},
  \bibinfo{person}{Zheng Zhang}, {and} \bibinfo{person}{Alexander~J Smola}.}
  \bibinfo{year}{2020}\natexlab{}.
\newblock \showarticletitle{An efficient neighborhood-based interaction model
  for recommendation on heterogeneous graph}. In
  \bibinfo{booktitle}{\emph{Proceedings of the 26th ACM SIGKDD International
  Conference on Knowledge Discovery \& Data Mining}}. \bibinfo{pages}{75--84}.
\newblock


\bibitem[\protect\citeauthoryear{Jing and Tang}{Jing and Tang}{2017}]%
        {jing2017guess}
\bibfield{author}{\bibinfo{person}{Xia Jing} {and} \bibinfo{person}{Jie Tang}.}
  \bibinfo{year}{2017}\natexlab{}.
\newblock \showarticletitle{Guess you like: course recommendation in MOOCs}. In
  \bibinfo{booktitle}{\emph{Proceedings of the International Conference on Web
  Intelligence}}. ACM, \bibinfo{pages}{783--789}.
\newblock


\bibitem[\protect\citeauthoryear{Kabbur, Ning, and Karypis}{Kabbur
  et~al\mbox{.}}{2013}]%
        {santosh2013fism}
\bibfield{author}{\bibinfo{person}{Santosh Kabbur}, \bibinfo{person}{Xia Ning},
  {and} \bibinfo{person}{George Karypis}.} \bibinfo{year}{2013}\natexlab{}.
\newblock \showarticletitle{{FISM:} factored item similarity models for top-N
  recommender systems}. In \bibinfo{booktitle}{\emph{The 19th {ACM} {SIGKDD}
  International Conference on Knowledge Discovery and Data Mining, {KDD} 2013,
  Chicago, IL, USA, August 11-14, 2013}}. \bibinfo{pages}{659--667}.
\newblock


\bibitem[\protect\citeauthoryear{Karimi, Derr, Huang, and Tang}{Karimi
  et~al\mbox{.}}{2020}]%
        {xz3}
\bibfield{author}{\bibinfo{person}{Hamid Karimi}, \bibinfo{person}{Tyler Derr},
  \bibinfo{person}{Jiangtao Huang}, {and} \bibinfo{person}{Jiliang Tang}.}
  \bibinfo{year}{2020}\natexlab{}.
\newblock \showarticletitle{Online Academic Course Performance Prediction Using
  Relational Graph Convolutional Neural Network.}
\newblock \bibinfo{journal}{\emph{International Educational Data Mining
  Society}} (\bibinfo{year}{2020}).
\newblock


\bibitem[\protect\citeauthoryear{Ke, Du, and Chen}{Ke et~al\mbox{.}}{2021}]%
        {ke2021cross}
\bibfield{author}{\bibinfo{person}{Gang Ke}, \bibinfo{person}{Hong-Le Du},
  {and} \bibinfo{person}{Yeh-Cheng Chen}.} \bibinfo{year}{2021}\natexlab{}.
\newblock \showarticletitle{Cross-platform dynamic goods recommendation system
  based on reinforcement learning and social networks}.
\newblock \bibinfo{journal}{\emph{Applied Soft Computing}}
  \bibinfo{volume}{104} (\bibinfo{year}{2021}), \bibinfo{pages}{107213}.
\newblock


\bibitem[\protect\citeauthoryear{Ke, Meng, Finley, Wang, Chen, Ma, Ye, and
  Liu}{Ke et~al\mbox{.}}{2017}]%
        {32}
\bibfield{author}{\bibinfo{person}{Guolin Ke}, \bibinfo{person}{Qi Meng},
  \bibinfo{person}{Thomas Finley}, \bibinfo{person}{Taifeng Wang},
  \bibinfo{person}{Wei Chen}, \bibinfo{person}{Weidong Ma},
  \bibinfo{person}{Qiwei Ye}, {and} \bibinfo{person}{Tie-Yan Liu}.}
  \bibinfo{year}{2017}\natexlab{}.
\newblock \showarticletitle{Lightgbm: A highly efficient gradient boosting
  decision tree}.
\newblock \bibinfo{journal}{\emph{Advances in neural information processing
  systems}}  \bibinfo{volume}{30} (\bibinfo{year}{2017}),
  \bibinfo{pages}{3146--3154}.
\newblock


\bibitem[\protect\citeauthoryear{Khalid, Lundqvist, and Yates}{Khalid
  et~al\mbox{.}}{2022}]%
        {21}
\bibfield{author}{\bibinfo{person}{Asra Khalid}, \bibinfo{person}{Karsten
  Lundqvist}, {and} \bibinfo{person}{Anne Yates}.}
  \bibinfo{year}{2022}\natexlab{}.
\newblock \showarticletitle{A literature review of implemented recommendation
  techniques used in Massive Open online Courses}.
\newblock \bibinfo{journal}{\emph{Expert Systems with Applications}}
  \bibinfo{volume}{187} (\bibinfo{year}{2022}), \bibinfo{pages}{115926}.
\newblock


\bibitem[\protect\citeauthoryear{Kingma and Ba}{Kingma and Ba}{2015}]%
        {DBLP:journals/corr/KingmaB14}
\bibfield{author}{\bibinfo{person}{Diederik~P. Kingma} {and}
  \bibinfo{person}{Jimmy Ba}.} \bibinfo{year}{2015}\natexlab{}.
\newblock \showarticletitle{Adam: {A} Method for Stochastic Optimization}. In
  \bibinfo{booktitle}{\emph{3rd International Conference on Learning
  Representations, {ICLR} 2015, San Diego, CA, USA, May 7-9, 2015, Conference
  Track Proceedings}}, \bibfield{editor}{\bibinfo{person}{Yoshua Bengio} {and}
  \bibinfo{person}{Yann LeCun}} (Eds.).
\newblock


\bibitem[\protect\citeauthoryear{Li, Hu, Shi, Song, and Lu}{Li
  et~al\mbox{.}}{2021}]%
        {41-zy}
\bibfield{author}{\bibinfo{person}{Chen Li}, \bibinfo{person}{Linmei Hu},
  \bibinfo{person}{Chuan Shi}, \bibinfo{person}{Guojie Song}, {and}
  \bibinfo{person}{Yuanfu Lu}.} \bibinfo{year}{2021}\natexlab{}.
\newblock \showarticletitle{Sequence-aware Heterogeneous Graph Neural
  Collaborative Filtering}. In \bibinfo{booktitle}{\emph{Proceedings of the
  2021 SIAM International Conference on Data Mining (SDM)}}. SIAM,
  \bibinfo{pages}{64--72}.
\newblock


\bibitem[\protect\citeauthoryear{Li, Ren, Chen, Ren, Lian, and Ma}{Li
  et~al\mbox{.}}{2017}]%
        {li2017nasr}
\bibfield{author}{\bibinfo{person}{Jing Li}, \bibinfo{person}{Pengjie Ren},
  \bibinfo{person}{Zhumin Chen}, \bibinfo{person}{Zhaochun Ren},
  \bibinfo{person}{Tao Lian}, {and} \bibinfo{person}{Jun Ma}.}
  \bibinfo{year}{2017}\natexlab{}.
\newblock \showarticletitle{Neural Attentive Session-based Recommendation}. In
  \bibinfo{booktitle}{\emph{Proceedings of the 2017 {ACM} on Conference on
  Information and Knowledge Management, {CIKM} 2017, Singapore, November 06 -
  10, 2017}}. \bibinfo{pages}{1419--1428}.
\newblock


\bibitem[\protect\citeauthoryear{Liao, Tang, and Zhao}{Liao
  et~al\mbox{.}}{2019}]%
        {xz2}
\bibfield{author}{\bibinfo{person}{Jinzhi Liao}, \bibinfo{person}{Jiuyang
  Tang}, {and} \bibinfo{person}{Xiang Zhao}.} \bibinfo{year}{2019}\natexlab{}.
\newblock \showarticletitle{Course drop-out prediction on MOOC platform via
  clustering and tensor completion}.
\newblock \bibinfo{journal}{\emph{Tsinghua Science and Technology}}
  \bibinfo{volume}{24}, \bibinfo{number}{4} (\bibinfo{year}{2019}),
  \bibinfo{pages}{412--422}.
\newblock


\bibitem[\protect\citeauthoryear{Liu, Tang, Li, Zhang, Ye, Chen, Guo, and
  Zhang}{Liu et~al\mbox{.}}{2018}]%
        {liu2018deep}
\bibfield{author}{\bibinfo{person}{Feng Liu}, \bibinfo{person}{Ruiming Tang},
  \bibinfo{person}{Xutao Li}, \bibinfo{person}{Weinan Zhang},
  \bibinfo{person}{Yunming Ye}, \bibinfo{person}{Haokun Chen},
  \bibinfo{person}{Huifeng Guo}, {and} \bibinfo{person}{Yuzhou Zhang}.}
  \bibinfo{year}{2018}\natexlab{}.
\newblock \showarticletitle{Deep reinforcement learning based recommendation
  with explicit user-item interactions modeling}.
\newblock \bibinfo{journal}{\emph{arXiv preprint arXiv:1810.12027}}
  (\bibinfo{year}{2018}).
\newblock


\bibitem[\protect\citeauthoryear{Liu, Wang, and Bhuiyan}{Liu
  et~al\mbox{.}}{2022a}]%
        {liu2022personalised}
\bibfield{author}{\bibinfo{person}{Xiangyong Liu}, \bibinfo{person}{Guojun
  Wang}, {and} \bibinfo{person}{Md~Zakirul~Alam Bhuiyan}.}
  \bibinfo{year}{2022}\natexlab{a}.
\newblock \showarticletitle{Personalised context-aware re-ranking in
  recommender system}.
\newblock \bibinfo{journal}{\emph{Connection Science}} \bibinfo{volume}{34},
  \bibinfo{number}{1} (\bibinfo{year}{2022}), \bibinfo{pages}{319--338}.
\newblock


\bibitem[\protect\citeauthoryear{Liu, Wang, and Zakirul Alam~Bhuiyan}{Liu
  et~al\mbox{.}}{2022b}]%
        {liu2022re}
\bibfield{author}{\bibinfo{person}{Xiangyong Liu}, \bibinfo{person}{Guojun
  Wang}, {and} \bibinfo{person}{Md Zakirul Alam~Bhuiyan}.}
  \bibinfo{year}{2022}\natexlab{b}.
\newblock \showarticletitle{Re-ranking with multiple objective optimization in
  recommender system}.
\newblock \bibinfo{journal}{\emph{Transactions on Emerging Telecommunications
  Technologies}} \bibinfo{volume}{33}, \bibinfo{number}{1}
  (\bibinfo{year}{2022}), \bibinfo{pages}{e4398}.
\newblock


\bibitem[\protect\citeauthoryear{Loni, Pagano, Larson, and Hanjalic}{Loni
  et~al\mbox{.}}{2019}]%
        {31}
\bibfield{author}{\bibinfo{person}{Babak Loni}, \bibinfo{person}{Roberto
  Pagano}, \bibinfo{person}{Martha Larson}, {and} \bibinfo{person}{Alan
  Hanjalic}.} \bibinfo{year}{2019}\natexlab{}.
\newblock \showarticletitle{Top-N recommendation with multi-channel positive
  feedback using factorization machines}.
\newblock \bibinfo{journal}{\emph{ACM Transactions on Information Systems
  (TOIS)}} \bibinfo{volume}{37}, \bibinfo{number}{2} (\bibinfo{year}{2019}),
  \bibinfo{pages}{1--23}.
\newblock


\bibitem[\protect\citeauthoryear{Lu, Zhou, Yu, and Jia}{Lu
  et~al\mbox{.}}{2019}]%
        {lu2019concept}
\bibfield{author}{\bibinfo{person}{Weiming Lu}, \bibinfo{person}{Yangfan Zhou},
  \bibinfo{person}{Jiale Yu}, {and} \bibinfo{person}{Chenhao Jia}.}
  \bibinfo{year}{2019}\natexlab{}.
\newblock \showarticletitle{Concept Extraction and Prerequisite Relation
  Learning from Educational Data}. In \bibinfo{booktitle}{\emph{Proceedings of
  the AAAI Conference on Artificial Intelligence}}, Vol.~\bibinfo{volume}{33}.
  \bibinfo{pages}{9678--9685}.
\newblock


\bibitem[\protect\citeauthoryear{Meng, Fan, Li, Wang, Zhang, Xu, Qi, and
  Bhuiyan}{Meng et~al\mbox{.}}{2022}]%
        {meng2022privacy}
\bibfield{author}{\bibinfo{person}{Shunmei Meng}, \bibinfo{person}{Shaoyu Fan},
  \bibinfo{person}{Qianmu Li}, \bibinfo{person}{Xinna Wang},
  \bibinfo{person}{Jing Zhang}, \bibinfo{person}{Xiaolong Xu},
  \bibinfo{person}{Lianyong Qi}, {and} \bibinfo{person}{Md~Zakirul~Alam
  Bhuiyan}.} \bibinfo{year}{2022}\natexlab{}.
\newblock \showarticletitle{Privacy-Aware Factorization-Based Hybrid
  Recommendation Method for Healthcare Services}.
\newblock \bibinfo{journal}{\emph{IEEE Transactions on Industrial Informatics}}
  \bibinfo{volume}{18}, \bibinfo{number}{8} (\bibinfo{year}{2022}),
  \bibinfo{pages}{5637--5647}.
\newblock


\bibitem[\protect\citeauthoryear{Omar, Bosri, Rahman, Begum, and Bhuiyan}{Omar
  et~al\mbox{.}}{2019}]%
        {omar2019towards}
\bibfield{author}{\bibinfo{person}{Abdullah~Al Omar}, \bibinfo{person}{Rabeya
  Bosri}, \bibinfo{person}{Mohammad~Shahriar Rahman}, \bibinfo{person}{Nasima
  Begum}, {and} \bibinfo{person}{Md~Zakirul~Alam Bhuiyan}.}
  \bibinfo{year}{2019}\natexlab{}.
\newblock \showarticletitle{Towards privacy-preserving recommender system with
  blockchains}. In \bibinfo{booktitle}{\emph{International Conference on
  Dependability in Sensor, Cloud, and Big Data Systems and Applications}}.
  Springer, \bibinfo{pages}{106--118}.
\newblock


\bibitem[\protect\citeauthoryear{Pan, Cai, Tang, Zhuang, and He}{Pan
  et~al\mbox{.}}{2019}]%
        {pan2019policy}
\bibfield{author}{\bibinfo{person}{Feiyang Pan}, \bibinfo{person}{Qingpeng
  Cai}, \bibinfo{person}{Pingzhong Tang}, \bibinfo{person}{Fuzhen Zhuang},
  {and} \bibinfo{person}{Qing He}.} \bibinfo{year}{2019}\natexlab{}.
\newblock \showarticletitle{Policy gradients for contextual recommendations}.
  In \bibinfo{booktitle}{\emph{The World Wide Web Conference}}.
  \bibinfo{pages}{1421--1431}.
\newblock


\bibitem[\protect\citeauthoryear{Pan, Li, Li, and Tang}{Pan
  et~al\mbox{.}}{2017a}]%
        {pan2017prerequisite}
\bibfield{author}{\bibinfo{person}{Liangming Pan}, \bibinfo{person}{Chengjiang
  Li}, \bibinfo{person}{Juanzi Li}, {and} \bibinfo{person}{Jie Tang}.}
  \bibinfo{year}{2017}\natexlab{a}.
\newblock \showarticletitle{Prerequisite relation learning for concepts in
  moocs}. In \bibinfo{booktitle}{\emph{Proceedings of the 55th Annual Meeting
  of the Association for Computational Linguistics (Volume 1: Long Papers)}}.
  \bibinfo{pages}{1447--1456}.
\newblock


\bibitem[\protect\citeauthoryear{Pan, Wang, Li, Li, and Tang}{Pan
  et~al\mbox{.}}{2017b}]%
        {pan2017course}
\bibfield{author}{\bibinfo{person}{Liangming Pan}, \bibinfo{person}{Xiaochen
  Wang}, \bibinfo{person}{Chengjiang Li}, \bibinfo{person}{Juanzi Li}, {and}
  \bibinfo{person}{Jie Tang}.} \bibinfo{year}{2017}\natexlab{b}.
\newblock \showarticletitle{Course concept extraction in MOOCs via
  embedding-based graph propagation}. In \bibinfo{booktitle}{\emph{Proceedings
  of the Eighth International Joint Conference on Natural Language Processing
  (Volume 1: Long Papers)}}. \bibinfo{pages}{875--884}.
\newblock


\bibitem[\protect\citeauthoryear{Pang, Jin, Zhang, and Zhu}{Pang
  et~al\mbox{.}}{2017}]%
        {25}
\bibfield{author}{\bibinfo{person}{Yanxia Pang}, \bibinfo{person}{Yuanyuan
  Jin}, \bibinfo{person}{Ying Zhang}, {and} \bibinfo{person}{Tao Zhu}.}
  \bibinfo{year}{2017}\natexlab{}.
\newblock \showarticletitle{Collaborative filtering recommendation for MOOC
  application}.
\newblock \bibinfo{journal}{\emph{Computer Applications in Engineering
  Education}} \bibinfo{volume}{25}, \bibinfo{number}{1} (\bibinfo{year}{2017}),
  \bibinfo{pages}{120--128}.
\newblock


\bibitem[\protect\citeauthoryear{Park, Kim, Choi, and Kim}{Park
  et~al\mbox{.}}{2012}]%
        {13}
\bibfield{author}{\bibinfo{person}{Deuk~Hee Park}, \bibinfo{person}{Hyea~Kyeong
  Kim}, \bibinfo{person}{Il~Young Choi}, {and} \bibinfo{person}{Jae~Kyeong
  Kim}.} \bibinfo{year}{2012}\natexlab{}.
\newblock \showarticletitle{A literature review and classification of
  recommender systems research}.
\newblock \bibinfo{journal}{\emph{Expert systems with applications}}
  \bibinfo{volume}{39}, \bibinfo{number}{11} (\bibinfo{year}{2012}),
  \bibinfo{pages}{10059--10072}.
\newblock


\bibitem[\protect\citeauthoryear{Peng, Li, Song, Yang, Ranjan, Yu, and He}{Peng
  et~al\mbox{.}}{2021a}]%
        {10.1145/3447585}
\bibfield{author}{\bibinfo{person}{Hao Peng}, \bibinfo{person}{Jianxin Li},
  \bibinfo{person}{Yangqiu Song}, \bibinfo{person}{Renyu Yang},
  \bibinfo{person}{Rajiv Ranjan}, \bibinfo{person}{Philip~S. Yu}, {and}
  \bibinfo{person}{Lifang He}.} \bibinfo{year}{2021}\natexlab{a}.
\newblock \showarticletitle{Streaming Social Event Detection and Evolution
  Discovery in Heterogeneous Information Networks}.
\newblock \bibinfo{journal}{\emph{ACM Trans. Knowl. Discov. Data}}
  \bibinfo{volume}{15}, \bibinfo{number}{5}, Article \bibinfo{articleno}{89}
  (\bibinfo{date}{may} \bibinfo{year}{2021}), \bibinfo{numpages}{33}~pages.
\newblock
\showISSN{1556-4681}


\bibitem[\protect\citeauthoryear{Peng, Wang, Du, Bhuiyan, Ma, Liu, Wang, Yang,
  Du, Wang, and Yu}{Peng et~al\mbox{.}}{2020}]%
        {PENG2020277}
\bibfield{author}{\bibinfo{person}{Hao Peng}, \bibinfo{person}{Hongfei Wang},
  \bibinfo{person}{Bowen Du}, \bibinfo{person}{Md~Zakirul~Alam Bhuiyan},
  \bibinfo{person}{Hongyuan Ma}, \bibinfo{person}{Jianwei Liu},
  \bibinfo{person}{Lihong Wang}, \bibinfo{person}{Zeyu Yang},
  \bibinfo{person}{Linfeng Du}, \bibinfo{person}{Senzhang Wang}, {and}
  \bibinfo{person}{Philip~S. Yu}.} \bibinfo{year}{2020}\natexlab{}.
\newblock \showarticletitle{Spatial temporal incidence dynamic graph neural
  networks for traffic flow forecasting}.
\newblock \bibinfo{journal}{\emph{Information Sciences}}  \bibinfo{volume}{521}
  (\bibinfo{year}{2020}), \bibinfo{pages}{277--290}.
\newblock
\showISSN{0020-0255}


\bibitem[\protect\citeauthoryear{Peng, Zhang, Dou, Yang, Zhang, and Yu}{Peng
  et~al\mbox{.}}{2021b}]%
        {peng2021reinforced}
\bibfield{author}{\bibinfo{person}{Hao Peng}, \bibinfo{person}{Ruitong Zhang},
  \bibinfo{person}{Yingtong Dou}, \bibinfo{person}{Renyu Yang},
  \bibinfo{person}{Jingyi Zhang}, {and} \bibinfo{person}{Philip~S Yu}.}
  \bibinfo{year}{2021}\natexlab{b}.
\newblock \showarticletitle{Reinforced neighborhood selection guided
  multi-relational graph neural networks}.
\newblock \bibinfo{journal}{\emph{ACM Transactions on Information Systems
  (TOIS)}} \bibinfo{volume}{40}, \bibinfo{number}{4} (\bibinfo{year}{2021}),
  \bibinfo{pages}{1--46}.
\newblock


\bibitem[\protect\citeauthoryear{Peng, Zhang, Li, Cao, Pan, and Yu}{Peng
  et~al\mbox{.}}{2022}]%
        {peng2022reinforced}
\bibfield{author}{\bibinfo{person}{Hao Peng}, \bibinfo{person}{Ruitong Zhang},
  \bibinfo{person}{Shaoning Li}, \bibinfo{person}{Yuwei Cao},
  \bibinfo{person}{Shirui Pan}, {and} \bibinfo{person}{Philip Yu}.}
  \bibinfo{year}{2022}\natexlab{}.
\newblock \showarticletitle{Reinforced, incremental and cross-lingual event
  detection from social messages}.
\newblock \bibinfo{journal}{\emph{IEEE Transactions on Pattern Analysis and
  Machine Intelligence}} (\bibinfo{year}{2022}).
\newblock


\bibitem[\protect\citeauthoryear{Perozzi, Al-Rfou, and Skiena}{Perozzi
  et~al\mbox{.}}{2014}]%
        {perozzi2014deepwalk}
\bibfield{author}{\bibinfo{person}{Bryan Perozzi}, \bibinfo{person}{Rami
  Al-Rfou}, {and} \bibinfo{person}{Steven Skiena}.}
  \bibinfo{year}{2014}\natexlab{}.
\newblock \showarticletitle{Deepwalk: Online learning of social
  representations}. In \bibinfo{booktitle}{\emph{Proceedings of the 20th ACM
  SIGKDD international conference on Knowledge discovery and data mining}}.
  ACM, \bibinfo{pages}{701--710}.
\newblock


\bibitem[\protect\citeauthoryear{Qiu, Tang, Liu, Gong, Zhang, Zhang, and
  Xue}{Qiu et~al\mbox{.}}{2016}]%
        {qiu2016modeling}
\bibfield{author}{\bibinfo{person}{Jiezhong Qiu}, \bibinfo{person}{Jie Tang},
  \bibinfo{person}{Tracy~Xiao Liu}, \bibinfo{person}{Jie Gong},
  \bibinfo{person}{Chenhui Zhang}, \bibinfo{person}{Qian Zhang}, {and}
  \bibinfo{person}{Yufei Xue}.} \bibinfo{year}{2016}\natexlab{}.
\newblock \showarticletitle{Modeling and predicting learning behavior in
  MOOCs}. In \bibinfo{booktitle}{\emph{Proceedings of the ninth ACM
  international conference on web search and data mining}}. ACM,
  \bibinfo{pages}{93--102}.
\newblock


\bibitem[\protect\citeauthoryear{Rabahallah, Mahdaoui, and Azouaou}{Rabahallah
  et~al\mbox{.}}{2018}]%
        {26}
\bibfield{author}{\bibinfo{person}{Kahina Rabahallah}, \bibinfo{person}{Latifa
  Mahdaoui}, {and} \bibinfo{person}{Fai{\c{c}}al Azouaou}.}
  \bibinfo{year}{2018}\natexlab{}.
\newblock \showarticletitle{MOOCs Recommender System using Ontology and
  Memory-based Collaborative Filtering}. In \bibinfo{booktitle}{\emph{ICEIS
  (1)}}. \bibinfo{pages}{635--641}.
\newblock


\bibitem[\protect\citeauthoryear{Rendle}{Rendle}{2012}]%
        {steffen2012fm}
\bibfield{author}{\bibinfo{person}{Steffen Rendle}.}
  \bibinfo{year}{2012}\natexlab{}.
\newblock \showarticletitle{Factorization Machines with libFM}.
\newblock \bibinfo{journal}{\emph{{ACM} {TIST}}} \bibinfo{volume}{3},
  \bibinfo{number}{3} (\bibinfo{year}{2012}), \bibinfo{pages}{57:1--57:22}.
\newblock


\bibitem[\protect\citeauthoryear{Rendle, Freudenthaler, Gantner, and
  Schmidt-Thieme}{Rendle et~al\mbox{.}}{2009}]%
        {steffen2012bpr}
\bibfield{author}{\bibinfo{person}{Steffen Rendle}, \bibinfo{person}{Christoph
  Freudenthaler}, \bibinfo{person}{Zeno Gantner}, {and} \bibinfo{person}{Lars
  Schmidt-Thieme}.} \bibinfo{year}{2009}\natexlab{}.
\newblock \showarticletitle{BPR: Bayesian Personalized Ranking from Implicit
  Feedback} \emph{(\bibinfo{series}{UAI '09})}. \bibinfo{publisher}{AUAI
  Press}, \bibinfo{address}{Arlington, Virginia, USA},
  \bibinfo{pages}{452–461}.
\newblock
\showISBNx{9780974903958}


\bibitem[\protect\citeauthoryear{Shen, Deng, Ray, and Jin}{Shen
  et~al\mbox{.}}{2018}]%
        {shen2018interactive}
\bibfield{author}{\bibinfo{person}{Yilin Shen}, \bibinfo{person}{Yue Deng},
  \bibinfo{person}{Avik Ray}, {and} \bibinfo{person}{Hongxia Jin}.}
  \bibinfo{year}{2018}\natexlab{}.
\newblock \showarticletitle{Interactive recommendation via deep neural memory
  augmented contextual bandits}. In \bibinfo{booktitle}{\emph{Proceedings of
  the 12th ACM Conference on Recommender Systems}}. \bibinfo{pages}{122--130}.
\newblock


\bibitem[\protect\citeauthoryear{Shi, Hu, Zhao, and Yu}{Shi
  et~al\mbox{.}}{2019}]%
        {shi2019heterogeneous}
\bibfield{author}{\bibinfo{person}{Chuan Shi}, \bibinfo{person}{Binbin Hu},
  \bibinfo{person}{Wayne~Xin Zhao}, {and} \bibinfo{person}{Philip~S. Yu}.}
  \bibinfo{year}{2019}\natexlab{}.
\newblock \showarticletitle{Heterogeneous information network embedding for
  recommendation}.
\newblock \bibinfo{journal}{\emph{IEEE Transactions on Knowledge and Data
  Engineering}} \bibinfo{volume}{31}, \bibinfo{number}{2}
  (\bibinfo{year}{2019}), \bibinfo{pages}{357--370}.
\newblock


\bibitem[\protect\citeauthoryear{Silver, Newnham, Barker, Weller, and
  McFall}{Silver et~al\mbox{.}}{2013}]%
        {silver2013concurrent}
\bibfield{author}{\bibinfo{person}{David Silver}, \bibinfo{person}{Leonard
  Newnham}, \bibinfo{person}{David Barker}, \bibinfo{person}{Suzanne Weller},
  {and} \bibinfo{person}{Jason McFall}.} \bibinfo{year}{2013}\natexlab{}.
\newblock \showarticletitle{Concurrent reinforcement learning from customer
  interactions}. In \bibinfo{booktitle}{\emph{International conference on
  machine learning}}. PMLR, \bibinfo{pages}{924--932}.
\newblock


\bibitem[\protect\citeauthoryear{Sun, Liu, Liu, Ren, Gan, and Nie}{Sun
  et~al\mbox{.}}{2020}]%
        {sun2020lara}
\bibfield{author}{\bibinfo{person}{Changfeng Sun}, \bibinfo{person}{Han Liu},
  \bibinfo{person}{Meng Liu}, \bibinfo{person}{Zhaochun Ren},
  \bibinfo{person}{Tian Gan}, {and} \bibinfo{person}{Liqiang Nie}.}
  \bibinfo{year}{2020}\natexlab{}.
\newblock \showarticletitle{LARA: Attribute-to-feature Adversarial Learning for
  New-item Recommendation}. In \bibinfo{booktitle}{\emph{Proceedings of the
  13th International Conference on Web Search and Data Mining}}.
  \bibinfo{pages}{582--590}.
\newblock


\bibitem[\protect\citeauthoryear{Sun, Yin, Liu, Chen, Cao, Shao, and
  Viet~Hung}{Sun et~al\mbox{.}}{2021}]%
        {sun2021heterogeneous}
\bibfield{author}{\bibinfo{person}{Xiangguo Sun}, \bibinfo{person}{Hongzhi
  Yin}, \bibinfo{person}{Bo Liu}, \bibinfo{person}{Hongxu Chen},
  \bibinfo{person}{Jiuxin Cao}, \bibinfo{person}{Yingxia Shao}, {and}
  \bibinfo{person}{Nguyen~Quoc Viet~Hung}.} \bibinfo{year}{2021}\natexlab{}.
\newblock \showarticletitle{Heterogeneous hypergraph embedding for graph
  classification}. In \bibinfo{booktitle}{\emph{Proceedings of the 14th acm
  international conference on web search and data mining}}.
  \bibinfo{pages}{725--733}.
\newblock


\bibitem[\protect\citeauthoryear{Sun and Han}{Sun and Han}{2012}]%
        {sun2012mining}
\bibfield{author}{\bibinfo{person}{Yizhou Sun} {and} \bibinfo{person}{Jiawei
  Han}.} \bibinfo{year}{2012}\natexlab{}.
\newblock \showarticletitle{Mining heterogeneous information networks:
  principles and methodologies}.
\newblock \bibinfo{journal}{\emph{Synthesis Lectures on Data Mining and
  Knowledge Discovery}} \bibinfo{volume}{3}, \bibinfo{number}{2}
  (\bibinfo{year}{2012}), \bibinfo{pages}{1--159}.
\newblock


\bibitem[\protect\citeauthoryear{Sun, Han, Yan, Yu, and Wu}{Sun
  et~al\mbox{.}}{2011}]%
        {sun2011pathsim}
\bibfield{author}{\bibinfo{person}{Yizhou Sun}, \bibinfo{person}{Jiawei Han},
  \bibinfo{person}{Xifeng Yan}, \bibinfo{person}{Philip~S. Yu}, {and}
  \bibinfo{person}{Tianyi Wu}.} \bibinfo{year}{2011}\natexlab{}.
\newblock \showarticletitle{Pathsim: Meta path-based top-k similarity search in
  heterogeneous information networks}.
\newblock \bibinfo{journal}{\emph{Proceedings of the VLDB Endowment}}
  \bibinfo{volume}{4}, \bibinfo{number}{11} (\bibinfo{year}{2011}),
  \bibinfo{pages}{992--1003}.
\newblock


\bibitem[\protect\citeauthoryear{Sutton, McAllester, Singh, and Mansour}{Sutton
  et~al\mbox{.}}{2000}]%
        {sutton2000policy}
\bibfield{author}{\bibinfo{person}{Richard~S Sutton}, \bibinfo{person}{David~A
  McAllester}, \bibinfo{person}{Satinder~P Singh}, {and}
  \bibinfo{person}{Yishay Mansour}.} \bibinfo{year}{2000}\natexlab{}.
\newblock \showarticletitle{Policy gradient methods for reinforcement learning
  with function approximation}. In \bibinfo{booktitle}{\emph{Advances in neural
  information processing systems}}. \bibinfo{pages}{1057--1063}.
\newblock


\bibitem[\protect\citeauthoryear{Van~Hasselt, Guez, and Silver}{Van~Hasselt
  et~al\mbox{.}}{2016}]%
        {van2016deep}
\bibfield{author}{\bibinfo{person}{Hado Van~Hasselt}, \bibinfo{person}{Arthur
  Guez}, {and} \bibinfo{person}{David Silver}.}
  \bibinfo{year}{2016}\natexlab{}.
\newblock \showarticletitle{Deep reinforcement learning with double
  q-learning}. In \bibinfo{booktitle}{\emph{Thirtieth AAAI Conference on
  Artificial Intelligence}}.
\newblock


\bibitem[\protect\citeauthoryear{Veličković, Cucurull, Casanova, Romero,
  Liò, and Bengio}{Veličković et~al\mbox{.}}{2018}]%
        {velickovic2017graph}
\bibfield{author}{\bibinfo{person}{Petar Veličković},
  \bibinfo{person}{Guillem Cucurull}, \bibinfo{person}{Arantxa Casanova},
  \bibinfo{person}{Adriana Romero}, \bibinfo{person}{Pietro Liò}, {and}
  \bibinfo{person}{Yoshua Bengio}.} \bibinfo{year}{2018}\natexlab{}.
\newblock \showarticletitle{Graph Attention Networks}. In
  \bibinfo{booktitle}{\emph{International Conference on Learning
  Representations}}.
\newblock


\bibitem[\protect\citeauthoryear{Vinyals, Babuschkin, Czarnecki, Mathieu,
  Dudzik, Chung, Choi, Powell, Ewalds, Georgiev, et~al\mbox{.}}{Vinyals
  et~al\mbox{.}}{2019}]%
        {vinyals2019grandmaster}
\bibfield{author}{\bibinfo{person}{Oriol Vinyals}, \bibinfo{person}{Igor
  Babuschkin}, \bibinfo{person}{Wojciech~M Czarnecki},
  \bibinfo{person}{Micha{\"e}l Mathieu}, \bibinfo{person}{Andrew Dudzik},
  \bibinfo{person}{Junyoung Chung}, \bibinfo{person}{David~H Choi},
  \bibinfo{person}{Richard Powell}, \bibinfo{person}{Timo Ewalds},
  \bibinfo{person}{Petko Georgiev}, {et~al\mbox{.}}}
  \bibinfo{year}{2019}\natexlab{}.
\newblock \showarticletitle{Grandmaster level in StarCraft II using multi-agent
  reinforcement learning}.
\newblock \bibinfo{journal}{\emph{Nature}} \bibinfo{volume}{575},
  \bibinfo{number}{7782} (\bibinfo{year}{2019}), \bibinfo{pages}{350--354}.
\newblock


\bibitem[\protect\citeauthoryear{Wan and Niu}{Wan and Niu}{2019}]%
        {36}
\bibfield{author}{\bibinfo{person}{Shanshan Wan} {and}
  \bibinfo{person}{Zhendong Niu}.} \bibinfo{year}{2019}\natexlab{}.
\newblock \showarticletitle{A hybrid e-learning recommendation approach based
  on learners’ influence propagation}.
\newblock \bibinfo{journal}{\emph{IEEE Transactions on Knowledge and Data
  Engineering}} \bibinfo{volume}{32}, \bibinfo{number}{5}
  (\bibinfo{year}{2019}), \bibinfo{pages}{827--840}.
\newblock


\bibitem[\protect\citeauthoryear{Wang, Ji, Shi, Wang, Ye, Cui, and Yu}{Wang
  et~al\mbox{.}}{2019}]%
        {DBLP:conf/www/WangJSWYCY19}
\bibfield{author}{\bibinfo{person}{Xiao Wang}, \bibinfo{person}{Houye Ji},
  \bibinfo{person}{Chuan Shi}, \bibinfo{person}{Bai Wang},
  \bibinfo{person}{Yanfang Ye}, \bibinfo{person}{Peng Cui}, {and}
  \bibinfo{person}{Philip~S. Yu}.} \bibinfo{year}{2019}\natexlab{}.
\newblock \showarticletitle{Heterogeneous Graph Attention Network}. In
  \bibinfo{booktitle}{\emph{The World Wide Web Conference, {WWW} 2019, San
  Francisco, CA, USA, May 13-17, 2019}},
  \bibfield{editor}{\bibinfo{person}{Ling Liu}, \bibinfo{person}{Ryen~W.
  White}, \bibinfo{person}{Amin Mantrach}, \bibinfo{person}{Fabrizio
  Silvestri}, \bibinfo{person}{Julian~J. McAuley}, \bibinfo{person}{Ricardo
  Baeza{-}Yates}, {and} \bibinfo{person}{Leila Zia}} (Eds.).
  \bibinfo{publisher}{{ACM}}, \bibinfo{pages}{2022--2032}.
\newblock


\bibitem[\protect\citeauthoryear{Wang, Liu, Han, and Shi}{Wang
  et~al\mbox{.}}{2021}]%
        {wang2021self}
\bibfield{author}{\bibinfo{person}{Xiao Wang}, \bibinfo{person}{Nian Liu},
  \bibinfo{person}{Hui Han}, {and} \bibinfo{person}{Chuan Shi}.}
  \bibinfo{year}{2021}\natexlab{}.
\newblock \showarticletitle{Self-supervised heterogeneous graph neural network
  with co-contrastive learning}. In \bibinfo{booktitle}{\emph{Proceedings of
  the 27th ACM SIGKDD Conference on Knowledge Discovery \& Data Mining}}.
  \bibinfo{pages}{1726--1736}.
\newblock


\bibitem[\protect\citeauthoryear{Welling and Kipf}{Welling and Kipf}{2016}]%
        {welling2016semi}
\bibfield{author}{\bibinfo{person}{Max Welling} {and} \bibinfo{person}{Thomas~N
  Kipf}.} \bibinfo{year}{2016}\natexlab{}.
\newblock \showarticletitle{Semi-supervised classification with graph
  convolutional networks}. In \bibinfo{booktitle}{\emph{J. International
  Conference on Learning Representations (ICLR 2017)}}.
\newblock


\bibitem[\protect\citeauthoryear{Williams}{Williams}{1992}]%
        {williams1992simple}
\bibfield{author}{\bibinfo{person}{Ronald~J Williams}.}
  \bibinfo{year}{1992}\natexlab{}.
\newblock \showarticletitle{Simple statistical gradient-following algorithms
  for connectionist reinforcement learning}.
\newblock \bibinfo{journal}{\emph{Machine learning}} \bibinfo{volume}{8},
  \bibinfo{number}{3-4} (\bibinfo{year}{1992}), \bibinfo{pages}{229--256}.
\newblock


\bibitem[\protect\citeauthoryear{Wu, Sun, Fu, Hong, Wang, and Wang}{Wu
  et~al\mbox{.}}{2019}]%
        {wu2019neural}
\bibfield{author}{\bibinfo{person}{Le Wu}, \bibinfo{person}{Peijie Sun},
  \bibinfo{person}{Yanjie Fu}, \bibinfo{person}{Richang Hong},
  \bibinfo{person}{Xiting Wang}, {and} \bibinfo{person}{Meng Wang}.}
  \bibinfo{year}{2019}\natexlab{}.
\newblock \showarticletitle{A neural influence diffusion model for social
  recommendation}. In \bibinfo{booktitle}{\emph{Proceedings of the 42nd
  international ACM SIGIR conference on research and development in information
  retrieval}}. \bibinfo{pages}{235--244}.
\newblock


\bibitem[\protect\citeauthoryear{Wu, Yang, Zhang, Hong, Fu, and Wang}{Wu
  et~al\mbox{.}}{2020}]%
        {wu2020joint}
\bibfield{author}{\bibinfo{person}{Le Wu}, \bibinfo{person}{Yonghui Yang},
  \bibinfo{person}{Kun Zhang}, \bibinfo{person}{Richang Hong},
  \bibinfo{person}{Yanjie Fu}, {and} \bibinfo{person}{Meng Wang}.}
  \bibinfo{year}{2020}\natexlab{}.
\newblock \showarticletitle{Joint item recommendation and attribute inference:
  An adaptive graph convolutional network approach}. In
  \bibinfo{booktitle}{\emph{Proceedings of the 43rd International ACM SIGIR
  conference on research and development in Information Retrieval}}.
  \bibinfo{pages}{679--688}.
\newblock


\bibitem[\protect\citeauthoryear{Xie, Zhang, Wang, Xia, and Lin}{Xie
  et~al\mbox{.}}{2021}]%
        {xie2021hierarchical}
\bibfield{author}{\bibinfo{person}{Ruobing Xie}, \bibinfo{person}{Shaoliang
  Zhang}, \bibinfo{person}{Rui Wang}, \bibinfo{person}{Feng Xia}, {and}
  \bibinfo{person}{Leyu Lin}.} \bibinfo{year}{2021}\natexlab{}.
\newblock \showarticletitle{Hierarchical Reinforcement Learning for Integrated
  Recommendation}. In \bibinfo{booktitle}{\emph{Proceedings of AAAI}}.
\newblock


\bibitem[\protect\citeauthoryear{Xu, Hu, Leskovec, and Jegelka}{Xu
  et~al\mbox{.}}{2019}]%
        {xu2018powerful}
\bibfield{author}{\bibinfo{person}{Keyulu Xu}, \bibinfo{person}{Weihua Hu},
  \bibinfo{person}{Jure Leskovec}, {and} \bibinfo{person}{Stefanie Jegelka}.}
  \bibinfo{year}{2019}\natexlab{}.
\newblock \showarticletitle{How Powerful are Graph Neural Networks?}. In
  \bibinfo{booktitle}{\emph{International Conference on Learning
  Representations}}.
\newblock


\bibitem[\protect\citeauthoryear{Yin, Wang, Zheng, Li, and Zhou}{Yin
  et~al\mbox{.}}{2020}]%
        {23}
\bibfield{author}{\bibinfo{person}{Hongzhi Yin}, \bibinfo{person}{Qinyong
  Wang}, \bibinfo{person}{Kai Zheng}, \bibinfo{person}{Zhixu Li}, {and}
  \bibinfo{person}{Xiaofang Zhou}.} \bibinfo{year}{2020}\natexlab{}.
\newblock \showarticletitle{Overcoming data sparsity in group recommendation}.
\newblock \bibinfo{journal}{\emph{IEEE Transactions on Knowledge and Data
  Engineering}} (\bibinfo{year}{2020}).
\newblock


\bibitem[\protect\citeauthoryear{Yu, Luo, Xiao, Zhong, Wang, Feng, Luo, Wang,
  Hou, Li, et~al\mbox{.}}{Yu et~al\mbox{.}}{2020b}]%
        {xz1}
\bibfield{author}{\bibinfo{person}{Jifan Yu}, \bibinfo{person}{Gan Luo},
  \bibinfo{person}{Tong Xiao}, \bibinfo{person}{Qingyang Zhong},
  \bibinfo{person}{Yuquan Wang}, \bibinfo{person}{Wenzheng Feng},
  \bibinfo{person}{Junyi Luo}, \bibinfo{person}{Chenyu Wang},
  \bibinfo{person}{Lei Hou}, \bibinfo{person}{Juanzi Li}, {et~al\mbox{.}}}
  \bibinfo{year}{2020}\natexlab{b}.
\newblock \showarticletitle{MOOCCube: a large-scale data repository for NLP
  applications in MOOCs}. In \bibinfo{booktitle}{\emph{Proceedings of the 58th
  Annual Meeting of the Association for Computational Linguistics}}.
  \bibinfo{pages}{3135--3142}.
\newblock


\bibitem[\protect\citeauthoryear{Yu, Wang, Luo, Hou, Li, Tang, and Liu}{Yu
  et~al\mbox{.}}{2019b}]%
        {JifanACL19}
\bibfield{author}{\bibinfo{person}{Jifan Yu}, \bibinfo{person}{Chenyu Wang},
  \bibinfo{person}{Gan Luo}, \bibinfo{person}{Lei Hou}, \bibinfo{person}{Juanzi
  Li}, \bibinfo{person}{Jie Tang}, {and} \bibinfo{person}{Zhiyuan Liu}.}
  \bibinfo{year}{2019}\natexlab{b}.
\newblock \showarticletitle{Course Concept Expansion in MOOCs with External
  Knowledge and Interactive Game}. In \bibinfo{booktitle}{\emph{Proceedings of
  the 57th Annual Meeting of the Association for Computational Linguistics}}.
  ACL, \bibinfo{pages}{4292--4302}.
\newblock


\bibitem[\protect\citeauthoryear{Yu, Wang, Zhong, Luo, Mao, Sun, Feng, Xu, Cao,
  Zeng, et~al\mbox{.}}{Yu et~al\mbox{.}}{2021}]%
        {xz5}
\bibfield{author}{\bibinfo{person}{Jifan Yu}, \bibinfo{person}{Yuquan Wang},
  \bibinfo{person}{Qingyang Zhong}, \bibinfo{person}{Gan Luo},
  \bibinfo{person}{Yiming Mao}, \bibinfo{person}{Kai Sun},
  \bibinfo{person}{Wenzheng Feng}, \bibinfo{person}{Wei Xu},
  \bibinfo{person}{Shulin Cao}, \bibinfo{person}{Kaisheng Zeng},
  {et~al\mbox{.}}} \bibinfo{year}{2021}\natexlab{}.
\newblock \showarticletitle{MOOCCubeX: A Large Knowledge-centered Repository
  for Adaptive Learning in MOOCs}. In \bibinfo{booktitle}{\emph{Proceedings of
  the 30th ACM International Conference on Information \& Knowledge
  Management}}. \bibinfo{pages}{4643--4652}.
\newblock


\bibitem[\protect\citeauthoryear{Yu, Liu, Ye, Cheng, Chen, and Ma}{Yu
  et~al\mbox{.}}{2020a}]%
        {17}
\bibfield{author}{\bibinfo{person}{Runlong Yu}, \bibinfo{person}{Qi Liu},
  \bibinfo{person}{Yuyang Ye}, \bibinfo{person}{Mingyue Cheng},
  \bibinfo{person}{Enhong Chen}, {and} \bibinfo{person}{Jianhui Ma}.}
  \bibinfo{year}{2020}\natexlab{a}.
\newblock \showarticletitle{Collaborative list-and-pairwise filtering from
  implicit feedback}.
\newblock \bibinfo{journal}{\emph{IEEE Transactions on Knowledge and Data
  Engineering}} (\bibinfo{year}{2020}).
\newblock


\bibitem[\protect\citeauthoryear{Yu, Shen, Zhang, Zeng, and Jin}{Yu
  et~al\mbox{.}}{2019a}]%
        {yu2019vision}
\bibfield{author}{\bibinfo{person}{Tong Yu}, \bibinfo{person}{Yilin Shen},
  \bibinfo{person}{Ruiyi Zhang}, \bibinfo{person}{Xiangyu Zeng}, {and}
  \bibinfo{person}{Hongxia Jin}.} \bibinfo{year}{2019}\natexlab{a}.
\newblock \showarticletitle{Vision-language recommendation via attribute
  augmented multimodal reinforcement learning}. In
  \bibinfo{booktitle}{\emph{Proceedings of the 27th ACM International
  Conference on Multimedia}}. \bibinfo{pages}{39--47}.
\newblock


\bibitem[\protect\citeauthoryear{Zhang, Hao, Chen, Li, Chen, and Sund}{Zhang
  et~al\mbox{.}}{2019}]%
        {zhang2019hierarchical}
\bibfield{author}{\bibinfo{person}{Jing Zhang}, \bibinfo{person}{Bowen Hao},
  \bibinfo{person}{Bo Chen}, \bibinfo{person}{Cuiping Li},
  \bibinfo{person}{Hong Chen}, {and} \bibinfo{person}{Jimeng Sund}.}
  \bibinfo{year}{2019}\natexlab{}.
\newblock \showarticletitle{Hierarchical Reinforcement Learning for Course
  Recommendation in MOOCs}.
\newblock \bibinfo{journal}{\emph{Psychology}} \bibinfo{volume}{5},
  \bibinfo{number}{4.64} (\bibinfo{year}{2019}), \bibinfo{pages}{5--65}.
\newblock


\bibitem[\protect\citeauthoryear{Zhang, Chen, Ming, Cui, Yin, and Xu}{Zhang
  et~al\mbox{.}}{2021}]%
        {zhang2021we}
\bibfield{author}{\bibinfo{person}{Sixiao Zhang}, \bibinfo{person}{Hongxu
  Chen}, \bibinfo{person}{Xiao Ming}, \bibinfo{person}{Lizhen Cui},
  \bibinfo{person}{Hongzhi Yin}, {and} \bibinfo{person}{Guandong Xu}.}
  \bibinfo{year}{2021}\natexlab{}.
\newblock \showarticletitle{Where Are We in Embedding Spaces?}
  \emph{(\bibinfo{series}{KDD '21})}. \bibinfo{publisher}{Association for
  Computing Machinery}, \bibinfo{address}{New York, NY, USA},
  \bibinfo{pages}{2223–2231}.
\newblock
\showISBNx{9781450383325}


\bibitem[\protect\citeauthoryear{Zhang, Chen, Tang, Stewart, and Sun}{Zhang
  et~al\mbox{.}}{2017}]%
        {zhang2017leap}
\bibfield{author}{\bibinfo{person}{Yutao Zhang}, \bibinfo{person}{Robert Chen},
  \bibinfo{person}{Jie Tang}, \bibinfo{person}{Walter~F Stewart}, {and}
  \bibinfo{person}{Jimeng Sun}.} \bibinfo{year}{2017}\natexlab{}.
\newblock \showarticletitle{LEAP: learning to prescribe effective and safe
  treatment combinations for multimorbidity}. In
  \bibinfo{booktitle}{\emph{proceedings of the 23rd ACM SIGKDD international
  conference on knowledge Discovery and data Mining}}.
  \bibinfo{pages}{1315--1324}.
\newblock


\bibitem[\protect\citeauthoryear{Zhao, Zhang, Zhang, Zheng, Bao, and Yan}{Zhao
  et~al\mbox{.}}{2020b}]%
        {zhao2020mahrl}
\bibfield{author}{\bibinfo{person}{Dongyang Zhao}, \bibinfo{person}{Liang
  Zhang}, \bibinfo{person}{Bo Zhang}, \bibinfo{person}{Lizhou Zheng},
  \bibinfo{person}{Yongjun Bao}, {and} \bibinfo{person}{Weipeng Yan}.}
  \bibinfo{year}{2020}\natexlab{b}.
\newblock \showarticletitle{MaHRL: Multi-goals Abstraction Based Deep
  Hierarchical Reinforcement Learning for Recommendations}. In
  \bibinfo{booktitle}{\emph{Proceedings of the 43rd International ACM SIGIR
  Conference on Research and Development in Information Retrieval}}.
  \bibinfo{pages}{871--880}.
\newblock


\bibitem[\protect\citeauthoryear{Zhao, Wang, Yang, Ye, Zhao, Chen, and
  Shen}{Zhao et~al\mbox{.}}{2019}]%
        {zhao2019leveraging}
\bibfield{author}{\bibinfo{person}{Wei Zhao}, \bibinfo{person}{Benyou Wang},
  \bibinfo{person}{Min Yang}, \bibinfo{person}{Jianbo Ye},
  \bibinfo{person}{Zhou Zhao}, \bibinfo{person}{Xiaojun Chen}, {and}
  \bibinfo{person}{Ying Shen}.} \bibinfo{year}{2019}\natexlab{}.
\newblock \showarticletitle{Leveraging long and short-term information in
  content-aware movie recommendation via adversarial training}.
\newblock \bibinfo{journal}{\emph{IEEE transactions on cybernetics}}
  \bibinfo{volume}{50}, \bibinfo{number}{11} (\bibinfo{year}{2019}),
  \bibinfo{pages}{4680--4693}.
\newblock


\bibitem[\protect\citeauthoryear{Zhao, Xia, Zhang, Ding, Yin, and Tang}{Zhao
  et~al\mbox{.}}{2018}]%
        {zhao2018deep}
\bibfield{author}{\bibinfo{person}{Xiangyu Zhao}, \bibinfo{person}{Long Xia},
  \bibinfo{person}{Liang Zhang}, \bibinfo{person}{Zhuoye Ding},
  \bibinfo{person}{Dawei Yin}, {and} \bibinfo{person}{Jiliang Tang}.}
  \bibinfo{year}{2018}\natexlab{}.
\newblock \showarticletitle{Deep reinforcement learning for page-wise
  recommendations}. In \bibinfo{booktitle}{\emph{Proceedings of the 12th ACM
  Conference on Recommender Systems}}. ACM, \bibinfo{pages}{95--103}.
\newblock


\bibitem[\protect\citeauthoryear{Zhao, Xia, Zou, Liu, Yin, and Tang}{Zhao
  et~al\mbox{.}}{2020a}]%
        {zhao2020whole}
\bibfield{author}{\bibinfo{person}{Xiangyu Zhao}, \bibinfo{person}{Long Xia},
  \bibinfo{person}{Lixin Zou}, \bibinfo{person}{Hui Liu},
  \bibinfo{person}{Dawei Yin}, {and} \bibinfo{person}{Jiliang Tang}.}
  \bibinfo{year}{2020}\natexlab{a}.
\newblock \showarticletitle{Whole-Chain Recommendations}. In
  \bibinfo{booktitle}{\emph{Proceedings of the 29th ACM International
  Conference on Information \& Knowledge Management}}.
  \bibinfo{pages}{1883--1891}.
\newblock


\bibitem[\protect\citeauthoryear{Zheng, Zhang, Zheng, Xiang, Yuan, Xie, and
  Li}{Zheng et~al\mbox{.}}{2018}]%
        {zheng2018drn}
\bibfield{author}{\bibinfo{person}{Guanjie Zheng}, \bibinfo{person}{Fuzheng
  Zhang}, \bibinfo{person}{Zihan Zheng}, \bibinfo{person}{Yang Xiang},
  \bibinfo{person}{Nicholas~Jing Yuan}, \bibinfo{person}{Xing Xie}, {and}
  \bibinfo{person}{Zhenhui Li}.} \bibinfo{year}{2018}\natexlab{}.
\newblock \showarticletitle{DRN: A deep reinforcement learning framework for
  news recommendation}. In \bibinfo{booktitle}{\emph{Proceedings of the 2018
  World Wide Web Conference on World Wide Web}}. International World Wide Web
  Conferences Steering Committee, \bibinfo{pages}{167--176}.
\newblock


\bibitem[\protect\citeauthoryear{Zhou, Dai, Chen, Zhang, Ren, Tang, He, and
  Yu}{Zhou et~al\mbox{.}}{2020}]%
        {zhou2020interactive}
\bibfield{author}{\bibinfo{person}{Sijin Zhou}, \bibinfo{person}{Xinyi Dai},
  \bibinfo{person}{Haokun Chen}, \bibinfo{person}{Weinan Zhang},
  \bibinfo{person}{Kan Ren}, \bibinfo{person}{Ruiming Tang},
  \bibinfo{person}{Xiuqiang He}, {and} \bibinfo{person}{Yong Yu}.}
  \bibinfo{year}{2020}\natexlab{}.
\newblock \showarticletitle{Interactive recommender system via knowledge
  graph-enhanced reinforcement learning}. In
  \bibinfo{booktitle}{\emph{Proceedings of the 43rd International ACM SIGIR
  Conference on Research and Development in Information Retrieval}}.
  \bibinfo{pages}{179--188}.
\newblock


\bibitem[\protect\citeauthoryear{Zhu, Lin, He, Wang, Guan, Liu, and Cai}{Zhu
  et~al\mbox{.}}{2019}]%
        {24}
\bibfield{author}{\bibinfo{person}{Yu Zhu}, \bibinfo{person}{Jinghao Lin},
  \bibinfo{person}{Shibi He}, \bibinfo{person}{Beidou Wang},
  \bibinfo{person}{Ziyu Guan}, \bibinfo{person}{Haifeng Liu}, {and}
  \bibinfo{person}{Deng Cai}.} \bibinfo{year}{2019}\natexlab{}.
\newblock \showarticletitle{Addressing the item cold-start problem by
  attribute-driven active learning}.
\newblock \bibinfo{journal}{\emph{IEEE Transactions on Knowledge and Data
  Engineering}} \bibinfo{volume}{32}, \bibinfo{number}{4}
  (\bibinfo{year}{2019}), \bibinfo{pages}{631--644}.
\newblock


\bibitem[\protect\citeauthoryear{Zou, Xia, Ding, Song, Liu, and Yin}{Zou
  et~al\mbox{.}}{2019}]%
        {zou2019reinforcement}
\bibfield{author}{\bibinfo{person}{Lixin Zou}, \bibinfo{person}{Long Xia},
  \bibinfo{person}{Zhuoye Ding}, \bibinfo{person}{Jiaxing Song},
  \bibinfo{person}{Weidong Liu}, {and} \bibinfo{person}{Dawei Yin}.}
  \bibinfo{year}{2019}\natexlab{}.
\newblock \showarticletitle{Reinforcement Learning to Optimize Long-term User
  Engagement in Recommender Systems}. In \bibinfo{booktitle}{\emph{Proceedings
  of the 25th ACM SIGKDD International Conference on Knowledge Discovery \&
  Data Mining}}. \bibinfo{pages}{2810--2818}.
\newblock


\bibitem[\protect\citeauthoryear{Zou, Xia, Du, Zhang, Bai, Liu, Nie, and
  Yin}{Zou et~al\mbox{.}}{2020}]%
        {zou2020pseudo}
\bibfield{author}{\bibinfo{person}{Lixin Zou}, \bibinfo{person}{Long Xia},
  \bibinfo{person}{Pan Du}, \bibinfo{person}{Zhuo Zhang}, \bibinfo{person}{Ting
  Bai}, \bibinfo{person}{Weidong Liu}, \bibinfo{person}{Jian-Yun Nie}, {and}
  \bibinfo{person}{Dawei Yin}.} \bibinfo{year}{2020}\natexlab{}.
\newblock \showarticletitle{Pseudo Dyna-Q: A reinforcement learning framework
  for interactive recommendation}. In \bibinfo{booktitle}{\emph{Proceedings of
  the 13th International Conference on Web Search and Data Mining}}.
  \bibinfo{pages}{816--824}.
\newblock


\end{thebibliography}

\clearpage

\end{document}